\documentclass[11pt]{article}
\usepackage{amsmath}
\usepackage{amsthm}
\usepackage{amssymb}
\usepackage{mathrsfs}
\usepackage[Symbol]{upgreek}
\usepackage{dsfont}
\usepackage[vcentermath]{youngtab}
\usepackage{textcomp}

\usepackage{latexsym}
\usepackage{graphicx}

\def\half{\frac{1}{2}}\def\qu{\frac{1}{4}}
\def\bsh{\backslash}

\newfont{\bbbold}{msbm10 scaled \magstep1}

\def\bbC{\mbox{\bbbold C}}

\def\bbF{\mbox{\bbbold F}}

\def\cL{{\cal L}}

\def\cN{{\cal N}}
\def\cO{{\cal O}}

\def\cV{{\cal V}}

\newfont{\goth}{eufm10 scaled \magstep1}

\def\ge{\mbox{\goth e}}

\def\gg{\mbox{\goth g}}

\def\gi{\mbox{\goth i}}

\def\gn{\mbox{\goth n}}
\def\go{\mbox{\goth o}}
\def\gp{\mbox{\goth p}}

\def\gs{\mbox{\goth s}}

\def\gu{\mbox{\goth u}}

\def\a{\alpha}\def\adt{\dot \alpha}
\def\b{\beta}\def\bdt{\dot \beta}
\def\c{\gamma}
\def\d{\delta}
\def\ve{\varepsilon}
\def\f{\phi}\def\F{\Phi}\def\vf{\varphi}
\def\h{\eta}

\def\l{\lambda}\def\L{\Lambda}
\def\m{\mu}

\def\s{\sigma}\def\S{\Sigma}

\def\th{\theta}

\def\be{\begin{equation}}\def\ee{\end{equation}}
\def\bea{\begin{eqnarray}}\def\eea{\end{eqnarray}}
\def\barr{\begin{array}}\def\earr{\end{array}}

\def\O{\Omega}
\def\del{\partial}

\def\xz{\times}

\def\nab{\nabla}

\def\hR{\hat{R}}

\def\hi{\hat{\imath}}\def\hj{\hat{\jmath}}
\def\hk{\hat{k}}\def\hu{\widehat{u}}

\def\hR{\widehat{R}}\def\hS{\widehat{S}}

\let\la=\label

\def\nn{\nonumber}
\def\bd{\begin{document}}
\def\ed{\end{document}}
\def\ba{\begin{array}}
\def\ea{\end{array}}
\def\bea{\begin{eqnarray}}
\def\eea{\end{eqnarray}}
\def\ft#1#2{{\frac{\scriptstyle #1}{\scriptstyle #2}}}
\def\fft#1#2{\frac{#1}{#2}}
\def\sst#1{{\scriptscriptstyle #1}}
\def\oneone{\rlap 1\mkern4mu{\rm l}}

\newcommand{\eq}[1]{(\ref{#1})}
\newcommand{\w}[1]{\\[0.#1cm]}
\def\eqs#1#2{(\ref{#1}-\ref{#2})}
\def\det{{\rm det\,}}
\def\tr{{\rm tr}}

\newcommand{\hoch}[1]{$\, ^{#1}$}
\newcommand{\imperial}{\it\small Theoretical Physics Group, Imperial College London\\ Prince Consort Road, London SW7 2AZ, UK}
\newcommand{\kings}
{\it\small Department of Mathematics, King's College, University of London\\ Strand, London WC2R 2LS, UK}
\newcommand{\uu}
{\it\small Department of Theoretical Physics, Uppsala, Sweden}
\newcommand{\hip}
{\it\small HIP-Helsinki Institute of Physics, P.O. Box 64 FIN-00014
University of Helsinki, Suomi-Finland}
\newcommand{\stock}
{\it\small Department of Theoretical Physics, Stockholm, Sweden}
\newcommand{\golm}
{\it\small AEI, Max Planck Institut f\"ur Gravitationsphysik\\ Am M\"{u}hlenberg 1, D-14476 Potsdam, Germany}
\makeatletter
\renewcommand\theequation{\thesection.\arabic{equation}}
\@addtoreset{equation}{section} \makeatother

\newcommand{\sa}{/ \hspace{-1.2ex}}
\newcommand{\saa}{/ \hspace{-1.4ex}}
\newcommand{\saaa}{\, / \hspace{-1.6ex}}
\newcommand{\Scal}[1]{\Bigl ({#1} \Bigr )}
\newcommand{\scal}[1]{\bigl ({#1} \bigr )}

\newcommand{\CR}{\nonumber \\*}

\newcommand{\trace}{\hbox {tr}~}
\newcommand{\traceS}{\hbox {tr}_{\scriptscriptstyle \mathfrak{S}}~}

\DeclareMathAlphabet{\mathpzc}{OT1}{pzc}{m}{it}
\def\BRST{\,\mathpzc{s}\,}
\def\aBRST{{\scriptstyle (\mathpzc{s})}}
\def\q{{{\scriptscriptstyle (Q)}}}
\def\qs{{\scriptscriptstyle (Q\mathpzc{s})}}
\def\Qsla{{\mathcal{S}_{\q}}}
\def\Slav{{\mathcal{S}_\aBRST}}
\def\epsilonb{{\overline{\epsilon}}}
\def\bulletup{{\scriptstyle \bullet}}

\newcommand{\gra}[2]{{\scriptscriptstyle (#1 , #2 )}}
\newcommand{\ord}[1]{{\scriptscriptstyle (#1)}}

\def\cL{{\cal L}}
\def\cN{\mathcal{N}}
\def\cO{\mathcal{O}}

\def\ie{{\it i.e.}\ }
\def\eg{{\it e.g.}\ }

\newcommand{\sfrac}[2]{{\scriptstyle \frac{#1}{#2}}}
\newcommand{\stfrac}[2]{{\scriptscriptstyle \frac{#1}{#2}}}

 \def\balpha{{\overline{\alpha}}}
 \def\bbeta{{\overline{\beta}}}
 \def\bgamma{{\overline{\gamma}}}
 \def\bdelta{{\overline{\delta}}}
 \def\bepsilon{{\overline{\epsilon}}}
 \def\bvarepsilon{{\overline{\varepsilon}}}
 \def\bzeta{{\overline{\zeta}}}
 \def\bareta{{\overline{\eta}}}
 \def\btheta{{\overline{\theta}}}
 \def\bvartheta{{\overline{\vartheta}}}
 \def\biota{{\overline{\iota}}}
 \def\bkappa{{\overline{\kappa}}}
 \def\blambda{{\overline{\lambda}}}
 \def\bmu{{\overline{\mu}}}
 \def\bnu{{\overline{\nu}}}
 \def\bxi{{\overline{\xi}}}
 \def\bpi{{\overline{\pi}}}
 \def\brho{{\overline{\rho}}}
 \def\bvarrho{{\overline{\varrho}}}
 \def\bsigma{{\overline{\sigma}}}
 \def\bvarsigma{{\overline{\varsigma}}}
 \def\btau{{\overline{\tau}}}
 \def\bphi{{\overline{\phi}}}
 \def\bvarphi{{\overline{\varphi}}}
 \def\bchi{{\overline{\chi}}}
 \def\bpsi{{\overline{\psi}}}
 \def\bomega{{\overline{\omega}}}

\def\thalf{{\textrm{\tiny\textonehalf}}}
\def\tquarter{{\textrm{\tiny\textonequarter}}}
\def\Ko{{\scriptscriptstyle K}}
\def\tKo{\scriptscriptstyle k }

\newcommand{\auth}{\large G.\ Bossard\footnote{email: bossard@aei.mpg.de}, P.S.\ Howe\footnote{email: paul.howe@kcl.ac.uk} and K.S.\ Stelle\footnote{email: k.stelle@imperial.ac.uk}}

\begin{document}

\renewcommand{\thefootnote}{\fnsymbol{footnote}}

\null
\begin{flushright}
{\small AEI-2009-007}\\
{\small kcl.mth-09-01}\\
{\small Imperial/TP/09/KSS/01}
\vskip 1.5 cm
\end{flushright}

\begin{center}
{\Large{\bf The ultra-violet question in maximally supersymmetric field theories}}
\vspace{.75cm}

\auth

\vspace{.5cm}

$^{\ast}$\golm
\\
\vskip 1 em
$^{\dagger}$\kings
\\
\vskip 1 em
$^{\ddagger}$\imperial

\vspace{1cm}

%%%%%%%%%%%%%%%%%%%%%%%%%
{\bf Abstract}
%%%%%%%%%%%%%%%%%%%%%%%%%%

\end{center}

We discuss various approaches to the problem of determining which supersymmetric invariants are permitted as counterterms in maximally supersymmetric super Yang--Mills and supergravity theories in various dimensions. We review the superspace non-renormalisation theorems based on conventional, light-cone, harmonic and certain non-Lorentz covariant superspaces, and we write down explicitly the relevant invariants. While the first two types of superspace admit the possibility of one-half BPS counterterms, of the form $F^4$ and $R^4$ respectively, the last two do not. This suggests that  UV divergences begin with one-quarter BPS counterterms, i.e. $d^2 F^4$ and $d^4 R^4$, and this is supported by an entirely different approach based on algebraic renormalisation. The algebraic formalism is discussed for non-renormalisable theories and it is shown how the allowable supersymmetric counterterms can be determined via cohomological methods. These results are in agreement with all the explicit computations that have been carried out to date. In particular, they suggest that maximal supergravity is likely to diverge at four loops in $D=5$ and at five loops in $D=4$, unless other infinity suppression mechanisms not involving supersymmetry or gauge invariance are at work.

\vspace{1cm}

\null

\renewcommand{\thefootnote}{\arabic{footnote}}
\setcounter{footnote}{0}

\pagebreak
\tableofcontents
\setcounter{page}{1}

%%%%%%%%%%%%%%%%%%%%%%%%%%%%%%%%%%%%%%%%%%%%%%%%%%%%%%%%%%%%%%%

\section{Introduction}

%%%%%%%%%%%%%%%%%%%%%%%%%%%%%%%%%%%%%%%%%%%%%%%%%%%%%%%%%%%%%%%%%%

 The derivation of an acceptable quantum theory of gravity remains one of the prime challenges facing fundamental theoretical physics. A basic problem in formulating such a theory was already recognised in the earliest approaches to the subject in the 1930s: the dimensional character of Newton's constant gives rise to ultraviolet divergent quantum correction integrals. In the 1970s, this was confirmed explicitly in the first Feynman diagram calculations of the radiative corrections to systems containing gravity plus matter \cite{'tHooft:1974bx}. The time lag between the general perception of the UV divergence problem and its first concrete demonstration was due to the complexity of Feynman diagram calculations involving gravity. The necessary techniques were an outgrowth of the long struggle to control, in a Lorentz-covariant manner, the quantisation of non-abelian Yang--Mills theories in the Standard Model of weak and electromagnetic interactions and in quantum chromodynamics.

With the advent of supergravity \cite{Freedman:1976xh,Deser:1976eh} in the mid 1970s, hopes rose that the specific combinations of quantum fields in supergravity theories might possibly tame the gravitational UV divergence problem. Indeed, it turns out that all irreducible supergravity theories in four-dimensional spacetime, {\it i.e.}\ theories in which all fields are irreducibly linked to gravity by supersymmetry transformations, have remarkable cancellations in Feynman diagrams with one or two internal loops \cite{Grisaru:1976nn}.

There is a sequence of such irreducible (or ``pure'') supergravity models, characterised by the number $\cN$ of local ({\it i.e.}\ spacetime-dependent) spinor parameters. In four-dimensional spacetime, minimal, or $\cN=1$, supergravity thus has four supersymmetries corresponding to the components of a single Majorana spinor transformation parameter. The maximal possible supergravity \cite{Cremmer:1979up} in four-dimensional spacetime has $\cN=8$ spinor parameters, {\it i.e.}\ 32 independent supersymmetries.

The hopes for ``miraculous'' UV divergence cancellations in supergravity were subsequently dampened by the realisation that the divergence-killing powers of supersymmetry most likely do not extend beyond the two-loop order for pure supergravity theories with $\cN=1,2$ supersymmetry \cite{Deser:1977nt,Deser:1978br}. The extension of this result to all $\cN$ had to await the development of the superspace formulation of the $\cN=8$ theory \cite{Brink:1979nt} with the aid of which it was easy to construct linearised  (and indeed fully non-linear) counterterms starting at the seven-loop order \cite{Howe:1980th,Kallosh:1980fi}
 although it proved somewhat more tricky to find a three-loop invariant \cite{Kallosh:1980fi,superactions}. The three-loop invariant, for all $\cN$, is quartic in  curvatures, and has a purely gravitational part given by the square of the Bel-Robinson tensor \cite{Deser:1977nt}.

The flowering of superstring theory in the 1980s and 1990s, in which the UV divergence problems of gravity are cured by a completely different mechanism which involves replacement of the basic field-theory point-particle states by extended relativistic object states, pushed the UV divergence properties of supergravity out of the limelight, leaving the supergravity UV problem in an unclear state.

Nonetheless, among some researchers a faint hope persisted that at least the maximal $\cN=8$ supergravity might have special UV properties. This hope was bolstered by the fact that the maximally supersymmetric Yang--Mills theory, which has $\cN=4$, {\it i.e.}\ 16-component supersymmetry, is completely free of ultraviolet divergences in four-dimensional spacetime \cite{Howe:1982tm,Mandelstam:1982cb,Brink:1982wv}. This was the first interacting UV-finite theory in four spacetime dimensions.

It is this possibility of ``miraculous'' UV divergence cancellations in maximal supergravity that has now been confirmed in a remarkable three-loop calculation by Z. Bern et al. \cite{Bern:2007hh}.  Performing such calculations at high loop orders requires a departure from textbook Feynman-diagram methods because the standard approaches can produce astronomical numbers of terms. Instead of following the standard propagator and vertex methods for the supergravity calculations, Bern et al.\ used another technique which goes back to Feynman: loop calculations can be performed using the unitarity properties of the quantum S-matrix. These involve cutting rules that reduce higher-loop diagrams to sums of products of leading-order ``tree'' diagrams without internal loops. This use of unitarity is an outgrowth of the optical theorem in quantum mechanics for the imaginary part of the S-matrix.

In order to obtain information about the real part of the S-matrix, an additional necessary element in the unitarity-based technique is the use of dimensional regularization to render UV divergent diagrams finite. In dimensional regularization, the dimensionality of spacetime is changed from $4$ to $4-\epsilon$, where $\epsilon$ is a small adjustable parameter. Traditional Feynman diagram calculations also often use dimensional regularization, but normally one just focuses on the leading $1/\epsilon$ poles in order to carry out a renormalization program. In the unitarity-based approach, all orders in $\epsilon$ need to be retained. This gives rise to logarithms in which real and imaginary contributions are related.

In the maximal $\cN=8$ supergravity theory, the complexity of the quantum amplitudes factorizes, with details involving the various field types occurring on the external legs of an amplitude multiplying a much simpler set of scalar-field Feynman diagrams. It is to the latter that the unitarity-based methods may be applied. Earlier applications \cite{Bern:1998ug} of the cutting-rule unitarity methods based on iterations of two-particle cuts gave an expectation that one might have cancellations for $D<10/L + 2$, where $D$ is the spacetime dimension and $L$ is the number of Feynman diagram loops (for $L>1$). Already, this gave an expectation that $D=4$ maximal supergravity would have cancellations of the UV divergences at the $L=3$ and $L=4$ loop orders\footnote{There is no available counterterm at $L=4$, so finiteness at this order is not a dramatic result \cite{Drummond:2003ex}}. This would leave the next significant test at $L=5$ loops. In the ordinary Feynman-diagram approach, a full calculation at this level would involve something like $10^{30}$ terms. Even using the unitarity-based methods, such a calculation would be a daunting, but perhaps not impossible, task.

\begin{figure}[htp]
\begin{center}
\includegraphics[scale=.5]{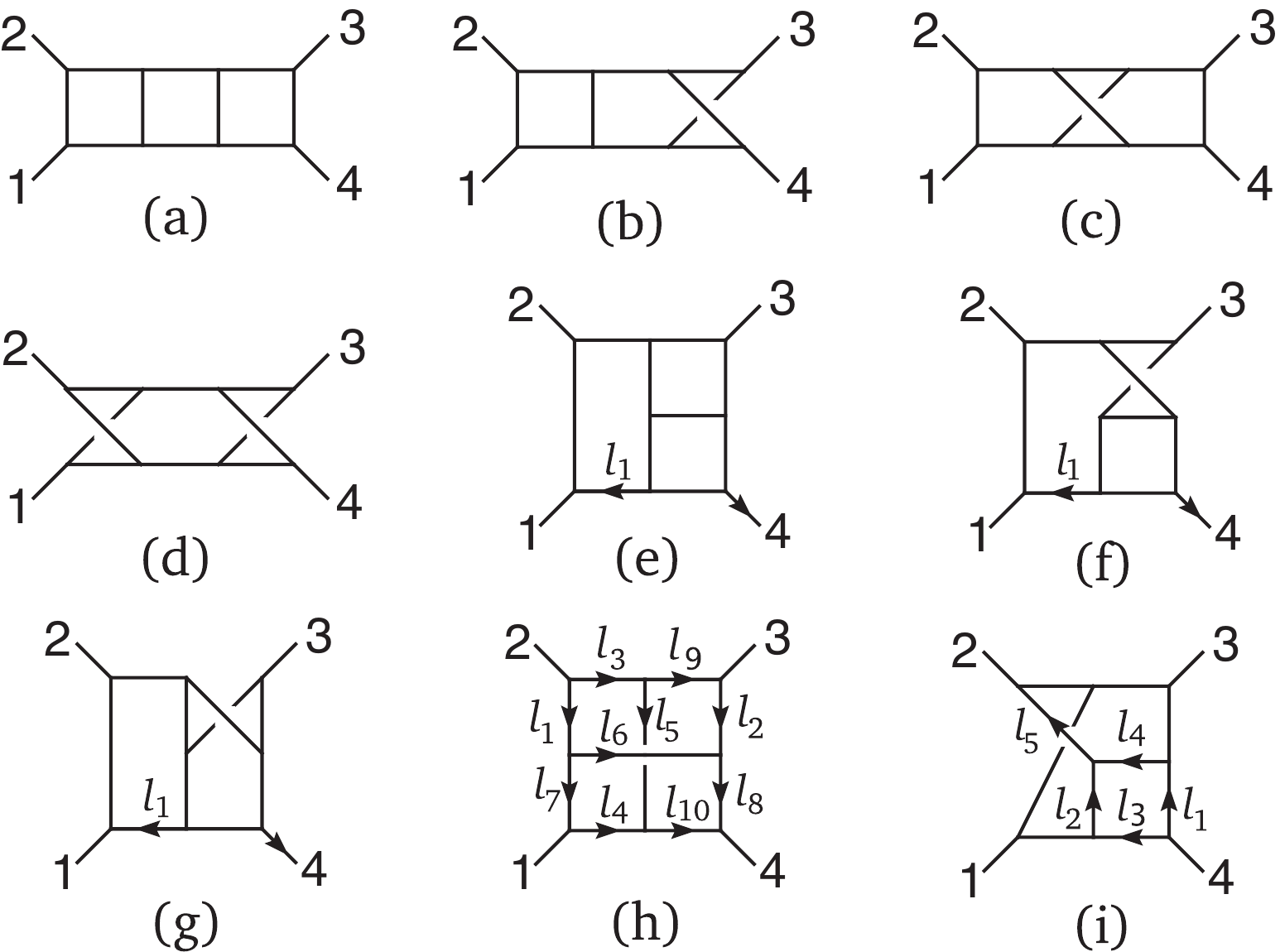} \caption{3-loop Feynman diagram types leading to unanticipated `superfiniteness' of maximal supergravity at this loop order. Diagrams (a)-(g) can be analysed using iterated 2-particle cuts, leading to an expectation of ultraviolet divergence cancellation. Diagrams (h) and (i) cannot be treated this way, but the result of summing all diagrams (a-i) is a deeper cancellation of the leading UV behaviour than anticipated.}\label{fig1}
\end{center}
\end{figure}

The impressive new elements in the 3-loop calculation of Bern et al are the completeness of their calculation and the unexpected further patterns of cancellations found. This could suggest a possibility of unexpected UV cancellations at yet higher loop orders. Although the various 3-loop diagram classes were already individually expected to be finite on the basis of the earlier work by Bern et al., the new results show that the remaining finite amplitudes display additional cancellations, rendering them ``superfinite''. In particular, the earlier work employed iterated 2-particle cuts and did not consider all diagram types. The new complete calculation displays further cancellations between diagrams that can be analysed using iterated 2-particle cuts and the additional diagrams that cannot be treated in this way. The set of three-loop diagrams is shown in Figure \ref{fig1}. The end result is that the sum of all diagram types is more convergent by two powers of external momentum than might otherwise have been anticipated. Yet more recent work has reorganised the calculation so that all diagram topologies give the final general structure in external momenta without the need for such cancellations \cite{Bern:2008pv}.

Does such a mechanism cascade in higher-order diagrams, rendering the maximal N=8 theory completely free of ultraviolet divergences? No one knows at present. Such a scenario might pose puzzling questions for the superstring programme, where it has been assumed that ordinary supergravity theories need string ultraviolet completions in order to form consistent quantum theories. On the other hand, there are hints from superstring theory \cite{Berkovits:2006vc} that, if extrapolated to the field theory limit \cite{Green:2006yu}, would suggest that these cancellations might continue up to nine loops in $\cN=8$ supergravity, and it has also been suggested, again from a string theory perspective \cite{Green:2006gt}, that $\cN=8$ supergravity might be finite to all orders, although it is not clear exactly what one can learn from superstring theory about purely perturbative field-theory divergences \cite{Green:2007zzb}.

One thing that seems clear is that ordinary Feynman diagram techniques coupled with the ``non-renormalization'' theorems of supersymmetry are unlikely to be able to explain finiteness properties of $\cN=8$ supergravity at arbitrary loop order. Earlier expectations \cite{Deser:1977nt,Deser:1978br,Kallosh:1980fi,superactions} were that the first loop order at which divergences that cannot be removed by field redefinitions would be three loops in all pure $D=4$ supergravities. A key element in this anticipation was the expectation that the maximal amount of supersymmetry that can be {\em linearly} realised in Feynman diagram calculations (aka ``off-shell supersymmetry'') is half the full supersymmetry of the theory, or 16 out of 32 supercharges for the maximal $\cN=8$ theory \cite{Howe:1983sr}.

Similarly to the way in which chiral integrals of $\cN=1$, $D=4$ supersymmetry achieve invariance  from integrals over less than the theory's full superspace, provided the integrand satisfies a corresponding BPS type constraint, there are analogous invariants involving integration over varying portions of an extended supersymmetric theory's full superspace \cite{superactions}. ``Half-BPS'' operators require integration over just half the full set of fermionic variables. And if half the full supersymmetry were the maximal amount that can be linearly realised (so giving strong results from the corresponding Ward identities), such operators would be the first to be allowed as UV counterterms. With the aid of harmonic superspace methods it can be shown that there are precisely three (linearised) BPS counterterms in $\cN=8$ supergravity \cite{Drummond:2003ex}; they arise at $L=3$ loops (half BPS), $L=5$ (one-quarter) and $L=6$ (one-eighth). The leading terms in spacetime are $R^4$ with 0,4 and 6 derivatives respectively.

The results of reference \cite{Bern:1998ug} show that the half-BPS expectation for the first allowed counterterms is too conservative in the case the maximal theory. But more recent advances in the understanding of supersymmetric non-renormalization theorems push the divergence-onset boundary out slightly for the maximal theory, so that half-BPS counterterms that require superspace integrals over half the 32 component superspace are now expected to be the last {\em disallowed} counterterms instead of the first {\em allowed} ones. The resulting current expectations for first divergences from a traditional Feynman diagram plus non-renormalization viewpoint are shown for various spacetime dimensions in Table \ref{tab1}.

\begin{table}[ht]
\centering
\begin{tabular}{|l|c|c|c|c|c|c|c|}
\hline
Dimension $D$&11&10&8&7&6&5&4\\
\hline
Loop order $L$&2&2&1&2&3&4&5\\
\hline
BPS degree\phantom{\Big|}&0&0&$\ft12$&$\ft14$&$\ft18$&$\ft18$&$\ft14$\\
\hline
General form\phantom{\Big|}&$\partial^{12} R^4$&$\partial^{10} R^4$&$R^4$&$\partial^4 R^4$&$\partial^6 R^4$&$\partial^6 R^4$&$\partial^4 R^4$\\
\hline
\end{tabular}
\caption{Current maximal supergravity divergence expectations from Feynman rules and non-renormalization theorems.
\label{tab1}}
\end{table}

The behaviour of maximally supersymmetric Yang--Mills theory in dimensions $D>4$ may be a model for what is happening. Contrary to earlier expectations of UV divergences at the 4-loop order in $D=5$ spacetime \cite{Howe:1983jm}, the unitarity-based methods indicate that this SYM onset should be postponed to the 6-loop order \cite{Bern:1998ug}. But here, the standard Feynman diagram methods have a comeback through the realisation that 4-loop finiteness could be explained using more sophisticated ``harmonic superspace'' methods \cite{Galperin:1984av,Galperin:1985va,Howe:2002ui}. In fact, maximal SYM theory admits a formulation for which twelve supersymmetries are linearly realised off-shell \cite{Galperin:1985uw,Delduc:1988cp}, so that the first allowed counterterm would be one-quarter BPS, not one-half \cite{Howe:2002ui}. Indeed, if there is an off-shell version of $\cN=8$ supergravity in $\cN=5$ harmonic superspace, then this would be enough to push the first allowed counterterm to five loops\footnote{For a short summary of the current UV divergence situation in supergravity, {\it cf.}\ \cite{Stelle:2007zz}.}  \cite{Howe:2002ui}. While this harmonic superspace possibility remains unclear in supergravity, we will discuss in section \ref{halfplusone} a finite-component formalism with 17 linearly realised supersymmetries that should also push the divergence limit for $D=4$ diagrams to 5 loops.

In the next section we review superspace non-renormalisation theorems and the implications for divergences that follow from them in conventional superspace. In section 3, we explicitly construct the relevant BPS counterterms using harmonic superspace. In section 4 we show how these can be rewritten as integrals over the whole of light-cone superspace thereby demonstrating that the light-cone formalism does not rule out even one-half BPS invariants as counterterms. In section 5 we give a discussion of the off-shell harmonic superspace formalism with twelve linearly realised supersymmetries for maximal SYM in arbitrary dimensions. This, in combination with non-renormalisation theorems, implies that such theories do not admit one-half BPS divergences in line with unitarity calculations. Another possible way of forbidding such divergences is to use formulations of maximal supersymmetric theories with one-half plus one supersymmetries linearly realised. This can be done with a finite number of auxiliary fields at the cost of manifest Lorentz covariance. This idea is discussed in section 6 in two dimensions where Lorentz symmetry can be maintained. In section 7 we discuss the algebraic approach to the analysis of non-linear supersymmetry Ward identities in non-renormalisable supersymmetric theories.  A generalisation of previous methods using Young tableaux leads to the conclusion that logarithmic divergences at $L>1$ loops are in contradiction to the supersymmetry Ward identities if and only if they are associated to supersymmetric counterterms involving $\ft12$ BPS operators. In this framework we are able to prove the absence of the four-loop and the two-loop logarithmic divergences of maximally supersymmetric Yang--Mills theory in five and six dimensions respectively. Although we have not fully worked out all the details needed to generalise this method rigourously to supergravity theories, similar arguments to those in Yang--Mills theory permit one to demonstrate the absence of logarithmic divergences at three-loops in maximal supergravity in four dimensions. Although the discussion is not yet complete, it gives further strong evidence in favour of the conjecture that the one-half BPS counterterms will not occur in maximal theories of this type.

%%%%%%%%%%%%%%%%%%%%%%%%%%%%%%%%%%%%%%%%%%%%%%%%%%%%

\section{Review of non-renormalisation theorems}

%%%%%%%%%%%%%%%%%%%%%%%%%%%%%%%%%%%%%%%%%%%%%%%%%

The prototype superspace non-renormalisation theorem states that the mass and interaction terms in the Wess-Zumino model do not receive infinite corrections. To establish this one first sets up Feynman rules in superspace, which is straightforward to do for chiral and gauge multiplets in $\cN=1,D=4$ supersymmetry. It can then be shown that all of the fermionic theta integrals except one can be carried out so that each contribution to the effective action in superspace is local in the odd coordinates, albeit generally non-local in the bosonic coordinates \cite{Grisaru:1979wc}. An important fact is that the Grassmann odd integration is over all four odd coordinates. Standard quantum field theory considerations imply that ultra-violet divergences are local in spacetime, and thus give rise to full superspace integrals in $\cN=1,D=4$ superspace, and this immediately rules out UV divergences corresponding to the chiral terms in the classical action.

The power of the superspace method can be increased by means of the background field method (BFM) in which the total field is split into a quantum part and a background part. The point of doing this is that one can then study 1PI diagrams which have only background fields on external lines. This is particularly useful in gauge theories because one can then use background gauge invariance instead of quantum BRST symmetry and this simplifies  the analysis of possible divergences. For supersymmetric gauge theories there is an additional advantage because the superspace gauge potentials are constrained while the quantum prepotentials, which appear as the solutions to these constraints, are not. It can be shown that it is possible to set up the Feynman rules in the superspace BFM in such a way that the background fields appearing on the external lines are constructed from the potentials rather than the prepotentials \cite{Grisaru:1982zh}. The relation between the two objects can be illustrated in $\cN=1,D=4$ abelian gauge theory where the superspace field-strength two-form $F=dA$ is constrained to satisfy

\be
 F_{\a\b}=F_{\adt\bdt}=F_{\a\bdt}=0\ ,
\la{2.1}
\ee

where the coordinates are $(x^a,\th^\a,\bar\th^{\adt})$ and where the fields are referred to the standard preferred basis given by the covariant derivatives $(\del_a,D_\a,\bar D_{\adt})$, with

\be
 [D_\a,\bar D_{\adt}]=i\del_{\a\adt}\ ,
\la{2.2}
\ee

as usual. In \eq{2.1} the third equation is a ``conventional'' constraint which allows one to solve for the vector component of the potential, $A_a$, in terms of the spinorial ones, while the first two are related by complex conjugation. These equations are solved by

\be
 A_\a=iD_\a V\ ,
\la{2.3}
\ee

where $V$ is the real, unconstrained superfield prepotential, and where we have made a partial choice of gauge in superspace which reduces the gauge parameter from a real scalar superfield to a chiral one.

A key point here is that the prepotential has lower dimension than the potential, so the fact that counterterms should be built from the latter will clearly lead to improved power-counting.

The above can be generalised to $\cN=2, D=4$ (or $\cN=1,D=6$) supersymmetry straightforwardly, although the technical details become much more complicated \cite{Grisaru:1982zh}. The case of $\cN=2,D=4$ supersymmetry was analysed completely in \cite{Howe:1983sr} and used to give a covariant superfield proof of the perturbative finiteness of $\cN=4,D=4$ SYM \cite{Howe:1982tm} and of a class of $\cN=2$ SYM-matter models \cite{Howe:1983wj}. The technical difficulties that arise are due to the low dimensionality of the SYM prepotential $(-2)$ in mass units) \cite{Mezincescu:1979af}, and the rather complicated nature of the off-shell version of the $\cN=2$ hypermultiplet. Indeed, in conventional superspace, the $\cN=2$ matter multiplet can only be formulated off-shell if the scalars are in the $1+3$ representations of $SU(2)$, although this restriction can be lifted in harmonic superspace.\footnote{Analysis of the $N=2$ non-renormalisation theorems in harmonic superspace has been carried out in \cite{Buchbinder:1997ib, Buchbinder:1997ya}.}

There is, however, a major obstruction to extending the formalism to arbitrary supersymmetric theories and that is because, in order to write down superspace Feynman diagrams, one must be able to represent supersymmetry linearly on unconstrained superfields, in other words, the theory in question must admit an off-shell formulation with an appropriate set of auxiliary fields. In general, however, this does not seem to be possible for theories with extended supersymmetry. It was shown in \cite{Siegel:1981dx} that $\cN=1,D=10$ SYM does not admit an off-shell version with a finite number of auxiliary fields, and a general study was carried out in \cite{Rivelles:1982gn}. The upshot is that a theory which has $Q$ supersymmetries on-shell will generically only admit a formulation with $q<Q$ linearly realised off-shell supersymmetries. For SYM theories in conventional Lorentz-covariant superspace, the maximal allowed value of $q$ is $8$, while for supergravity theories $q$ cannot be larger than $16$ \cite{Howe:1982mt}; in other words, in conventional superspace, only half of the supersymmetries of maximally supersymmetric Yang--Mills or supergravity theories can be realised off-shell. This is clearly relevant to non-renormalisation theorems because they require the existence of an off-shell formalism. On the other hand, the first UV divergence that is encountered must correspond to a counterterm which becomes invariant under the full non-linear supersymmetry transformations when the classical equations of motion are imposed. This follows from consideration of the Ward Identities for a non-linearly realised symmetry which state that, in the absence of anomalies, the first UV divergence will be invariant up to a term (coming from the fact that a non-linear symmetry must itself be renormalised) which is proportional to the equations of motion of the classical theory.

Putting this all together, we can state the non-renormalisation theorem for supersymmetric theories as follows: the first non-zero UV divergence corresponds to a counterterm which can be expressed as an integral over $q$ thetas, where $q$ is the number of linearly realised supersymmetries, which is constructed from the potentials of the off-shell formalism, which is gauge-invariant and which is invariant under the full non-linear $Q$ supersymmetries modulo the classical equations of motion. There is a final caveat, which is that the theorem does not apply at one loop due to technical difficulties that arise with gauge-fixing in the background field method.

It is clear that this theorem will only be restrictive for potential counterterms which can only be expressed as subsuperspace integrals in the full $Q$-theta superspace. We shall refer to these counterterms as BPS because their integrands correspond to shortened representations of supersymmetry; clearly the non-BPS counterterms, which correspond to integrals over the full $Q$-susy superspace, cannot be protected by non-renormalisation theorems of the above type.

In the context of non-renormalisation theorems in conventional Lorentz covariant superspace we know that $q=\half Q$ for maximal supersymmetric theories which implies that the first allowed counterterms should be one-half BPS; in SYM this translates to spacetime integrals of $F^4$, while in supergravity $R^4$ would be permitted.

For maximal SYM, these one-half BPS counterterms correspond to $L=4$ loops in $D=5$ and $L=2$ loops in $D=6$; in $D=7$ this counterterm does not arise for dimensional reasons and the first allowed divergence is the one-quarter BPS one which occurs at two loops. In $D=8$ $F^4$ occurs at one loop where it is not protected, while SYM diverges at two loops ($d^6F^4$) and one loop ($d^2F^4$), the first allowed divergences, in $D=9$ and $10$ respectively.

In $D=6$ it turns out \cite{Howe:1983jm} that gauge invariance protects the putative $L=2$ $F^4$ divergence. This works as follows: the theory is quantised in terms of $\cN=1,D=6$ superfields, i.e. $q=8$, but the $\cN=1$ SYM multiplet has no scalars. This means that the lowest allowed counterterm (beyond one loop) in the $\cN=1$ SYM theory is in fact of structure $d^2 F^4$ (the full superspace integral of four spinorial superfields) which corresponds to a three-loop divergence. The independent scalar multiplet cannot alter this result and so the conclusion that $F^4$ is protected in six dimensions follows. In $D=5$, on the other hand, the $\cN=1$ SYM multiplet does contain a scalar and so the $F^4$  invariant would be allowed in this case.

The above results are consistent with the calculations of reference \cite{Marcus:1984ei}, and with almost all of the more recent unitarity computations \cite{Bern:1998ug}. The only mismatch is the $F^4$ counterterm at four loops in $D=5$ which would be allowed to be divergent, as we have just seen, but which turns out to have a zero coefficient. A possible explanation for this would be that the five-dimensional theory somehow knows about gauge invariance in six dimensions, but it seems difficult to justify this in any convincing way. An alternative explanation, which we shall discuss later on, is given by harmonic superspace considerations because this formalism allows $q=12$.

For maximal supergravity the situation is that non-renormalisation theorems in conventional superspace would allow $R^4$ divergences at three loops in $D=4$ and at two loops in $D=5$. This divergence also occurs at one loop in $D=8$ where there is no protection in any case. In the remaining dimensions, other than six, maximal supergravity diverges at two loops with $R^4$ counterterms together with $4,8,10$ and $12$ derivatives respectively for $D=7,9,10,11$. In $D=6$ the first allowed divergence is at three loops, $d^6 R^4$, and is known to have a non-vanishing coefficient. Note that this discussion makes use of the fact that there is no $d^2 R^4$ invariant consistent with all of the required symmetries.

The supergravity predictions of conventional superspace are in agreement with calculations carried out so far except for the $R^4$ divergences in $D=4,5$ which are known to have vanishing coefficients.

Various non-renormalisation theorems can also be derived in the component formalism, by making use of the full non-linear supersymmetry Ward identities within the Batalin--Vilkovisky formalism. Although these methods have only been applied to renormalisable theories up until now, they have permitted one to obtain theorems independently of the renormalisation scheme, which avoids the open problem of how to define a regularisation that preserves both supersymmetry and gauge invariance. A first attempt to prove the superconformal invariance of $\cN=4$ super Yang--Mills by exhibiting the absence of superconformal anomalies was proposed in \cite{WhiteComform}. A rigourous proof of the vanishing of the $\beta$ function was then been developed in term of twisted variables in Euclidean space \cite{Sor,Sor4,GuillaumeN4}. This proof was subsequently extended to the superconformal $\cN=2$ Yang--Mills theories defined on Minkowski spacetime with the whole global symmetry kept manifest \cite{N2}.

%%%%%%%%%%%%%%%%%%%%%%%%%%%%%%%%%%%%%%%%%%%%%%%%%%%%%%%%%%%%%%%%

\section{Counterterms}

%%%%%%%%%%%%%%%%%%%%%%%%%%%%%%%%%%%%%%%%%%%%%%%%%%%%%%%%%%%%%%%%%

In this section we shall give the explicit forms for the relevant BPS counterterms in maximal supergravity and SYM theories, although we shall only be concerned with the linearised theory in the supergravity case. These counterterms are most simply expressed in harmonic superspace, so we start off with a brief review of this idea \cite{Galperin:1984av,Rosly:1982,Karlhede:1984vr}.

%%%%%%%%%%%%%%%%%%%%%%%%%

\subsection{Harmonic superspace}

%%%%%%%%%%%%%%%%%%%%%%%%%%

In $D=4$, $\cN$-extended superspace has coordinates $(x^{\a\adt},\th^{\a i},\bar\th^{\adt}_i)$, where $i=1,\ldots \cN$ and the thetas (thetabars)  transform under the $\cN\,(\bar \cN)$ representations of $U(\cN)$.\footnote{For the maximal theories the internal symmetry group is $SU(\cN)$.} The superspace covariant derivatives $D_{\a i},\bar D_{\adt}^i$ satisfy the standard anticommutation relations

\bea
 [D_{\a i},\bar D_{\bdt}^j]&=& i\d_i{}^j \del_{\a\bdt} \nn\w1
 [D_{\a i},D_{\b j}]&=&[\bar D_{\adt}^i,\bar D_{\bdt}^j]=0\ .
\la{3.0}
\eea

We define a Grassmann analyticity (GA) structure of type $(p,q)$ to be a set of $p$ $D$s and $q$ $\bar D$s which are mutually anticommuting. Such a set is specified by a rank $p$ matrix $u_r{}^i$, $r=1,\ldots p$, and a rank $q$ matrix $v_i{}^{r'}$, $r'=\cN-q+1,\ldots \cN$, such that

\be
 u_r{}^i v_i{}^{r'}=0\ .
\la{3.1}
\ee

Such a pair of matrices determines a $p$-plane $V_p$ inside an $(\cN-q)$-plane $V_{\cN-q}$ in $\bbC^\cN$ called a flag of type $(p,\cN-q)$. The space of all $(p,q)$ GA structures for $\cN$-extended supersymmetry is the flag manifold $\bbF_{p,\cN-q}(\cN)$ which we shall denote simply by $\bbF$ in the following unless it is necessary to be more specific. It is a compact complex manifold and can be represented as the coset space $(U(p)\xz U(\cN-(p+q))\xz U(q))\bsh U(\cN)$ (in the maximal theories $U(\cN)$ is replaced by $SU(\cN)$ and the isotropy group is modified accordingly). In this context it is convenient to regard the matrices $u_r{}^i$ and $v_i{}^{r'}$ as submatrices of an $(S)U(\cN)$ group element $u$ and its inverse respectively.

We define $(\cN,p,q)$ harmonic superspace to be the product of ordinary $\cN$-extended superspace $M_\cN$ with $\bbF_{p,\cN-q}(\cN)$ \cite{Hartwell:1994rp,Howe:1995md}. Fields on this space can be expanded in harmonics on $\bbF$ with coefficients that are ordinary superfields. We shall use the so-called harmonic formalism \cite{Galperin:1984av} in which fields on $\bbF$ are given as equivariant fields on $U(\cN)$, so that their dependence on the isotropy directions is specified by their transformation properties under this group. We put $u=u_I{}^i=(u_r{}^i,u_R{}^i,u_{r'}{}^i)$, where the indices $(r,R,r')$ label the fundamental representations of the three factors in the isotropy group. We can use $u$ and its inverse to refer tensors to $U(\cN)$ or isotropy group bases, for example, we can define

\be
 D_{\a I}:=u_I{}^i D_{\a i};\qquad \bar D_{\adt}^I:=(u^{-1})_i{}^I \bar D_{\adt}^i\ .
 \la{3.1.1}
\ee

The right-invariant vector fields on $U(\cN)$ are $D_I{}^J$; they act as follows:

\be
 D_I{}^J u_K{}^k=\d_J{}^K u_I{}^k\ ,
 \la{3.1.2}
\ee

and satisfy

\be
 [D_I{}^J,D_K{}^L]=\d_K{}^J D_I{}^L-\d_I{}^L D_K{}^J\ .
 \la{3.1.2.1}
\ee

This set of derivatives divide into three subsets: the isotropy subset, $\{D_r{}^s,D_R{}^S,D_{r'}{}^{s'}\}$, the subset corresponding to the $\bar\del$ operator on $\bbF$, $\{D_r{}^S,D_r{}^{s'},D_R{}^{s'}\}$, and the conjugate subset, $\{D_R{}^s,D_{r'}{}^s,D_{r'}{}^S\}$.

In harmonic superspace one defines Grassmann (G-) analytic fields to be those that are annihilated by $D_{\a r}$ and $\bar D_{\adt}^{r'}$, and harmonic, or H-analytic fields to be those which are annihilated by the $\bar\del$ operator on $\bbF$. The fields we are interested in are both G- and H-analytic, which we will refer to as being simply analytic. Since $\bbF$ is compact complex, such fields have short harmonic expansions. We shall refer to them as BPS multiplets; that is, a BPS multiplet in the harmonic framework is by definition one that is annihilated by some fraction of the set of superspace covariant derivatives. Thus a one-half BPS multiplet is annihilated by half of the odd derivatives and so on. In general one would expect that, for example, a one-half BPS multiplet in $\cN=4,D=4$ susy would have an expansion up to eighth order in the odd coordinates, but there are special multiplets which are ultra-short. These include the field strength and supercurrent multiplets in $\cN=4$ SYM.

Invariants are constructed by integrating analytic fields with respect to an appropriate measure. For $(\cN,p,q)$ superspace this is

\be
 d\m^\cN_{p,q}:= d^4x\,du\,[D_{p+1}\ldots D_\cN \bar D^1\ldots \bar D^{\cN-q}]^2\ ,
 \la{3.1.3}
\ee

where $du$ denotes the standard Haar measure on the coset.

It is sometimes useful to use a larger coset as this allows more flexibility. The largest possible arises when the isotropy group is the maximal torus $(U(1)^{\cN-1}$ in $SU(\cN)$. It is the space of full flags, i.e. $V_1\subset V_2\ldots \subset V_{\cN-1}\subset \bbC^\cN$. In this case a group element $u$ can be written $(u_1{}^i, u_2{}^i,\ldots u_\cN{}^i)$ where each of the numerical subscripts is acted on by the corresponding $U(1)$ subgroup with the condition that an object carrying all $\cN$ indices has charge zero.

%%%%%%%%%%%%%%%%%%%%%%%%%%%%%%%

\subsection{Invariants in $D=4$}

%%%%%%%%%%%%%%%%%%%%%%%%%%%%%%%%

At the linearised level the $\cN=4$ SYM multiplet is described by a scalar superfield $W_{ij}$ which transforms under the real six-dimensional representation of $SU(4)$. It satisfies

\bea
 D_{\a i} W_{jk}&=& D_{\a [i} W_{jk]}\nn\w1
 \bar D_{\adt}^i W_{jk}&=& -\frac{2}{3}\d_{[j}{}^i \bar D_{\adt}^l W_{k]l}\
\la{3.2}
\eea

as well as the reality condition $\bar W^{ij}=\half \ve^{ijkl} W_{kl}$. It can be described as an analytic field on $(4,2,2)$ superspace, $W_{12}:=u_1{}^i u_2{}^j W_{ij}$, which is an $SU(2)$ singlet but which carries a charge under the $\gu(1)$ part of the isotropy algebra $\gs\gu(2)\oplus\gs\gu(2)\oplus\gu(1)$; it is annihilated by $D_{\a r},r=1,2$ and $\bar D_{\adt}^{r'}, r'=3,4$. It can also be described as an analytic field on $(4,1,1)$ superspace, $W_{1R}:=u_1{}^i u_R{}^j W_{ij},\ R=2,3$, which transforms under the $\gs\gu(2)$ of the isotropy algebra $\gs\gu(2)\oplus\gu(1)\oplus\gu(2)$, as well as carrying $U(1)$ charges. In this version it is annihilated by $D_{\a 1}$ and $\bar D_{\adt}^4$.

In the interacting theory the derivatives appearing in \eq{3.2} have to be replaced by the corresponding gauge-covariant derivatives, but since the invariants are constructed from gauge-invariant quantities, this fact does not play any significant role in their enumeration.

There are two one-half BPS invariants; they are

\bea
 I_{\half}&=& \int\, d\m^4_{2,2}\, \tr (W_{12})^4 \nn\w1
 I'_{\half}&=& \int\, d\m^4_{2,2}\, \tr(W_{12})^2\tr(W_{12})^2\nn\w1
\la{3.3}
\eea

There is a double-trace one-quarter BPS invariant given by

\be
 I_{\frac{1}{4}}=\int\, d\m^4_{1,1}\, \tr(W_{1R}W_{1S})\tr(W_1{}^R W_1{}^S)\ ,
 \la{3.5}
\ee

where the $SU(2)$ index $R$ is raised by means of $\ve^{RS}$ Notice that \eq{3.5} vanishes in the Maxwell case. There is also a single-trace invariant which looks as if it might be one-quarter BPS but which turns out to be expressible as a full superspace integral:

\bea
 I_\Ko &=& \int\,d\m^4_{1,1} \tr((\ve^{RS} W_{1R} W_{1S})^2)\nn\w1
 &=&\int\,d^4x\,d^{16}\th \,K
 \la{3.7}
\eea

where $K:=\tr(W_{ij}\bar W^{ij})$ is the Konishi operator. Both of these terms integrate up to give spacetime contributions of the form $\int\,dx\,( d^2 F^4 + F^5 + \ldots)$ \cite{Drummond:2003ex}.

A feature of these invariants is that they can be understood as arising as products of the energy-momentum supermultiplet $T$. The symmetric product of two of these contains a one-half BPS multiplet, the integrand of $I'_{\half}$, the one-quarter BPS integrand of $I_{\qu}$, a long multiplet and a shortened but not BPS multiplet whose leading component is, like the energy-momentum multiplet, a set of scalars in the real $20'$ representation of $SU(4)$.\footnote{This multiplet is known to be a protected operator in SCFT \cite{Eden:2000bk}; it obeys a second-order fermionic derivative constraint \cite{Heslop:2001dr}.} $T$ itself is a one-half BPS operator which is ultra-short; it can be integrated over four thetas to give the on-shell SYM action,

\be
 S_{SYM}=\int\, d^4x\, du\, [D_3 D_4]^2\, \tr(W_{12})^2\ .
 \la{3.8}
\ee

The maximal supergravity theory in $D=4$ is very similar, at least in the linearised theory. The field strength superfield is described by a scalar superfield $W_{ijkl}$ which transforms under the real seventy-dimensional representation of $SU(8)$. It satisfies

\bea
 D_{\a i} W_{jklm}&=& D_{\a [i} W_{jklm]}\nn\w1
 \bar D_{\adt}^i W_{jklm}&=& -\frac{4}{5}\d_{[j}{}^n \bar D_{\adt}^l W_{klm]n}\ ,
\la{3.9}
\eea

and satisfies an obvious reality condition. It can be represented by an analytic superfield $W:=W_{1234}:=u_1{}^iu_2{}^ju_3{}^ku_4{}^l W_{ijkl}$ in $(8,4,4)$ superspace. The three-loop one-half BPS invariant is given by \cite{Hartwell:1994rp}

\be
 I_{\half}=\int\, d\m^8_{4,4}\, W^4\ .
 \la{3.10}
\ee

An interesting feature of this integral is that it is also invariant under linearised $E_7$ transformations. We recall that the Lie algebra of $E_7$ is a sum of the $\gs\gu(8)$ sub-algebra together with seventy generators which, at the linearised level, act on the scalars by a shift. To see that this is a symmetry of $I_{\half}$ note that, for any positive integer $k$, $W^k$ is one-half BPS, but that it is ultra-short\footnote{This follows from the facts that $W$ itself has top components of the form $\th^4$ or $\bar\th^4$ and that we can only have at most eight powers of $\th$ or $\bar\th$ in $(8,4,4)$ superspace.} for the special values of $k=1,2,3$. The translational symmetry can be written $\d W= L$ where $L$ is an analytic superfield constructed in the same way as $W$ but where $L_{ijkl}$ is a set of seventy constant parameters, i.e. independent of all of the (conventional) superspace coordinates.  Clearly the variation of the integrand in \eq{3.10} will be proportional to $W^3$, but since this is ultra-short, integrating it over sixteen thetas will give zero.

All of the possible $SU(8)$-invariant BPS integrals were written down in \cite{Drummond:2003ex}; it turns that there are only two more: a one-quarter BPS invariant, which integrates to $d^4 R^4$, and a one-eighth invariant which integrates to $d^6 R^4$. It is not possible to construct a $d^2 R^4$ invariant, which would have corresponded to a four-loop counterterm. The one-quarter BPS invariant is

\be
 I_{\qu}=\int\,d\m^8_{2,2}\, (\ve^{RTSU} W_{12 RS} W_{12 TU})^2\ ,
 \la{3.11}
\ee

where $R\in\{3,4,5,6\}$. The one-eighth invariant is

\be
 I_{\frac{1}{8}} =\int\,d\m^8_{1,1}\,\ve^{R_1 \ldots R_6} \ve^{S_1\ldots S_6} W_{1R_1 R_2 R_3} W_{1R_4 R_5 S_1} W_{1R_6 S_2 S_3} W_{1S_4 S_5 S_6}\ ,
 \la{3.12}
\ee

where $R\in\{2,3,4,5,6,7\}$.

%%%%%%%%%%%%%%%%%%%%%%

\subsection{Harmonic superspaces in $D>4$}

%%%%%%%%%%%%%%%%%%%%%%

We shall be concerned with $D=5,6,7$ where the internal symmetry groups are symplectic. For maximal SYM we have $Sp(2)$ in $D=5$, $Sp(1)$ in $D=7$ and $Sp(1)\xz Sp(1)$ in $D=6$, while for maximal supergravity we have $Sp(4)$ in $D=5$, $Sp(2)$ in $D=7$ and $Sp(2)\xz Sp(2)$ in $D=6$.\footnote{For some discussions of harmonic superspaces in $D>4$ see, for example, \cite{Howe:1985ar,Howe:1998jw,Ferrara:2000xg,Kuzenko:2005sz}.} In each case the supersymmetry algebra takes the form

\be
 [D_{\a i},D_{\b j}]=i\h_{ij} (\c^a)_{\a\b}\del_a\ ,
 \la{3.13}
\ee

where $(\c^a)_{\a\b}$ is antisymmetric on its  spinor indices, and $\h_{ij}$ is the symplectic invariant matrix. In the $D=6$ case there are two copies with opposite chirality spinors. Internal indices are lowered or raised by $\h_{ij}$ and its tensorial inverse $\h^{ij}$ according to the rules $v_i=v^j\h_{ji}$, $v^i=\h^{ij} v_j$.

In order to have Grassmann analyticity with $p$ mutually anticommuting derivatives, say in $Sp(n)$ where the group acts on $\bbC^{2n}$, we need to specify an isotropic $p$-plane in $\bbC^{2n}$. The isotropy group of such a plane is $U(p)\xz Sp(n-p)$, and therefore the space of such planes, $\bbF_p(2n)$, is the coset space of $Sp(n)$ with this isotropy group. The various harmonic superspaces are then formed by taking the product of ordinary superspace with this harmonic coset. Since $\bbF_p(2n)$ is again compact and complex, the field strength superfields of interest will be analytic with respect to both types of analyticity as in the four-dimensional case.

%%%%%%%%%%%%%%

\subsubsection{$D=5$}

%%%%%%%%%%%%%

Spinors in five dimensions are four-component, but when there are an even number of them they can be taken to satisfy a pseudo-Majorana constraint which makes use of the symplectic invariant $\h_{ij}$. The R-symmetry group for maximal SYM (SG) is $Sp(2)$ ($Sp(4)$). The superspace measures are

\be
 d\m^n_p:=d^5 x\,du\,[D_{p+1}\ldots D_{2n}]^4\ .
 \la{3.13.1}
\ee

The SYM field strength $W_{ij}$ is in the real five-dimensional representation of $Sp(2)$, so it is antisymmetric and traceless with respect to $\h$. It satisfies

\be
 D_{\a i} W_{jk}=D_{\a [i} W_{jk]} -\frac{2}{3} \h_{i[j} D_\a^l W_{k]l}\ .
 \la{3.15}
\ee

To write down a one-half BPS invariant it is necessary to pick out two anticommuting derivatives $D_{\a r}=u_r{}^i D_{\a i}, r=1,2$. We then need to choose $\h_{rs}=\h_{r's'}=0$, $\h_{rs'}=\d_{rs'}$, where $r'=3,4$, and where the matrix $u=(u_r{}^i,u_{r'}{}^i)$ is an element of $Sp(2)$; the isotropy group is $U(2)$. It is easy to see that $W_{12}:=u_1{}^i u_2{}^j W_{ij}$ is analytic, so the one-half BPS invariants can be written

\be
 I_{\half}=\int\, d\m^2_2\, <W_{12}^4>\ ,
 \la{3.16}
\ee

where the brackets indicate either the single or double-trace group structures.

For the one-quarter BPS invariants we need to specify only one derivative $D_{\a 1}:=u_1{}^i D_{\a i}$, say, so that the isotropy group is $U(1)\xz SU(2)$. In this case it is simpler to take $\h$ to be block diagonal with non-vanishing components in the $(12)$ and $(34)$ planes. The superfield $W_{1R}:=u_1{}^i u_R{}^j W_{ij},\,R\in\{3,4\}$ is one-quarter BPS analytic and the true invariant is

\be
 I_{\qu}=\int\,d\m^2_1\,\tr(W_{1R} W_{1S})\tr(W_1{}^R W_1{}^S)\ ,
 \la{3.17}
\ee

where $R$ is raised by $\ve^{RS}$. Note that the $R$ index is acted on only by $Sp(1)$ and so carries no separate $U(1)$ charge.

In maximal supergravity the field strength superfield is $W_{ijkl}$, $i=1,\ldots 8$. It is totally antisymmetric, symplectic traceless and real. It satisfies the differential constraint

\be
 D_{\a i} W_{jklm}=D_{\a [i}W_{jklm]}-\frac{3}{5}\h_{i[j} D_\a^n W_{klm]n}\ .
 \la{3.18}
\ee

The one-half BPS fields are annihilated by four of the derivatives, $D_{\a r}$, say, and the appropriate isotropy group is therefore $U(4)$. The superfield $W_{1234}$, defined in the obvious way by harmonic projection, is clearly one-half BPS and the $R^4$ invariant is

\be
 I_{\half}=\int\,d\m^4_4\, (W_{1234})^4\ .
 \la{3.19}
\ee

For the one-quarter BPS case we can choose two anticommuting derivatives $D_1,D_2$ and isotropy group $U(2)\xz Sp(2)$. The $d^4R^4$ invariant is

\be
 I_{\qu}=\int\, d\m^4_2\, (W_{12}{}^{RS} W_{12RS})^2 \,
 \la{3.19.1}
\ee

where $R\in\{5,6,7,8\}$ is raised or lowered by the symplectic matrix restricted to this subspace.

The above invariant does not arise as a possible counterterm, but the one-eighth BPS one ($d^6 R^4$) does. It is given by

\be
 I_{\frac{1}{8}}=\int\,d\m^4_1\, W_{1 RST} W_1{}^{RSU} W_{1 UVW} W_1{}^{TVW}\ ,
 \la{3.19.2}
\ee

where $R\in\{3,4,5,6,7,8\}$.

%%%%%%%%%%%

\subsubsection{$D=6$}

%%%%%%%%%%%%%

The spinors in $D=6$ are pseudo-Majorana-Weyl; for the maximal theories we have pairs of both chiralities. In the SYM case the R-symmetry group is $Sp(1)\xz Sp(1)$ and the field strength $W_i{}^{\hi}$ satisfies

\be
 D_{\a (i} W_{j)}{}^{\hk}=D^{\a (\hi} W_k{}^{\hj)}=0\ ,
 \la{3.20}
\ee

where $\a=1,\ldots 4$ and where both $i$ and $\hi$ can take on two values. The one-half BPS harmonic space in this case is made up of two copies of $U(1)\bsh SU(2)$. The analytic field strength, annihilated by $D_{\a 1}$ and $D^{\a \hat{1}}$, is $W_1{}^{\hat{1}}$. The two invariants are

\be
 I_{\half}=\int\,d^6x\,du\,d\hu\,[D_2 D^{\hat{2}}]^4\,<(W_1{}^{\hat{1}})^4>\ ,
 \la{3.21}
\ee

where the brackets indicate the two group invariants as before. To get a one-quarter BPS invariant we need to harmonise only one of the $Sp(1)$s, say the unhatted one. The superfield $W_{1}{}^{\hi}$ is annihilated by $D_{\a 1}$. The true one-quarter BPS invariant is

\be
 I_{\qu}=\int\,d^6x\,du\, [D_2 D^{\hat{1}} D^{\hat{2}}]^4\, \tr( W_{1\hj} W_{1\hk})\,\tr(W_1{}^{\hj} W_1{}^{\hk})\ .
 \la{3.22}
\ee

We can obtain a second one-quarter BPS invariant by harmonising the hatted sector.

In supergravity the R-symmetry group is $Sp(2)\xz Sp(2)$. There are twenty-five scalar fields which appear as the leading component of the field-strength superfield $W_{ij}{}^{\hi\hj}$. Each pair of indices transforms as a five under $Sp(2)$ and obeys a constraint similar to that obeyed by the $D=5$ SYM field strength. For the one-half BPS case, the isotropy group is $U(2)\xz U(2)$, the field strength is $W:=W_{12}{}^{\hat{1}\hat{2}}$, and the invariant is

\be
 I_{\half}=\int\,d^6x\,du\,d\hu\, [D_3 D_4 D^{\hat{3}}D^{\hat{4}}]^4\, W^4\ .
 \la{3.23}
\ee

There appear to be two possibilities for one-quarter BPS invariants. The first involves constructing fields which are annihilated by two $D$s of the same type, say $D_{\a r}$, and the second involves fields which are annihilated by one $D$ of each type. In the first case the field is $W_{12}{}^{\hi\hj}$ and the invariant is

\be
 I_{\qu}=\int\,d^6x\,du\,[D_3 D_4 D^{\hat{1}}\ldots D^{\hat{4}}]^4 (W_{12}\cdot W_{12})^2
 \la{3.24}
\ee

with the obvious scalar product on the hatted indices. In the second case, the field is $W_{1R}{}^{\hat{1}\hR}$, where $R\in\{3,4\}$ and similarly for $\hR$, and the invariant is

\be
 I'_{\qu}=\int\,d^6x\,du\,d\hu\,[D_2 D_3 D_4 D^{\hat{2}} D^{\hat{3}} D^{\hat{4}}]^4 (W\cdot W)^2\ ,
 \la{3.25}
\ee

where

\be
 W\cdot W:=\ve^{RS} \ve_{\hR\hS} W_{1R}{}^{1\hR} W_{1S}{}^{1\hS}\ .
 \la{3.26}
\ee

Although this is mildly interesting it is not relevant to the UV problem because this invariant does not come into play by power counting in six dimensions. The one that does is one-eighth BPS. The superfield is $W_{1R}{}^{\hi\hj}$, $R\in\{3,4\}$, and the invariant is

\be
 \int\,d^6x\,du\,[D_2\ldots D_4 D^{\hat{1}}\ldots D^{\hat{4}}]^4 W_{1R}\cdot W_{1S} W_1{}^R\cdot W_1{}^S\ ,
 \la{3.27}
\ee

where the scalar product is on the  hatted indices and $\ve^{RS}$ is used to raise indices. This is the $d^6R^4$ invariant that is known to be divergent at three loops in $D=6$. As in the SYM case, we can form a second such invariant by harmonising the hatted sector.

%%%%%%%%%%%

\subsubsection{$D=7$}

%%%%%%%%%%%

The R-symmetry group for maximal SYM in $D=7$ is $Sp(1)$ and the field strength $W_{ij}$ is in the triplet representation. It obeys the differential constraint

\be
 D_{\a (i} W_{jk)}=0\ .
 \la{3.28}
\ee

The only possibility for Lorentz-invariant harmonics is given by the two-sphere $U(1)\bsh Sp(1)$. The field $W_{11}:=u_1{}^i u_1{}^j W_{ij}$ is analytic. The one-half BPS invariant, which is again not relevant to the UV question, is

\be
 I_{\half}=\int\,d^7x\,du (D_2)^8 <W_{11}^4>\ .
 \la{3.29}
\ee

The one-quarter BPS invariant cannot be obtained by these means and we postpone a discussion of it until after supergravity has been dealt with.

The field strength tensor for $D=7$ maximal supergravity transforms under the fourteen-dimensional representation of the R-symmetry group $Sp(2)$. It can be written
as a symmetric traceless $SO(5)$-tensor $W_{IJ}$, or alternatively as $W_{ij,kl}$, $i,j=1\ldots 4$, with the obvious symmetries. The differential constraints it obeys are

\be
 D_{\a i} W_{jk,lm}=\left(D_{\a[i} W_{jk],lm} -\frac{2}{3} \h_{i[j} D_\a^n W_{k]n,lm}\right) \ + (jk\leftrightarrow lm)\ .
 \la{3.30}
\ee

One-half BPS superfields are annihilated by two derivatives, the field strength is $W:=W_{12,12}$, and the $R^4$ integral invariant is

\be
 \int\,d^7 x\,du\, [D_3 D_4]^8 W^4 \ .
 \la{3.31}
\ee

It is not relevant to the counterterm discussion, but the one-quarter BPS one is relevant since it is known to be divergent at two loops. In this case, the field strength $W_{1R,1S}, R,S=3,4$ is annihilated by $D_1$ and the invariant is

\be
 I_{\qu}=\int\,d^7x\,du\,[D_2D_3D_4]^8 (W_{1R,1S} W_{1\ ,1}^{\,R\ S})^2\ .
 \la{3.32}
\ee

We return now to the question of the one-quarter BPS invariant in $D=7$ SYM. It will be convenient to write the scalars in $10-n$ dimensional SYM as a vector $W_I$ of $SO(n)$. The supercurrent is

\be
 T_{IJ}:=\tr \Scal{ W_I W_J-\frac{1}{n}\d_{IJ}W^K W_K}\
 \la{3.33}
\ee

The double-trace multiplets we are interested in occur in the product of two $T$s. For $D=4,5,6$ this gives rise to a one-half BPS multiplet which includes $(\tr F^2)^2$, a one-quarter BPS multiplet, which is the one we are interested in, a long multiplet which begins with a singlet scalar, and a multiplet whose leading component is in the same representation as $T$ and which satisfies a second-order, non-BPS differential constraint. However, in $D=7$, this multiplet does not separate from the one-quarter BPS one, at least not at lowest order. If one examines the lowest-order scalars in the product of two $T$s one sees that, for $D=4,5,6$, there are four representations of the R-symmetry group whereas, in $D=7$, there are only three. There is a singlet, the leading component of a long supermultiplet, a quintuplet, the leading component of the one-half BPS multiplet, and a triplet. The latter field starts off a multiplet which combines all the states of the lower-dimensional one-quarter BPS and non-BPS shortened multiplets; however, it is not clear if this is reducible in $D=7$ or whether the larger Lorentz symmetry group prevents this.

%%%%%%%%%%%%%%%%%%%%%%%%%%%%%%%%%%%%%%%%%%%%%%%%%%%%%%%%%%%%%%%%%%%%%%%%%%%

\section{BPS invariants in light-cone superspace}

%%%%%%%%%%%%%%%%%%%%%%%%%%%%%%%%%%%%%%%%%%%%%%%%%%%%%%%%%%%%%%%%%%%%%%%%%

In this section we show how the invariants described above can be rewritten in light-cone superspace for the case of $D=4$. The coordinates of ordinary superspace, in light-cone notation, are as follows:

\bea
 x^{\a\adt}&=&(x^{++},x^{--},x^{+-},x^{-+})\nn\w1
 \th^{\a i}&=&(\th^{+ i},\th^{- i})\nn\w1
 \bar\th^{\adt}_i&=&(\bar\th^+_i,\th^-_i)\ ,
 \la{4.1}
\eea

where the plus and minus indices indicate the transformation properties under the $SO(1,1)$ subgroup of the Lorentz group. The transverse coordinate $x^{-+}$ is the complex conjugate of $x^{+-}$. Light-cone superspace is the subspace defined by setting $\th^{- i}=\bar\th^-_i=0$. Note that the full supersymmetry algebra is not realised on this space, only the light-cone subalgebra. We shall denote the light-cone covariant derivatives by small letters,

\be
 d_{+ i}=D_{+ i}|_{\th^-=0};\qquad \bar d^-_i=\bar D_-^i|_{\th^-=0}\ .
 \la{4.2}
\ee

Note that the R-symmetry group is still manifest in this approach so that we can apply harmonic superspace techniques here as well. In the light-cone formalism it is permissible to divide by $\del_{++}$, which is regarded as an algebraic operation, and this allows one to eliminate all but the physical degrees of freedom. In particular, given a massless free fermion $\l_\a$, one can use the Dirac equation to write $\l_-$ in terms of $\l_+$,

\be
 \l_-=\frac{\del_{-+}}{\del_{++}}\l_+\ .
 \la{4.3}
\ee

$\cN=4$ SYM was formulated in light-cone superspace some time ago \cite{Brink:1982pd}
and used to give proofs of the UV finiteness of that theory \cite{Mandelstam:1982cb,Brink:1982wv}. Use is made of the light-cone gauge in which $A_{++}=0$ and $A_{--}$ is regarded as a dependent field. The multiplet of physical fields can then be packaged in a single light-cone chiral superfield which we shall discuss below. A similar formalism exists for $\cN=8, D=4$ supergravity.

One would expect that allowable counterterms in this approach would be integrals over light-cone superspace of local functions of the chiral superfield, with the proviso that inverse powers of $\del_{++}$ are allowed. In addition, one would demand that at least the counterterm corresponding to the first UV divergence should be invariant under the full Lorentz-covariant supersymmetry transformations. In \cite{Kallosh:2008mq} it was argued that these two requirements cannot be simultaneously satisfied and therefore that $\cN=8$ supergravity should be UV finite. In the following we shall show that this argument is not correct, even for BPS invariants, by explicitly rewriting the covariant counterterms as light-cone superspace integrals.

We begin with a simple example - an on-shell chiral field in $\cN=1,D=4$ supersymmetry. We shall show that a chiral invariant can be rewritten as a light-cone one.

The $\cN=1$ chiral superfield $\f$, satisfying $\bar D_{\adt}\f=0$, can be written

\be
 \f=-i\bar D_+ D_+ \frac{\f}{\del_{++}}
 \la{4.4}
\ee

so that a chiral Lagrangian of the form $\f^n$ can be written

\be
 \f^n=-i\bar D_+(\f^{n-1} D_+ \frac{\f}{\del_{++}})\ .
 \la{4.5}
\ee

Thus the chiral invariant is

\bea
 \int\,d^4x\,D^2\,\f^n:=\int\,D_+D_- \f^n&\sim&\int\,D_+\bar D_+ D_-(\f^{n-1} D_+ \frac{\f}{\del_{++}})\nn\w2
 &\sim&\int\,D_+\bar D_+ (\f^{n-2}D_-\f D_+ \frac{\f}{\del_{++}})\nn\w2
 &\sim&\int\,D_+\bar D_+(\f^{n-2} \del_{-+} D_+ \frac{\f}{\del_{++}} D_+ \frac{\f}{\del_{++}})\ ,
 \la{4.6}
\eea

where the $\sim$ sign means up to constants and spacetime derivatives and where we have used the free field equation $D_+ D_-\f=0$. In the final expression the odd integration is only over $\th^+,\bar\th^+$, so that any terms involving the minus coordinates must be total derivatives in spacetime and can therefore be dropped. Thus the integral can be written as

\be
 \int\,d^4x\,D^2\,\f^n=\int\,d_+\bar d_+ (\vf^{n-2} \del_{-+} d_+ \frac{\vf}{\del_{++}} d_+ \frac{\vf}{\del_{++}})\ ,
 \la{4.7}
\ee

where $\vf:=\f|_{\th^-=0}$ is the light-cone chiral superfield.

\vskip 1cm

The next example is the $\cN=2$ hypermultiplet, $\f_i,i=1,2$. It obeys the free on-shell constraints

\be
 D_{\a (i} \f_{j)}=\bar D_{\adt (i} \f_{j)}=0\ ,
 \la{4.8}
\ee

where the $SU(2)$ indices are raised or lowered with the epsilon tensor. In harmonic superspace, with coset $U(1)\bsh SU(2)$, the field $\f_1:=u_1{}^i\f_i$ is analytic,

\be
 D_{\a 1}\f_1=\bar D_{\adt}^2 \f_1=0
 \la{4.9}
\ee

as well as being harmonic analytic. Similar considerations apply to $\bar\f_1$. Thus we can write

\bea
 \f_1&=&-i\bar D_+^2 D_{+2} \frac{\f_1}{\del_{++}}\nn\w1
 &=&-iD_{+1}\bar D_+^2 \frac{\f_2}{\del_{++}}\ ,
 \la{4.10}
\eea

where $\f_2:=u_2{}^i \f_i$.

There is a one-half BPS integral invariant for the hypermultiplet given by

\be
 I=\int\,d^4x\,du\,[D_2\bar D^1]^2 (\f_1\bar\f_1)^2\ .
 \la{4.11}
\ee

Using \eq{4.10} we can write the integrand as

\be
 (\f_1\bar\f_1)^2=-iD_{+1}\bar D_+^2\left( \frac{\f_2}{\del_{++}}\f_1\bar\f_1\bar\f_1\right)\ .
 \la{4.12}
\ee

The invariant can thus be expressed as an integral over light-cone superspace with integrand given by

\be
 D_{-2}\bar D_-^1\left(\frac{\f_2}{\del_{++}}\f_1\bar\f_1\bar\f_1\right)\ .
 \la{4.13}
\ee

This can then be written in terms of light-cone superfields by expressing the minus fermionic derivatives in terms of plus ones with the aid of the fermionic equations of motion.

We now consider the $\cN=4$ SYM one-half BPS invariant (in the linearised case). The $\cN=4$ superfield $W_{12}$ (harmonics understood) is annihilated by $D_1,D_2,\bar D^3,\bar D^4$. It can be written

\bea
 W_{12}&=&iD_{+1}\bar D_+^3 \frac{W_{23}}{\del_{++}}\nn\w1
 &=&-iD_{+2}\bar D_+^4 \frac{W_{14}}{\del_{++}}\ .
 \la{4.14}
\eea

So the $\cN=4$ one-half BPS Lagrangian $(W_{12})^4$ can be written

\be
 (W_{12})^4\sim D_{+1}D_{+2}\bar D_+^3\bar D_+^4\left((W_{12})^2 \frac{W_{23}}{\del_{++}}\frac{W_{14}}{\del_{++}}\right)\ ,
 \la{4.15}
\ee

which therefore allows us to write the invariant as a light-cone superspace integral:

\be
 I=\int\,d^4x\,du\,(d_+)^4 (\bar d_+)^4 \left[ D_{-3} D_{-4} \bar D_-^1\bar D_-^2
 ((W_{12})^2 \frac{W_{23}}{\del_{++}}\frac{W_{14}}{\del_{++}})\right]_{\th^-=0}\ .
 \la{4.15.1}
\ee

To see that this can be expressed in terms of the light-cone chiral superfield let us define

\be
 w_{12}:=W_{12}|_{\th^-=0}\ .
 \la{4.16}
\ee

This satisfies

\be
 d_{+1}w_{12}=d_{+2}w_{12}=0\ ,
 \la{4.17}
\ee

and so can be written as

\be
 w_{12}=d_{+1}d_{+2} \vf_{--}\ .
 \la{4.18}
\ee

Since $\bar d_+^3 w_{12}=\bar d_+^4 w_{12}=0$, and since the harmonic dependence of $w_{12}$ is already taken care of by the derivatives in \eq{4.18}, it follows that $\vf_{--}$ can be taken to be chiral, $\bar d_+^i \vf_{--}=0$. Furthermore, due to the reality condition on $W$ we have

\be
 d_{+1}d_{+2}\vf_{--}=\bar d_+^3 \bar d_+^4 \bar\vf_{--}\ .
 \la{4.18.1}
\ee

We can therefore identify $\vf_{--}$ with the chiral superfield of reference \cite{Brink:1982pd}. The integrand of \eq{4.15.1} can be converted into the desired form by first acting with the $D_-$s on the fields, then rewriting the results in terms of $D_+$s by using the equations of motion and finally by evaluating the result at $\th^-=0$. This can be expressed in terms of $w_{ij}$ and hence in terms of $\vf_{--}$.

The situation in $\cN=8$ supergravity is very similar. We shall again consider only the one-half BPS $R^4$ integral. The field-strength superfield $W_{1234}$, which is annihilated by four $D$s and four $\bar D$s can be used to define a light-cone field strength $w_{1234}:=W_{1234}|_{\th^-=0}$. Because this is annihilated by $d_{+r},r=1\ldots 4$ it can be written in the form

\be
 w_{1234}=d_{+1} d_{+2} d_{+3} d_{+4} \vf_{----}\ ,
 \la{4.19}
\ee

where $\vf_{----}$ can be taken to be chiral in view of the $\bar d_+$ constraints. This is the $\cN=8$ light-cone superfield \cite{Brink:2008qc}. It also satisfies the reality constraint

\be
 d_{+1}d_{+2}d_{+3}d_{+4}\vf_{----}=\bar d_+^1 \bar d_+^2 \bar d_+^3 \bar d_+^4 \bar\vf_{----}\ .
 \la{4.20}
\ee

The one-half BPS invariant is an integral over sixteen thetas of $(W_{1234})^4$. This integrand can be written

\be
 (W_{1234})^4\sim D_{+1}D_{+2}D_{+3}D_{+4}\bar D_+^5 \bar D_+^6 \bar D_+^7 \bar D_+^8 \left(\frac{W_{2345}}{\del_{++}}\frac{W_{1346}}{\del_{++}}
 \frac{W_{1247}}{\del_{++}}\frac{W_{1238}}{\del_{++}}\right)\ ,
 \la{4.21}
\ee

from which we can easily show, using the same argument as above, that the invariant can indeed be recast as an allowed counterterm in light-cone superspace.

The above type of argument can easily be adapted to other integrals, with the conclusion that light-cone superspace considerations do not place any restrictions on the allowed counterterms, not even the one-half BPS ones. This is not altogether surprising since the formalism preserves only half of the full supersymmetry manifestly.

%%%%%%%%%%%%%%%%%%%%%%%%%%%%%%%%%%%%%%%%%%%%%%%%%%%%%%%%%%%%%%%%%%%%%%%%%%

\section{Off-shell harmonic formalism}

%%%%%%%%%%%%%%%%%%%%%%%%%%%%%%%%%%%%%%%%%%%%%%%%%%%%%%%%%%%%%%%%%%%%%%%%%%

In this section we review the off-shell formalism for $\cN=3,D=4$ SYM \cite{Galperin:1985uw,Delduc:1988cp} and extend it to $D>4$. On-shell this theory is the same as the maximal theory so that this approach allows us to preserve $q=12$ supersymmetries manifestly off-shell; a naive application of the superspace non-renormalisation theorem would then lead to the result that the one-half BPS counterterms are protected; in particular, it would predict that maximal SYM should be finite at four loops in $D=5$.

The gauge potential $A$ is a Lie algebra-valued superspace one-form with gauge transformation

\be
 A\rightarrow gAg^{-1} +dg g^{-1}, \qquad g\in G\ ,
 \la{5.1}
\ee

and we assume that the exterior derivative acts from the right. The field strength $F=dA + A^2$ transforms under the adjoint representation. In index notation we have

\be
 [\nab_A,\nab_B]=-t_{AB}{}^C\nab_C -F_{AB}\ ,
 \la{5.2}
\ee

where $A$ is a super-index and where it is understood that as an operator a field such as $F$ acts on a $\gg$-valued field via the commutator; $t_{AB}{}^C$ is the (flat) superspace torsion.

We shall use the following indices: $a=0,1,2,3$, vector index for $D=4$, $i=1,2,3$ $U(3)$ (anti-) fundamental representation index, $\a,\adt=1,2$ $D=4$ two-component spinor indices. Complex conjugation raises(or lowers) a $U(3)$ index.

%%%%%%%%%%%%%%%%%%%%%%%%%%%%%%%%%%%%%%%%%%%%%%%%%%%%%%%%%%%%%%%%%%%%%%%%%%%

\subsection{The on-shell theory}

%%%%%%%%%%%%%%%%%%%%%%%%%%%%%%%%%%%%%%%%%%%%%%%%%%%%%%%%%%%%%%%%%%%%%%%%

The basic constraints defining the theory are

\bea
 [\nab_{\a i},\bar\nab_{\bdt}^j]&=& i\d_i{}^j \nab_{\a\bdt}\nn\w1
 [\nab_{\a i},\nab_{\b j}]&=&i\ve_{ijk} \bar Z^k\nn\w1
 [\bar\nab_{\adt}^i,\bar\nab_{\bdt}^j]&=&i\ve^{ijk} Z_k\ ,
 \la{5.3}
\eea

the third being the (hermitean) conjugate of the second. The object $Z_i$, and its conjugate $\bar Z^i$, is in general a linear combination of a covariant derivative in the extra dimensions and a scalar field. So for $D=4$, $Z_i$ represents the three physical complex scalar fields while in $D=10$ $Z_i$ is a (covariant) derivative in the extra six dimensions regarded as three complex ones. In these two extreme cases the theory has $SO(1,3)\xz U(3)$ symmetry, but in the intermediate cases $U(3)$ is broken down to a subgroup. Nevertheless, we can treat the theory in all dimensions at once by regarding fields as operators as mentioned above.

The consequences of the constraints are analysed by means of the Bianchi identities. At dimension three-halves (taking $F_{ab}$ to have dimension two) one finds that the only fields allowed are the physical fermions $\l_\a$ and $\chi_{\a}^i$ and their conjugates. At dimension two one finds the supersymmetry variations of the fermions in terms of the dimension two field-strengths. The latter include $F_{ab}$, the additional components of the field strength in $D>4$, derivatives of scalars and scalar commutators. There are no new fields at this dimension. At dimension five-halves one finds the supersymmetry variations of the dimension-two field strengths. If one then applies the commutator of two fermionic derivatives to a fermion field and uses the information from the Bianchis one finds the field equations for the fermions. As an example, we find

\be
 \nab^\b{}_{\adt}\l_\b + [\bar Z^i,\bar\chi_{\adt i}]=0
 \la{5.4}
\ee

It is clear that since there are no non-physical fields the theory must be on-shell, by supersymmetry, and one can check this explicitly if one desires. The important point here is that restricting the theory to only twelve manifest supersymmetries does not lead to an off-shell theory given the basic constraints \eq{2.1}, as one would expect, given that it is well-known that this happens in $D=4$.

%%%%%%%%%%%%%%%%%%%%%%%%%%%%%%%%%%%%%%%%%%%%%%%%%%%%%%%%%%%%

\subsection{Off-shell with harmonics}

%%%%%%%%%%%%%%%%%%%%%%%%%%%%%%%%%%%%%%%%%%%%%%%%%%%%%%%%%%%%

In order to go off-shell we shall use harmonic superspace. The formalism is basically the same as in $D=4$. The harmonic variables parametrise the (full) flag space $\bbF_{1,2}(3)=H\bsh U(3)$, where the isotropy group $H=U(1)\xz U(1)\xz U(1)$. We use the same notation as in section 3, so numerical indices $1,2,3$ transform under the three different $U(1)$s in the isotropy group. The $\bar\del$-operator on $\bbF$, in the equivariant formalism, is $(D_1{}^2,D_1{}^3,D_2{}^3)$, while G-analytic fields are annihilated by $(D_{\a 1},\bar D_{\adt}^3)$.

The basic constraints \eq{5.3} are equivalent to

\be
 [\nab_{\a 1},\nab_{\b 1}]=[\bar \nab_{\adt}^3,\bar \nab_{\bdt}^3]=[\nab_{\a 1},\bar \nab_{\bdt}^3]=0\ .
 \la{5.7}
\ee

This is because the fields of the theory do not depend on $u$ so that these variables can be factored out in equations \eq{5.7} which therefore imply \eq{5.3}.

In order to discuss the off-shell theory it will be convenient to introduce some new notation. We set $d_g:=E^{\a 1} D_{\a 1} -\bar E^{\adt}_3 \bar D_{\adt}^3$ and let $d_h$ denote the $\bar\del$ operator on $\bbF$. Clearly we have

\be
 d_g^2=d_h^2=d_g d_h + d_h d_g=0\ .
 \la{5.8}
\ee

The field equations can be reinterpreted as the statement that there is a flat partial connection $A_g$ satisfying $d_h A_g=0$ \cite{Roslyi:1985hn}. Gauge transformations are also independent of the harmonic coordinates, $d_h g=0$. Since $A_g$ is flat it can be written in pure gauge form

\be
 A_g=d_g V V^{-1}
 \la{5.9}
\ee

where the group-valued function $V$ depends on $u$ in a restricted fashion, since $d_h A_g=0$. We can now make a generalised gauge transformation with a $u$-dependent gauge group element to set $A_g=0$. This induces a gauge field in the harmonic direction, $A_h= d_h V^{-1} V$, and the residual gauge invariance now consists of G-analytic transformations. So we have shown that the original flat partial gauge field $A_g$ is equivalent to a pure gauge connection $A_h$. Clearly, given such an $A_h$ we can go back to the original $A_g$. The theory can be taken off-shell by allowing $A_h$ to be an arbitrary partial gauge field satisfying $d_g A_h=0$ and subject to Grassmann-analytic gauge transformations. The Chern-Simons action

\be
 S=\int\,d^D x\, du\, d y\,[D_2 D_3\bar D^1\bar D^2]^2\, Q(A_h) \ ,
 \la{5.10}
\ee

where $Q(A_h)$ is the Chern-Simons three-form in the harmonic directions, leads to the equation of motion $F_{hh}=0$. This implies that $A_h$ is locally pure gauge, but since $\bbF$ is a non-trivial space, this need not be the case globally. So the action \eq{5.10} leads to the correct equations of motion only when $A_h$ is restricted to belong to the class of trivial gauge fields. Note that this construction is really a variation of the Ward observation which relates self-dual Yang--Mills gauge fields to holomorphic vector bundles on twistor space that are trivial on each twistor line, the analogue of the latter being the flag space $\bbF_{1,2}(3)$. The necessity of restricting the fields in this way was first pointed out in \cite{Roslyi:1985hn}.

In \eq{5.10}, $du$ is the standard measure on $\bbF$ and $y$ denotes the additional spacetime coordinates. The $U(1)$ charges in the measure are exactly matched by those of the integrand provided that the $y$ integration is neutral. One might worry that the three dimensions are complex but there is a real structure which one can introduce on $\bbF$ which guarantees the reality of the action.

The integrand is G-analytic, i.e. annihilated by $D_{\a 1}, \bar D_{\adt}^3$. Now in flat superspace, as can be seen from \eq{2.1}, the covariant derivatives involve derivatives with respect to all the spacetime coordinates multiplied by linear factors of $\theta$, so that G-analytic functions will depend on shifted $x,y$ variables. For example, consider $D=10$. The extra coordinates can be taken to be three complex ones $y^i$ together with their conjugates $y_i$. In this case we can define $y_I$ and $y^I$ by multiplying by factors of $u$ and its inverse. The G-analytic shifted coordinates are given by

\be
 \hat y^I:=y^I-\th^{\a 1} D_{\a 1} y^I +\bar\th^{\adt}_3\bar D_{\adt}^3 y^I\ ,
 \la{5.11}
\ee

and similarly for $y_I$ (as well as for $x$). We can chose from among these variables in intermediate dimensions where not all of the six additional $y$ coordinates will be non-zero.

%%%%%%%%%%%%%%%%%%%%%%%%%%%%%%%%%%%%%%%%%

\subsection{Application to SYM divergences}

%%%%%%%%%%%%%%%%%%%%%%%%%%%%%%%%%%%%

The Feynman rules for the off-shell $\cN=3,D=4$ SYM theory were written down in \cite{Galperin:1985uw,Delduc:1988cp}. If we assume that this can be repeated for $D>4$, then we can use this formulation as an off-shell version of maximal SYM with $q=12$ linearised supersymmetries. Manifest Lorentz invariance is lost in $D>4$, but this need not be a problem as long as it can be shown that the first UV divergence does indeed correspond to a fully covariant counterterm. There does not seem to be a problem with this in $D=5,6$ where the one-quarter BPS invariants are given as integrals over twelve thetas, but in $D=7$ life is a little more complicated because it is more difficult to write this invariant in this way.

An immediate consequence of the off-shell harmonic formalism is that one-half BPS invariants cannot occur as putative counterterms in maximal SYM. Instead, the first possible divergences that can arise are the one-quarter BPS ones. If the assumptions made above are correct, this would bring the non-renormalisation theorems into full agreement with the computations. The only problem with this is that the formalism naively looks too strong in $D=7$ where it is not immediately obvious how to write the one-quarter BPS invariant as a twelve-theta integral. It is probable that this is just a technicality.

%%%%%%%%%%%%%%%%%%%%%%%%%%%%%%%%%

\subsection{Supergravity}

%%%%%%%%%%%%%%%%%%%%%%%%%%%%%%

An obvious question to ask is whether the above construction can be made to work in supergravity. However, it is not at all easy to see how to do this. One possible idea, in the linearised theory, is to try taking the square of the $\cN=3$ abelian SYM theory.
Since the states of maximal supergravity can be obtained by squaring the states of $\cN=4$ SYM, and since the latter is the same as the $\cN=3$ theory on-shell, one might wonder if squaring the $\cN=3$ theory off-shell might lead to an off-shell version of $\cN=8$ supergravity with $\cN=6$ supersymmetries linearly realised, i.e. $q=24$. It is possible to do this formally by mimicking the GIKOS construction described above, but the resulting off-shell theory seems to be an $\cN=6$ conformal supergravity theory rather than the desired Poincar\'e one. It is usually thought that conformal supergravity theories do not exist for $\cN>4$, but this theory presumably has an infinite number of physical fields arising from the infinite number of auxiliary fields in the SYM theory. Moreover, there is no guarantee that an interacting version exists. It is perhaps not so surprising that this construction does not work since the known off-shell formulations of supergravity theories are always made up of more than one supermultiplet; in $D=4$ one can always view off-shell Poincar\'e theories as being comprised of a Weyl supermultiplet together with one or more compensators. From this point of view the squaring construction might have more chance of working out for $\cN=5$ where one can construct multiplets with maximal spin $3/2$ as well as spin $2$ by ``multiplying'' an $\cN=3$ vector multiplet with an $\cN=2$ matter multiplet.

Indeed, $\cN=5$ would be sufficient to account for the currently known computational results. In particular, it would rule out one-half BPS counterterms. In $D=4$, the existence of such a formalism would postpone the predicted onset of UV divergences to five loops, because there is no candidate four-loop counterterm.

%%%%%%%%%%%%%%%%%%%%%%%%%%%%%%%%%%%%%%%%%%%%%%%%%%%%%%%%%%%%%%%%%

\section{One-half susy plus one}\label{halfplusone}

%%%%%%%%%%%%%%%%%%%%%%%%%%%%%%%%%%%%%%%%%%%%%%%%%%%%%%%%%%%%%%%%%%%%

The reason why one is forced to harmonic superspace and an infinite number of auxiliary fields in order to construct off-shell versions of supersymmetric theories with a large amount of supersymmetry is that off-shell representations with a finite number of fields are not compatible with the bosonic, in particular Lorentz, symmetries. However, if one is prepared to reduce the bosonic symmetry group, it is possible to find finite sets of auxiliary fields. The first example of this was given in \cite{Baulieu:2007ew} where an off-shell version of $D=10$ SYM was written down with $SO(1,1)\xz Spin(7)$ symmetry\footnote{Ref  \cite{Baulieu:2007ew} was inspired by earlier work \cite{octonions,BKS,BBT}.}. This version of the theory has nine supersymmetries and only a seven-plet of dimension two auxiliary scalars. This is of great interest in the context of UV divergences since off-shell formulations with one more than half of the total number of supersymmetries could be expected to rule out
one-half BPS counterterms.

In this section we shall discuss one-half susy plus one formulations of maximal SYM and SG theories reduced to two dimensions. The reduction allows us to maintain Lorentz symmetry, while invariants can still be studied even though they do not occur as counterterms in this setting.

Let us begin with the linearised theories. For SYM in $D=2$ we have eight scalars and eight left- and right-moving fermions. The on-shell multiplet, with $(8,8)$ supersymmetry is given by a scalar superfield $W_a$ satisfying

\bea
 D_{\a +} W_a&=& (\s_a)_{\a \adt} \psi_{\adt +} \nn\w1
 D_{\adt -} W_a&=& (\s_a)_{\adt \a} \psi_{\a -}\ ,
 \la{6.1}
\eea

where $a,\a,\adt$ are indices for the $(8_v,8_s,8_c)$ representations of $Spin(8)$, and $\s_a$ are the spin matrices. The supersymmetry algebra is

\bea
 [D_\a,D_\b]&=&2i\d_{\a\b} \del_{++}\nn\w1
 [D_{\adt},D_{\bdt}]&=&2i \d_{\adt\bdt} \del_{--}\nn\w1
 [D_\a,D_{\bdt}]&=& 0\ ,
 \la{6.2}
\eea

where the spacetime coordinates are $(x^{++},x^{--})$. We can go off-shell by reducing the supersymmetry from $(8,8)$ to $(8,1)$.

In this case we get

\bea
 D_\a W_a&=& (\s_a)_{\a\adt} \psi_{\adt +}\nn\w1
 D_- W_a&=& \psi_{a -}\ .
 \la{6.3}
\eea

We then find that

\be
 D_-\psi_{\adt}:=G_{\adt}
 \la{6.4}
\ee

defines a superfield whose leading component is a set of dimension-two auxiliary fields. There are no other independent component fields and so we find a representation with $(16+16)$ components off-shell. This theory still has $SO(8)$ symmetry, but it has to be reduced to $Spin(7)$ in the gauge theory because one of the components of $G_{\adt}$ becomes identified with the YM field strength.

Equations \eq{6.3} can easily be modified to give an off-shell multiplet with $(16,1)$ supersymmetry which corresponds to linearised maximal supergravity in $D=2$. We have

\bea
 D_{i +} W_\a &=& (\S_i)_{\a\adt} \psi_{\adt +}\nn\w1
 D_-W_\a &=& \psi_{\a -}\
 \la{6.5}
\eea

where $i=1,\ldots 16$ is an $SO(16)$ vector index and $\a,\adt=1,\ldots 128$ are Weyl spinor indices in $Spin(16)$. There are $128$ auxiliary fields are defined by

\be
 D_-\psi_{\adt +}:=G_{\adt}\ .
 \la{6.6}
\ee

In this case the full non-linear theory still has $SO(16)$ symmetry as we shall see.

%%%%%%%%%%%%%%%%%%%%%%%%%%%%%%%%%%%%%%%%

\subsection{Maximal SYM with $(8,1)$ supersymmetry}

%%%%%%%%%%%%%%%%%%%%%%%%%%%%%%%%%%%%%%%%%

As noted above, in order to accommodate a gauge field it is necessary to reduce the R-symmetry group form $SO(8)$ to $Spin(7)$. We shall use $a=1,\ldots 8$ to denote a $Spin(7)$ spinor index and $i=1,\ldots 7$ to denote an $SO(7)$ vector index. The constraints on the superspace field strength tensor are

\be
 F_{a +,b+}= F_{--}=0\ ,
 \la{6.7}
\ee

while the scalar superfield $W_a$ is equal to $F_{a +,-}$. From the Bianchi identities we find

\bea
 F_{++,+}&=&-i\psi_+\nn\w1
 F_{--,a +}&=&-i\psi_{a -}\nn\w1
 F_{++,--}&=&\frac{1}{8}\nab_{a +}\psi_{a -}=-\nab_- \psi_+\ ,
 \la{6.8}
\eea

and
\bea
 \nab_{a+} W_b&=&\d_{ab} \psi_+ + \nab_{[a +} W_{b]}\nn\w1
 \nab_- W_a&=& \psi_{a -}\ .
 \la{6.9}
\eea

However, these constraints do not completely define the desired multiplet; it is necessary to impose a secondary constraint \cite{Baulieu:2007ew}

\be
 \nab_{[a+} W_{b]}=(\c^i)_{ab} \l_{i+}\ ,
 \la{6.10}
\ee

where $\c_i$ denotes the gamma matrices for $Spin(7)$, so that $\psi_+$ and $\l_{i+}$ together give the eight left-moving physical fermions. The auxiliary fields are defined by $G_i:=\nab_-\l_{i +}$; there are only seven as the field count is completed by the off-shell gauge field.

The solution to these constraints is easy enough to find at the linearised level; in fact, one needs two prepotentials, $\L_{i -7}$ which has dimension $-\frac{5}{2}$, and $M_{-6}$ which has dimension $-3$.

%%%%%%%%%%%%%%%%%%%%%%%%%%%%%%%%%%%%%%%%%

\subsection{Maximal supergravity in $D=2$ with $(16,1)$ supersymmetry}

%%%%%%%%%%%%%%%%%%%%%%%%%%%%%%%%%%%%%%%%%%

The theory will be described in a curved superspace. The structure group is taken to be $Spin(1,1)\xz SO(16)$, reflecting the fact that the tangent bundle splits into even and odd components. A set of basis forms is given by $E^A:=(E^a,E^{\a +}, E^{-})$, with $E^a=(E^{++},E^{--})$. The index $\a=1\ldots 16$ is a vector index for $SO(16)$ while the pluses and minuses denote $Spin(1,1)$ representations. We introduce a  connection one-form $\O_A{}^B$ taking its values in $\gs\gp\gi\gn(1,1)\oplus \gs\go(16)$. The non-zero components of the connection are

\bea
 \O_{++}{}^{++}&=& 2A \nn\w1
 \O_{--}{}^{--}&=& -2A \nn\w1
 \O_{-}{}^{-}&=&-A\nn\w1
 \O_{\a +}{}^{\b +}&=& \O_\a{}^\b + \d_\a{}^\b A\ ,
 \la{6.11}
\eea

where $A$ is the Lorentzian connection and $\O_\a{}^\b$ is the  $\gs\go(16)$  connection. The curvature two-form $R_A{}^B$ has a similar decomposition; we denote the Lorentz curvature by $F$ and the $\gs\go(16)$ curvature by $R_\a{}^\b$. The torsion and curvature tensors are defined in the usual manner

\bea
 T^A&=& DE^A\ := d E^A + E^B \O_B{}^A \nn\w1
 R_A{}^B&:=& d\O_A{}^B + \O_A{}^C \O_C{}^B\ .
 \la{6.12}
\eea

The Bianchi identities are

\bea
 DT^A-E^B R_B{}^A&=& 0\nn\w1
 DR_A{}^B&=&0\ .
 \la{6.13}
\eea

It is worth noting that there are no Dragon identities in $d=1,2$, in particular, the curvature tensor is not determined in terms of the torsion from the Bianchi identities.

The physical fields of the supergravity multiplet consist of 128 scalars and 128 spinors which we shall assume to be described by an $SO(16)\bsh E_8$ sigma model. A basis for the Lie algebra $\ge_8$ can be split into a set of $\gs\go(16)$ generators $M_{\a\b}=-M_{\b\a}$ together with a set of 128 coset generators $N_I$ transforming under one of the two Weyl spinor representations of $\gs\gp\gi\gn(16)$. The $\ge_8$ algebra is given by

\bea
 [M_{\a\b},M^{\c\d}]&=& 4 \d_{[\a}{}^{[\c} M_{\b]}{}^{\d]} \nn\w1
 [M_{\a\b},N_I]&=& (\S_{\a\b})_I{}^J N_j\nn\w1
 [N_I,N_J]&=& k(\S^{\a\b})_{IJ}M_{\a\b}\ ,
 \la{6.14}
\eea

where $\S_\a$ denotes the $Spin(16)$ sigma matrices, $\S_{\a\b}:=\S_{[\a}\S_{\b]}$, and the number $k$ is a real constant. The sigma model is formulated in terms of an element $\cV$ of $E_8$. The Maurer-Cartan form $\F$ splits into an $\ge_8$-valued component $P$ and an $\gs\go(16)$-valued component which will be identified with the $\gs\go(16)$ part of the superspace connection,

\be
 \F= d\cV\,\cV^{-1}:= P + \O\ ,
 \la{6.15}
\ee

The fact that $d\F + \F^2=0$ implies that

\bea
 DP&=& 0 \nn\w1
 R&=&-P^2\ ,
 \la{6.16}
\eea

where $R:=\half R^{\a\b} M_{\a\b}$.

To analyse the above equations we shall assume that only the fields of the off-shell supergravity multiplet are present and then check the Bianchi identities to ensure that the system is consistent. In addition to the scalars, contained in $\cV$, and the spinors, there is also a set of dimension-one auxiliary scalars $G_{I'}$, where the primed index denotes the second Weyl spinor representation of $\gs\gp\gi\gn(16)$.

The only non-zero components of the dimension-zero torsion are

\bea
 T_{\a +,\b +}{}^{++}&=&-2i \d_{\a\b} \nn\w1
 T_{-,-}{}^{--}&=&-2i\ ,
 \la{6.17}
\eea

where here, and below, commas are used to separate indices. At dimension one-half all components of the torsion must vanish as the spinor fields transform according to the spinor representations of $Spin(16)$. On the other hand, the dimension one-half components of $P$ are given by

\bea
 P_{\a + I}&=&i(\S_\a)_{IJ'} \L_{+ J'}\nn\w1
 P_{- I}&=&i \L_{-} \ .
 \la{6.18}
\eea

The second of these equations is a definition, but the first is a constraint. Using \eq{6.18} in the identity $DP=0$ at dimension one one finds

\bea
 \nab_{\a +} \L_+&=& \S_\a P_{++}\ ,\qquad \nab_- \L_{\a +}=G \nn\w1
 \nab_{\a +} \L_-&=&-\S_\a G\ ,\ \qquad\nab_- \L_-\ \,= P_{--}\ ,
 \la{6.19}
\eea

where $\nab$ denotes the covariant derivative with respect to both groups. In \eq{6.19} the $SO(16)$ spinor indices are not explicitly indicated, and we shall use this convention in the following whenever there is no possibility of confusion.

At dimension one the $SO(16)$ curvature is determined as a bilinear in the sigma model fields by the Maurer-Cartan equation, and we may choose $T_{ab}{}^c=0$. We can then complete the determination of the dimension-one torsion by means of the first Bianchi identity. The non-zero dimension-one torsions are found to be:

\bea
 T_{++,\b +}{}^{\c +}&=&ik A_{++\b}{}^\c\nn\w1
 T_{--,\b +}{}^{\c +}&=&-ik A_{--\b}{}^\c\nn\w1
 T_{++,-}{}^{\c +}&=&-2ik B^\c\nn\w1
 T_{--,\b +}{}^-&=&2ik B_\b\ ,
 \la{6.20}
\eea

where $A$ and $B$ are given in \eq{6.22}. The non-zero dimension-one curvatures are

\bea
 F_{\a +,-}&=&-2k B_\a\nn\w1
 R_{\a +,\b +,\c\d}&=& 2k\left(4 \d_{(\a[\c} A_{++\b)\d]}-\d_{\a\b} A_{++\c\d}\right)\nn\w1
 R_{\a +,-,\c\d}&=&-2k B_{\a\c\d}-4k \d_{\a[\c} B_{\d]}\nn\w1
 R_{-,-,\c\d}&=&-2k A_{--\c\d}\ .
 \la{6.21}
\eea

The bilinears $A$ and $B$ are

\bea
 A_{++\a\b}&=&\L_+\S_{\a\b}\L_+ \nn\w1
 A_{--\a\b}&=&\L_-\S_{\a\b}\L_- \nn\w1
 B_\a&=&\L_+\S_\a \L_-\nn\w1
 B_{\a\b\c}&=&\L_+ \S_{\a\b\c}\L_-\ .
 \la{6.22}
\eea

The dimension-three-halves torsions are

\bea
 T_{++,--}{}^{\a +}&:=&\Psi^{\a +}=2k(P_{--}\S^\a\L_+-G\S^\a\L_-)\nn\w1
 T_{++,--}{}^{-}&:=&\Psi^-=-2k ( P_{++}\L_- + G \L_+)\ ,
 \la{6.23}
\eea

while the non-zero dimension-three-halves curvatures are

\bea
 F_{--,\a +}&=&-i\Psi_\a^+ \nn\w1
 F_{++,-}&=&-i\Psi^-\nn\w1
 R_{++,\b +,\c\d}&=&2ik P_{++}\S_{\c\d}\S_\b\L_+ \nn\w1
 R_{--,\b +,\c\d}&=&2ik P_{--}\S_{\c\d}\S_\b\L_+ \nn\w1
 R_{++,-,\c\d}&=&2ik P_{++}\S_{\c\d} \L_-\nn\w1
 R_{--,-,\c\d}&=&2ik P_{--}\S_{\c\d} \L_-
 \la{6.24}
\eea

Finally, at dimension two, the curvatures are

\bea
 F_{++,--}&=&2k\left(P_{++} P_{--}+ \nab_{++}\L_- \L_- + \nab_{--}\L_+ \L_+ + G^2-\frac{1}{4} A_{++\a\b} A_{--}{}^{\a\b}\right)\nn\w1
 R_{++,--,\c\d}&=& 2k P_{++}\S_{\c\d} P_{--}\ .
 \la{6.25}
\eea

We have explicitly checked that the Bianchi identities are satisfied up to and including those at dimension two. It is worthwhile pointing out that some of the constraints imposed on the torsion at dimension one-half partially determine the dimension one-half $SO(16)$ connection, while this quantity is also completely specified by the Maurer-Cartan equation for the sigma model. It is not obvious that these two constraints are compatible but the fact that the Bianchis are satisfied confirms that they are. Indeed, to verify the dimension-two Bianchi identities it is necessary to make use of some lengthy $SO(16)\ \S$-matrix gymnastics.

It will be seen from the above results that the only independent component fields are those of the off-shell supergravity multiplet, the Lorentzian curvatures being determined in terms of them as composites. Since the two-dimensional zweibein is pure gauge up to a conformal factor it follows that the latter is determined in terms of the sigma model fields, and similarly for the gravitini. This implies that the formalism presented here is not superconformal. If desired one could remedy this situation by making a super-Weyl transformation, but this would also have an effect on the Maurer-Cartan equation.

Finally, it is easy to use the above formalism to derive the action. The Lorentzian curvature $F$ is a closed two-form which has the right dimension to be a Lagrangian two-form in the ectoplasm approach \cite{ectoplasm,purectoplasm}. The action is given by

\be
 S=\int\, d^2 x\, \ve^{\mu\nu} F_{\mu\nu}(x,0) \ ;
 \la{6.26}
\ee

its invariance under local supersymmetry transformations follows from the fact that $F$ is closed as a superform. If this is rewritten in a preferred basis the leading term in the Lagrangian is given by the determinant of the component zweibein multiplied by the leading component (in a theta-expansion) of $F_{++,--}$. The latter is given in the first equation in \eq{3.9} which we see has the correct form for a sigma model action.

%%%%%%%%%%%%%%%%

\subsection{Discussion}

%%%%%%%%%%%%%%%

We have seen that the constraints to the $(8,1)$ off-shell version of maximal SYM can be solved at the linearised level; provided that this solution can be extended to the full theory, we would expect to be able to use the standard superspace non-renormalisation theorems in this theory. It is not difficult to see that it is not possible to construct the one-half $F^4$ BPS invariant as an integral over the full 9 odd dimensional superspace involving only the background potentials and gauge fields. Moreover, one can lift the $D=2$ analysis given here to higher dimensions at the cost of manifest Lorentz covariance.

In the $(16,1)$ version of maximal supergravity things are slightly different. Although one would expect to be able to solve the constraints straightforwardly, the analysis does not lift to higher dimensions quite so easily. This is due to extra off-shell degrees of freedom which are not present in the special case of $D=2$. So more work remains to be done in this case, but there does not seem to be any fundamental obstruction to formulating maximal supergravity theories with seventeen supersymmetries in higher-dimensional spacetimes. Provided that this programme can be implemented, one would again conclude that one-half BPS invariants are forbidden as counterterms.

%%%%%%%%%%%%%%%%%%%%%%%%%%%%%%%%%%%%%%%%%%%%%%%%%%%%%%%%%%%%%%%%%%%

\section{The algebraic method}

%%%%%%%%%%%%%%%%%%%%%%%%%%%%%%%%%%%%%%%%%%%%%%%%%%%%%%%%%%%%%%%%%%%

As we shall now see, it is also possible to derive essentially equivalent non-renormalisation theorems making use of the full non-linear supersymmetry Ward identities\footnote{These should perhaps more properly be called Slavnov--Taylor identities since we will also encompass gauge theories. But we will refer to them here generically as Ward identities.}  Since the full supersymmetry closes only modulo the classical equations of motion, it has often been dubbed ``on-shell'' supersymmetry. In addition to being nonlinear, the lack of off-shell closure seriously complicates the analysis of the related Ward identities. But, using the Batalin-Vilkovisky formalism, it is still possible \cite{HoweBV}. The basis of the method was first derived in \cite{Sor,Sor4} in the context of $\cN=2$ and $\cN=4$ super-Yang--Mills theory in four dimensions. We will follow a more modern version of the method which does not involve the use of twisted variables \cite{N2}.  For this purpose, one considers the quantum field theory in component formalism; then the non-linear supersymmetry Ward identities become Ward identities which require one to introduce sources for the supersymmetry transformations of the fields.

The algebraic renormalisation proof goes in two main steps. The first step consists in using the Callan--Symanzik equation to relate the beta function corresponding to the first logarithmic divergence for a given counterterm operator to the anomalous dimension describing the mixing of that operator with the classical Lagrangian operator, both considered as local operator insertions into the generating functional of 1PI diagrams. The second step consists in using the descent equations of the supersymmetry Ward identities in order to relate the mixing under renormalisation of various operators within the same chain of operators. The basic result is that any counterterms in the supermultiplet of a $\ft12$ BPS operator defines an irreducible cocycle of the descent equations. The classical action of maximally supersymmetric Yang--Mills theory corresponds to the only such cocycle that stops at form-degree $D-5$, and, as such, all the counterterms associated to $\ft12$ BPS operators are protected in maximally supersymmetric Yang--Mills theories. It turns out that super Yang--Mills counterterms are forbidden by the supersymmetry Ward identities if and only if they are associated to $\ft12$ BPS operators. As we will see, the same argument extends to the case of maximal supergravity in four dimensions, and gives the result that the three-loop invariant is not allowed by the supersymmetry Ward identities. The extension of our understanding of the length of the cocycle in Yang--Mills theory to Einstein theory suggests that the $D=4$ three-loop invariant is the only supersymmetry invariant that can be shown to be disallowed by the Ward identity using this method. Since all counterterms that are not disallowed by Ward identities usually correspond to actual divergences if these are otherwise allowed by power counting, the hypothetical finiteness of maximal supergravity would require the existence of some yet undiscovered hidden mechanism of the theory.

As far as is known at present, the algebraic method and the background field method in superspace give the same divergence predictions. However, we have not completely proven that the cohomology of the cocycles in maximally supersymmetric supergravity does not have some very peculiar behaviour that could give stronger results. Note that the computation of the representative cocycles in supergravity is essentially the same as that of the ``ectoplasmic'' cocycle \cite{ectoplasm,purectoplasm} in superspace. The key statement translates in this language into the property that the component of the superform associated to a counterterm of lowest degree on the body must be of higher degree than the one associated to the classical action. Nevertheless, some components of the superforms vanish in the Wess--Zumino gauge of the component formalism, and it is not yet clear how this would affect the cohomology.

In order to handle the supersymmetry Ward identities, it turns out to be very useful to introduce commuting spinors. This permits one to restrict the number of sources by defining the supersymmetry transformations via a nilpotent differential as in the case of the BRST formalism. In a supersymmetric gauge theory, the non-linear representation of the supersymmetry algebra on the fields closes only up to field-dependent gauge transformations. J.\ Dixon solved this problem by introducing a single extended BRST operator for both supersymmetry and gauge transformations \cite{Slavnov,Taylor}. However, in order to distinguish the Ward identities associated to rigid symmetries of the theory from the BRST symmetry associated to gauge symmetry, we prefer to introduce distinct operators. This requires the introductions of extra fields in the theory, the so-called shadow fields \cite{shadow}.

The basis of the algebraic renormalisation method is the quantum action principle, which states that the derivative of the 1PI generating functional $\Gamma$ with respect to a source or a parameter is equal to the insertion of a local functional into the 1PI generating functional \cite{PS}. The theorem is also valid for a Slavnov--Taylor like operator, that is, a quadratic functional operator containing one derivative with respect to a field and another with respect to a source.
\be \frac{ \partial\, }{\partial \lambda} \Gamma = \Bigl[ \int d^D x \, F_\lambda ( \varphi , \partial \varphi ) \cdot \Gamma \Bigr] \hspace{10mm}\int d^D x   \frac{\delta  \Gamma}{\delta \varphi (x) } \frac{ \delta \Gamma}{ \delta V(x) } =    \Bigl[ \int d^D x  \, A ( \varphi , \partial \varphi ) \cdot \Gamma \Bigr] \ee
This permits one to prove that the derivatives of the 1PI generating functional with respect to either the coupling constant $g$ or to the renormalisation scale $\mu$ entering the renormalisation group equation are given at first order and up to BRST-exact terms by local functionals of the fields invariant under the action of the classical non-linear supersymmetry. This is almost all that we really need to prove our non-renormalisation theorem, and therefore we will not introduce explicitly here all the sources and the Slavnov--Taylor operators, although their introduction was necessary to prove the above lemma.

Before studying the theories that we are interested in, we will illustrate the formalism by a simple example, namely the massless Wess--Zumino model in four dimensions with one single supermultiplet. In this case, we will introduce all the needed sources and we will write down explicitly the Ward identities.

\subsection{A simple example: the Wess--Zumino model}

The supermultiplet of the theory is composed of one scalar field $\phi$, one pseudo-scalar $\phi_5$, and a Majorana spinor $\lambda$. We use the convention that ${\gamma_5}^2 = - 1$. Although one can define a linear realisation of supersymmetry on these fields by introducing auxiliary fields, we will not do so here in order to exhibit the fact that the absence of auxiliary fields is not an obstacle within the Batalin--Vilkovisky formalism. For simplicity, we will omit mass terms from the WZ model.

We renormalise the fields by a factor linear in the coupling constant $g$, in such a way that the supersymmetry transformations do not depend upon $g$. They are given by
\begin{gather}
\delta \phi = \scal{  \overline{\epsilon} \lambda } \hspace{10mm} \delta \phi_5 = \scal{ \overline{\epsilon} \gamma_5 \lambda} \CR
\delta \lambda =  - i \sa \partial ( \phi + \phi_5 \gamma_5 ) \epsilon + \frac{1}{2}  ( \phi + \phi_5 \gamma_5 )^2  \epsilon  \end{gather}
We introduce the associated action
\begin{multline}  \Sigma = -\frac{1}{2 g^2} \int d^4 x \biggl( \partial_\mu \phi \partial^\mu \phi + \partial_\mu \phi_5 \partial^\mu \phi_5 + i \scal{ \overline{\lambda} \sa \partial \lambda } + \scal{ \overline{\lambda} \phi \lambda } -  \scal{ \overline{\lambda} \phi_5 \gamma_5  \lambda } + \frac{1}{4} \scal{ \phi^2 + \phi^2_5 }^2 \biggr)\\* + \int d^4 x \biggl( \phi^\q \scal{ \overline{\epsilon} \lambda} + \phi_5^\q \scal{ \overline{\epsilon} \gamma_5 \lambda}  - \scal{ \overline{\lambda}^\q \bigl[ - i \sa \partial  ( \phi + \phi_5 \gamma_5 ) + \sfrac{1}{2} ( \phi + \phi_5 \gamma_5 )^2 \bigr] \epsilon} \biggr)\\* + \frac{g^2}{4} \int d^4 x ( \overline{\epsilon} \gamma^\mu \epsilon ) \scal{ \overline{\lambda}^\q \gamma_\mu \lambda^\q } \hspace{30mm} \end{multline}
where the fields with a $(Q)$ superscript are sources for the supersymmetry transformations of the fields and $\epsilon$ is a {\em commuting} Majorana spinor parameter. This action satisfies the Ward identity
\be \int d^4 x \left ( \frac{ \delta^R \Sigma }{ \delta \phi} \frac{ \delta^L \Sigma }{ \delta \phi^\q } + \frac{ \delta^R \Sigma }{ \delta \phi_5} \frac{ \delta^L \Sigma }{ \delta \phi^\q_5 } + \frac{ \delta^R \Sigma }{ \delta \lambda} \frac{ \delta^L \Sigma }{ \delta \overline{\lambda}^\q} \right) = 0 \ee
In the absence of a non-trivial anomaly, which is the case for supersymmetry, one can prove that there exist a renormalisation scheme such that the quantum generating functional of 1PI graphs also satisfies the supersymmetry Ward identity.

Although there is no gauge invariance in the Wess--Zumino model, it is nonetheless useful to introduce a BRST operator. This is trivial at the classical level, but it permits one to define the one-to-one correspondence between the BRST cohomology classes of local functionals of fields in the functional formalism and the composite operators in the operator formalism. Indeed, the fields in the functional formalism are arbitrary whereas the operators in the operator formalism satisfy the equations of motion. The linearised Slavnov--Taylor operator of a theory without gauge invariance defines the Kozul--Tate differential associated to the equations of motion, and its cohomology is isomorphic to the set of functionals of the fields satisfying the equations of motion \cite{Henneaux}.

In order to define consistently the two needed Ward identities, one introduces sources for each field with respect to its supersymmetry transformation, its BRST transformation, and also the successive action of the supersymmetry and the BRST transformation.

The supersymmetry Ward identity finally reads
\begin{multline} \Qsla(\Gamma) \equiv  \int d^4 x \left ( \frac{ \delta^R \Gamma }{ \delta \phi} \frac{ \delta^L \Gamma }{ \delta \phi^\q } + \frac{ \delta^R \Gamma }{ \delta \phi_5} \frac{ \delta^L \Gamma }{ \delta \phi^\q_5 } + \frac{ \delta^R \Gamma }{ \delta \lambda} \frac{ \delta^L \Gamma }{ \delta \overline{\lambda}^\q} \right . \\* \left . - \phi^\aBRST \frac{\delta^L \Gamma}{\delta \phi^\qs}   - \phi_5^\aBRST \frac{\delta^L \Gamma}{\delta \phi_5^\qs}  - \overline{\lambda}^\aBRST  \frac{\delta^L \Gamma}{\delta \overline{\lambda}^\qs}  \right)
  = 0 \end{multline}
Note that the additive source component in this identity is completely trivial in this case since there is no gauge invariance in the model. Nevertheless, these additional sources can appear if we consider insertions of composite operators.

In this framework, the Ward identity requires the logarithmic divergences to be left invariant by the linearised supersymmetry Slavnov--Taylor operator
\begin{multline} \Qsla_{|\Gamma} \, \equiv  \int d^4 x \left ( \frac{ \delta^R \Gamma }{ \delta \phi} \frac{ \delta^L \,  }{ \delta \phi^\q } + \frac{ \delta^R \Gamma }{ \delta \phi_5} \frac{ \delta^L \, }{ \delta \phi^\q_5 } + \frac{ \delta^R \Gamma }{ \delta \lambda} \frac{ \delta^L \, }{ \delta \overline{\lambda}^\q}  - \frac{ \delta^R \Gamma }{ \delta \phi^\q} \frac{ \delta^L \,  }{ \delta \phi } \right .  \\*\left . -\frac{ \delta^R \Gamma }{ \delta \phi_5^\q} \frac{ \delta^L \, }{ \delta \phi } - \frac{ \delta^R \Gamma }{ \delta \lambda^\q} \frac{ \delta^L \, }{ \delta \overline{\lambda }}- \phi^\aBRST \frac{\delta^L \,}{\delta \phi^\qs}   - \phi_5^\aBRST \frac{\delta^L \,}{\delta \phi_5^\qs}  - \overline{\lambda}^\aBRST  \frac{\delta^L \,}{\delta \overline{\lambda}^\qs}  \right)
 \end{multline}
Although we have not introduced auxiliary fields, the combinations of parameters and sources $g^2 \scal{ \overline{\epsilon} \lambda^\q}$ and $g^2  \scal{ \overline{\epsilon} \gamma_5  \lambda^\q}$ behave like auxiliary fields with respect with the linearised Slavnov--Taylor operator
\begin{gather}
 \Qsla_{|\Sigma} \,    g^2 \scal{ \overline{\epsilon} \lambda^\q}  =  - \scal{\epsilonb [ i \sa \partial + \phi - \phi_5 \gamma_5 ] \lambda } \hspace{10mm} \Qsla_{|\Sigma} \,    g^2 \scal{ \overline{\epsilon} \gamma_5 \lambda^\q}  =  - \scal{\epsilonb \gamma_5  [ i \sa \partial + \phi - \phi_5 \gamma_5 ] \lambda }
 \CR
  \Qsla_{|\Sigma} \, \phi = \scal{  \overline{\epsilon} \lambda } \hspace{10mm}  \Qsla_{|\Sigma} \, \phi_5 = \scal{ \overline{\epsilon} \gamma_5 \lambda} \CR
 \Qsla_{|\Sigma} \, \lambda =  - i \sa \partial ( \phi + \phi_5 \gamma_5 ) \epsilon + \frac{1}{2}  ( \phi + \phi_5 \gamma_5 )^2  \epsilon  + g^2 \scal{ \overline{\epsilon} \lambda^\q} \epsilon +  g^2  \scal{ \overline{\epsilon} \gamma_5  \lambda^\q} \gamma_5 \epsilon
 \end{gather}
 with ${\Qsla_{|\Sigma}}^2 = - i ( \epsilonb \gamma^\mu \epsilon) \partial_\mu $. This property extends to the case for which one cannot define a linear realisation of supersymmetry by the introduction of auxiliary fields. The right count of degrees of freedom is then enforced by the constraints on the combinations of sources and commuting spinor parameters as implied by the  Fierz identities \cite{N2}.

The linearised Slavnov--Taylor operator is similarly defined as
\begin{multline} \Slav_{|\Gamma} \, \equiv  \int d^4 x \left ( \frac{ \delta^R \Gamma }{ \delta \phi} \frac{ \delta^L \,  }{ \delta \phi^\aBRST } + \frac{ \delta^R \Gamma }{ \delta \phi_5} \frac{ \delta^L \, }{ \delta \phi^\aBRST_5 } + \frac{ \delta^R \Gamma }{ \delta \lambda} \frac{ \delta^L \, }{ \delta \overline{\lambda}^\aBRST}  - \frac{ \delta^R \Gamma }{ \delta \phi^\aBRST} \frac{ \delta^L \,  }{ \delta \phi }\right .  \\*\left . -\frac{ \delta^R \Gamma }{ \delta \phi_5^\aBRST} \frac{ \delta^L \, }{ \delta \phi } - \frac{ \delta^R \Gamma }{ \delta \lambda^\aBRST} \frac{ \delta^L \, }{ \delta \overline{\lambda }} +\phi^\q \frac{\delta^L \,}{\delta \phi^\qs}   + \phi_5^\q \frac{\delta^L \,}{\delta \phi_5^\qs}  + \overline{\lambda}^\q  \frac{\delta^L \,}{\delta \overline{\lambda}^\qs}  \right)
 \end{multline}
Any functional of the physical fields alone (\ie any functional not depending on the sources) is then trivially BRST invariant, and any functional of the fields linear in the equation of motions can be written as a BRST-exact functional with respect with the linearised BRST Slavnov--Taylor operator such that it does not appear in the set of physical observables. The precise statement is that the insertion of a BRST-exact functional in the 1PI generating functionals vanishes once the equations of motion of the fields have been enforced.

The Callan--Symanzik functional operator $\mathcal{C}$ acts as the derivative with respect with the renormalisation scale $\frac{d\, }{d \mu}$ as follows
\begin{multline}  \mathcal{ C} \equiv \frac{\partial \, }{\partial \mu} + \beta \frac{\partial\, }{\partial g} + \gamma_\phi  \int d^4x  \left( \phi \frac{ \delta^L \, }{\delta \phi} - \phi^\q \frac{ \delta^L \, }{\delta \phi^\q} - \phi^\aBRST \frac{ \delta^L \, }{\delta \phi^\aBRST } - \phi^\qs \frac{ \delta^L \, }{\delta \phi^\qs } \right) \\*+ \gamma_{\phi_5}  \int d^4x   \left( \phi_5 \frac{ \delta^L \, }{\delta \phi_5} - \phi_5^\q \frac{ \delta^L \, }{\delta \phi_5^\q} - \phi_5^\aBRST \frac{ \delta^L \, }{\delta \phi_5^\aBRST } - \phi_5^\qs \frac{ \delta^L \, }{\delta \phi_5^\qs } \right)  \\*- \gamma_\lambda   \int d^4x \left( \overline{\lambda} \frac{ \delta^L \, }{\delta \overline{\lambda}} - \overline{\lambda}^\q \frac{ \delta^L \, }{\delta \overline{\lambda}^\q} - \overline{\lambda}^\aBRST \frac{ \delta^L \, }{\delta \overline{\lambda}^\aBRST } - \overline{\lambda}^\qs \frac{ \delta^L \, }{\delta \overline{\lambda}^\qs } \right)  \end{multline}
where the terms including sources are fixed by the condition that on any functional
\be \Qsla_{| \mathscr{F}} \,  \mathcal{C} \mathscr{F} - \mathcal{C} \Qsla(\mathscr{F}) = 0 \hspace{10mm}  \Slav_{| \mathscr{F}} \,  \mathcal{C} \mathscr{F} - \mathcal{C} \Slav(\mathscr{F}) = 0 \ee

By virtue of the quantum action principle, the derivative of the 1PI generating functional with respect to the coupling constant is given by the insertion of a local functional of canonical dimension four\footnote{This follows from the fact that the theory is strictly renormalisable. If there were a mass term, the corresponding insertion would be of canonical dimension less than or equal to four.} into $\Gamma$. Then using the supersymmetry Ward identity, one has
\be 0 =  \Qsla_{| \Gamma } \,  \frac{\partial \Gamma}{\partial g} - \frac{\partial\, }{\partial g}  \Qsla(\Gamma) =  \Qsla_{|\Gamma} \,  \frac{\partial \Gamma}{\partial g}  = 0\ee
By uniqueness of the supersymmetry invariant, one gets that
\be  \frac{\partial \Gamma}{\partial g}  = - \frac{2 \, a(g)}{g^3} \Bigl[  \int d^4 x  \cL^\ord{\rm c} \, \cdot \Gamma \Bigr] + \Slav_{| \Gamma}  \Bigl[  \Uppsi^\ord{1} \, \cdot \Gamma \Bigr] \ee
where $a(g)$ is a formal series in $g^2$ of the form $a(g) = 1 + \mathcal{ O}(g^2)$, $\cL^\ord{\rm c}$ is the density associated to the classical action which is left invariant by the linearised Slavnov--Taylor operator up to a pure divergence,
\begin{multline}
\cL^\ord{\rm c} = -\left( \frac{1}{2} \partial_\mu \phi \partial^\mu \phi + \frac{1}{2} \partial_\mu \phi_5 \partial^\mu \phi_5 + \frac{i}{2} \scal{ \overline{\lambda} \sa \partial \lambda } +  \frac{1}{2} \scal{ \overline{\lambda} \phi \lambda } -   \frac{1}{2}\scal{ \overline{\lambda} \phi_5 \gamma_5  \lambda } + \frac{1}{8} \scal{ \phi^2 + \phi^2_5 }^2\right) \\* + \frac{g^4}{4} \scal{  \overline{\epsilon} \lambda^\q}^2 + \frac{g^4}{4} \scal{  \overline{\epsilon} \gamma_5 \lambda^\q}^2 + \frac{1}{4} \partial_\mu \partial^\mu \scal{ \phi^2 + \phi_5^2 } - \frac{i g^2}{2} \partial_\mu \scal{ \epsilonb \gamma^\mu [ \phi - \phi_5 \gamma_5 ] \lambda^\q }
\end{multline}
and the last term corresponds to the trivial terms in the sources that could also contribute. Note that we have added pure divergence terms in $\cL^\ord{\rm c}$ which will become meaningful later on.

Let us now consider the commutator of the derivative with respect with the coupling constant and the Callan--Symanzik operator
\begin{multline}  \biggl[ \mathcal{C} \, ,\, \frac{\partial \, }{\partial g} \biggr] = - \frac{\partial \beta }{\partial g} \frac{\partial\, }{\partial g} - \frac{\partial \gamma_\phi}{\partial g}   \int d^4x  \left( \phi \frac{ \delta^L \, }{\delta \phi} - \phi^\q \frac{ \delta^L \, }{\delta \phi^\q} - \phi^\aBRST \frac{ \delta^L \, }{\delta \phi^\aBRST } - \phi^\qs \frac{ \delta^L \, }{\delta \phi^\qs } \right) \\* - \frac{\partial  \gamma_{\phi_5}}{\partial g}   \int d^4x   \left( \phi_5 \frac{ \delta^L \, }{\delta \phi_5} - \phi_5^\q \frac{ \delta^L \, }{\delta \phi_5^\q} - \phi_5^\aBRST \frac{ \delta^L \, }{\delta \phi_5^\aBRST } - \phi_5^\qs \frac{ \delta^L \, }{\delta \phi_5^\qs } \right)  \\*+ \frac{\partial  \gamma_\lambda}{\partial g}    \int d^4x \left( \overline{\lambda} \frac{ \delta^L \, }{\delta \overline{\lambda}} - \overline{\lambda}^\q \frac{ \delta^L \, }{\delta \overline{\lambda}^\q} - \overline{\lambda}^\aBRST \frac{ \delta^L \, }{\delta \overline{\lambda}^\aBRST } - \overline{\lambda}^\qs \frac{ \delta^L \, }{\delta \overline{\lambda}^\qs } \right) \ .\end{multline}
Applying this to $\Gamma$ we get
\bea \biggl[ \mathcal{C} \, ,\, \frac{\partial \, }{\partial g} \biggr] \, \Gamma &=& - \frac{\partial \beta }{\partial g} \frac{\partial \Gamma }{\partial g}  + \Slav_{| \Gamma} \Qsla_{| \Gamma} \int d^4x \biggl( \gamma_\phi \phi \phi^\qs + \gamma_{\phi_5} \phi_5 \phi_5^\qs - \gamma_\lambda \overline{\lambda} \lambda^\qs  \biggr) \CR
&=&  \frac{\partial \beta }{\partial g}  \,  \frac{2 \, a(g)}{g^3} \Bigl[  \int d^4 x  \cL^\ord{\rm c} \, \cdot \Gamma \Bigr] + \Slav_{| \Gamma}  \Bigl[  \Uppsi^\ord{2} \, \cdot \Gamma \Bigr]  \label{Diff1}\eea
On the other hand, making use of the Callan--Symanzik equation
\be \mathcal{ C} \, \Gamma = 0 \ee
one gets
\begin{multline}  \biggl[ \mathcal{C} \, ,\, \frac{\partial \, }{\partial g} \biggr] \, \Gamma = - \beta \frac{ \partial\, }{\partial g}  \biggl(   \frac{2 \, a(g)}{g^3} \biggr) \,  \Bigl[  \int d^4 x  \cL^\ord{\rm c} \, \cdot \Gamma \Bigr] - \frac{2 \, a(g)}{g^3} \, \mathcal{C}  \Bigl[  \int d^4 x  \cL^\ord{\rm c} \, \cdot \Gamma \Bigr] + \Slav_{| \Gamma}  \Bigl[  \Uppsi^\ord{1} \, \cdot \Gamma \Bigr]   \label{Diff2} \end{multline}
Again by uniqueness of the supersymmetry invariant $\cL^\ord{\rm c}$,
\be  \mathcal{C}  \Bigl[  \int d^4 x  \cL^\ord{\rm c} \, \cdot \Gamma \Bigr] = \gamma^\ord{2}  \Bigl[  \int d^4 x  \cL^\ord{\rm c} \, \cdot \Gamma \Bigr] + \Slav_{| \Gamma}  \Bigl[  \Uppsi^\ord{3} \, \cdot \Gamma \Bigr]  \label{Diff3} \ee
where $\gamma^\ord{2}$ is the anomalous dimension of the composite operator $\cL^\ord{\rm c}$ corresponding to its diagonal renormalisation by itself. Using the fact that this insertion is not trivial one finally gets from (\ref{Diff1}), (\ref{Diff2}) and (\ref{Diff3}) taken together that
\be  \frac{ \partial\, }{\partial g}  \biggl(  \beta  \frac{a(g)}{g^3} \biggr) = - \gamma^\ord{2} \,  \frac{ a(g)}{g^3} \ee
This differential equation is the first main step of the proof; it relates the $\beta$ function and the anomalous dimension of the Lagrange density considered as a composite operator insertion.

We will now relate this anomalous dimension to that of a chiral operator by use of the decent equations. One verifies that
\be  \Qsla_{|\Sigma} \,    \cL^\ord{\rm c} = \partial^\mu  \cL^\ord{\rm c}_\mu \ee
with
\begin{multline}
   \cL^\ord{\rm c}_\mu = - \frac{i}{2} \Scal{ \epsilonb \gamma_\mu \Bigl[ - i \sa \partial ( \phi + \phi_5 \gamma_5 ) + \frac{1}{2} \scal{ \phi - \phi_5 \gamma_5}^2 - g^2 \scal{ \overline{\epsilon} \lambda^\q} + g^2 \scal{ \overline{\epsilon} \gamma_5 \lambda^\q} \gamma_5 \Bigr] } \\*
 + \frac{1}{2} \partial_\mu \scal{ \epsilonb [ \phi + \phi_5 \gamma_5 ] \lambda} - \frac{ig^2}{4} ( \epsilon \gamma^\nu \epsilon) \scal{ \overline{\lambda} \gamma_\nu \gamma_\mu \lambda^\q } -  \frac{ig^2}{2} \left( \epsilonb \gamma_\mu [ \phi - \phi_5 \gamma_5 ] \frac{\delta^L \Sigma}{\delta \overline{\lambda}} \right) \\* + \partial^\nu \scal{ \epsilonb \gamma_{\mu\nu} [ \phi + \phi_5 \gamma_5 ] \lambda } \hspace{20mm}
 \end{multline}
where we have added a pure divergence in order to get an irreducible solution of the next descent equation
\be \Qsla_{|\Sigma} \,    \cL^\ord{\rm c}_\mu = - i ( \epsilonb \gamma_\mu \epsilon) \cL^\ord{\rm c} + \partial^\nu \cL^\ord{\rm c}_{\nu\mu} \ee
that is
\be \cL^\ord{\rm c}_{\mu\nu} =  - \frac{1}{2} \scal{ \epsilonb [ \phi + \gamma_5 \phi_5 ]^3 \epsilon} \ .\ee
By irreducible we mean that we cannot use the freedom in the choice of $\cL^\ord{\rm c}$ and $\cL^\ord{\rm c}_{\mu}$ in order to cancel this last component. $\cL^\ord{\rm c}_{\mu\nu}$ satisfies the last descent equation
\be  \Qsla_{|\Sigma} \,   \cL^\ord{\rm c}_{\mu\nu} = -  i  ( \epsilonb \gamma_\mu \epsilon) \cL^\ord{\rm c}_\nu +   i  ( \epsilonb \gamma_\nu \epsilon) \cL^\ord{\rm c}_\mu \ .\ee
Note that the usual complex chiral scalar fields can be defined such that
\be \Phi \equiv  \phi + i \phi_5 \hspace{10mm} \bar \Phi \equiv  \phi - i \phi_5 \ee
in terms of which
\be \cL^\ord{\rm c}_{\mu\nu}  = - \frac{1}{2} \scal{\epsilonb_+ \gamma_{\mu\nu} \epsilon_+} \, \Phi^3 - \frac{1}{2} \scal{\epsilonb_- \gamma_{\mu\nu} \epsilon_-} \, \bar \Phi^3 \ .\ee
Let us define the extended form
\be \tilde\cL^\ord{\rm c} \equiv \frac{1}{24} \varepsilon_{\mu\nu\sigma\rho} \, \cL^\ord{\rm c} \, dx^\mu_{\, \wedge} dx^\nu_{\, \wedge} dx^\sigma_{\, \wedge} dx^\rho - \frac{1}{6} {\varepsilon_{\mu\nu\sigma}}^\rho \, \cL^\ord{\rm c}_\rho \, dx^\mu_{\, \wedge} dx^\nu_{\, \wedge} dx^\sigma + \frac{1}{2} {\varepsilon_{\mu\nu}}^{\sigma\rho} \, \cL^\ord{\rm c}_{\sigma\rho} \, dx^\mu_{\, \wedge} dx^\nu \ .\ee
Then the descent equations can be written in closed form as
\be \scal{ d +  \Qsla_{|\Sigma} + i_{i(\epsilonb \gamma \epsilon)}} \, \tilde \cL^\ord{\rm c} = 0 \ee
where $ \scal{ d +  \Qsla_{|\Sigma} + i_{i(\epsilonb \gamma \epsilon)}}$ defines a nilpotent differential which extends at the quantum level to
\be  \scal{ d +  \Qsla_{|\Gamma} + i_{i(\epsilonb \gamma \epsilon)}}^2 = 0 \ .\ee
In order to understand this we must introduce sources for the various composite operators defining the cocycle. The minimal way to consider the coupling of the Lagrange density to a source,  preserving the supersymmetry Ward identity, is in fact to couple the whole set of forms defining the corresponding cocycle.
\be \Sigma[u] = \Sigma + \int \Scal{ u \, \cL^\ord{\rm c}_4 + u_{1\, \wedge} \cL^\ord{\rm c}_{3} + u_{2\, \wedge} \cL^\ord{\rm c}_{2} } + {\cal O} ( u^2) \ .\ee
Since the density $\cL^\ord{\rm c}_4$ depends explicitly on the sources, it is necessary to add terms quadratic in the sources in order for the action to be a solution of the Ward identity. Nevertheless, we are only interested in insertions of one single composite operator in this discussion, and these higher order terms can thus be disregarded. It is actually convenient to define the extended form source $\tilde u \equiv u + u_1  + u_2 $, in such way that the coupling to the sources can be written
\be \Sigma[u] = \Sigma + \int \tilde u_{\, \wedge} \tilde \cL^\ord{\rm c} + {\cal O} (\tilde u^2) \label{SourceCoupling} \ee
with the Berezin prescription that only the form of maximal degree $4$ of the wedge product $\tilde u_{\, \wedge} \tilde \cL^\ord{\rm c} $ gives rise to a non-zero integral. One has to define the transformations of the sources by the action of the linearised Ward identity in such way that the complete action coupled to the sources still satisfies the Ward identity
\be \Qsla(\Sigma[u]) = \int \Scal{\Qsla_{|\Sigma}\tilde u_{\, \wedge } \tilde \cL^\ord{\rm c} - \tilde u_{\, \wedge}\scal{ d + i_{i(\epsilonb \gamma \epsilon)}} \tilde \cL^\ord{\rm c} } + {\cal O} (\tilde u^2) \label{Sinv} \ee
Integrating by part the right-hand-side of (\ref{Sinv}), one obtains that the sources must transform as a cocycle
\be  (d + \Qsla_{|\Sigma}+ i_{i(\epsilonb \gamma \epsilon)} ) \tilde u = 0 \ee
It follows that in order for the Callan--Symanzik operator to commute with $\Qsla$, the whole cocycle $\cL^\ord{\rm c}$ can only mix with extended forms that also define a cocycle of the extended differential.

If the $4$-component of a cocycle of this differential is zero, one can show that the whole cocycle is trivial by use of the algebraic Poincar\'e lemma. Since $\cL^\ord{\rm c}$ defines the unique supersymmetric density, it follows that all the cocycles are cohomologically equivalent to $\tilde \cL^\ord{\rm c}$ .

We thus conclude that
\be   \mathcal{C}  \Bigl[ \tilde \cL^\ord{\rm c} \, \cdot \Gamma \Bigr] = \gamma^\ord{2}  \Bigl[   \tilde \cL^\ord{\rm c} \, \cdot \Gamma \Bigr] + \scal{ d +  \Qsla_{|\Gamma} + i_{i(\epsilonb \gamma \epsilon)} } \Bigl[  \tilde \Xi \, \cdot \Gamma \Bigr]   \ee
Since the commuting spinors are just parameters, it follows that the anomalous dimension of the Lagrange density $\gamma^\ord{2}$ is also the anomalous dimension $\gamma_{\Phi^3}$ of the dimension three chiral operator $ \Phi^3$
\be  \mathcal{C}  \Bigl[   \Phi^3  \, \cdot \Gamma \Bigr] = \gamma_{\Phi^3}  \Bigl[   \Phi^3  \, \cdot \Gamma \Bigr] \ .\ee
We have thus derived within the component formalism the well known result that the $\beta$ function of the Wess--Zumino model is related to the anomalous dimension of the chiral operator $\Phi^3$
\be \frac{ \partial\, }{\partial g}  \biggl(  \beta  \frac{a(g)}{g^3} \biggr) = - \gamma_{\Phi^3}  \,  \frac{ a(g)}{g^3} \ee
In fact we know from the superspace non-renormalisation theorem that both the $\beta$ function and $\gamma_{\Phi^3}$ are zero. Let us only assume that $\gamma_{\Phi^3}=0$ has been proven to be zero. Then using the form of the $\beta$ function $\beta = \beta_1 g^3 + \mathcal{ O}(g^5)$, and the fact that the formal series $a(g)= 1 + \mathcal{O}(g^2)$ can be inverted, we obtain that
\be \beta = a(g)^{-1} g^3 \beta_1 \ee
It then follows from the one-loop computation that $\beta=0$ at all order in perturbation theory.

So far, the reader might be excused for thinking that this procedure is a rather involved formalism for proving results which have been known for many years. However, as we shall see in the following, this method extends nicely to strictly non-renormalisable theories and will provide non-trivial results for maximal super Yang--Mills and maximal supergravity.

\subsection{Supersymmetric Yang--Mills theory in higher dimensions}

Now that we have explained the strategy in a simple model, let us consider  a class of non-renormalisable theories, namely maximally supersymmetric Yang--Mills theory in higher dimensions. We will renormalize the fields by multiplication with the coupling constant $g$ in such way that these new fields have canonical dimension one (or $\frac{3}{2}$ for the fermions) as in four dimensions.  We define $n_\ord{\thalf}$ as the loop order at which the counterterms associated to $\ft12$ BPS invariants $I_{\stfrac{1}{2}}$ and $I_{\stfrac{1}{2}}^\prime$ can occur as logarithmic divergences in $D$ dimensional spacetime. In the same way $n_\ord{\tquarter}$ will be the corresponding loop order for the logarithmic divergences associated to the invariants $I_\Ko$ and $I_{\stfrac{1}{4}}$, while $n_\ord{\thalf\tquarter}$ will be the loop order at which these latter invariants can be needed to renormalize insertions of the $\ft12$ BPS invariants.

\begin{table}[ht]
\centering
\begin{tabular}{|lc|c|c|c|c|c|}
\hline
Dimension & $D$ & \ \  5 \ \ & \ \ 6 \ \ & \ \  7 \ \  & \ \ 8\ \ & \ 10 \  \   \\
\hline
$I_{\stfrac{1}{2}}\  ,\  I^\prime_{\stfrac{1}{2}} \rightarrow I_\Ko\  ,\  I_{\stfrac{1}{4}} \ \  $& $n_{\ord{\thalf\tquarter}}$ & 2 &1 &$\emptyset$ &$\emptyset$ &$\emptyset$ \\
\hline
$S  \rightarrow I_{\stfrac{1}{2}}\  ,\  I^\prime_{\stfrac{1}{2}} \ \ $& $n_\ord{\thalf}$ & 4 & 2 & $\emptyset$ & 1 & $\emptyset$  \\
\hline
$S \rightarrow I_\Ko\  ,\  I_{\stfrac{1}{4}} \ \  $& $n_\ord{\tquarter}$ & 6 & 3 & 2&$\emptyset$ &1 \\
\hline
\end{tabular}
\caption{Loop orders of anticipated logarithmic divergences for various BPS operators.
\label{tab2}}
\end{table}

The logarithmic divergences have the property of introducing a dependance of the renormalized coupling constants on the unphysical renormalization scale $\mu$. It is convenient to redefine the dimensionful coupling constants in term of dimensionless parameters by rescaling them by appropriate powers of $\mu$. We will nevertheless consider the dimensionful coupling constants, but with a dependence on the renormalization scale given at tree level by their canonical dimensions. The logarithmic divergences will modify perturbatively this dependence.

We consider the classical action
\be \Sigma = \frac{1}{g^2} S + z_\ord{\thalf} I_{\stfrac{1}{2}} + z^{\prime}_\ord{\thalf}  I^\prime_{\stfrac{1}{2}} + z_\ord{\tKo} I_\Ko + z_\ord{\tquarter} I_{\stfrac{1}{4}} + \cdots \ee
where the $\cdots$ stand for higher order counterterms, completions of the considered counterterms as well as the gauge-fixing action and the source terms for the BRST and supersymmetry transformations.

Power counting and supersymmetry constrain the renormalization scale dependence of the renormalized coupling constants as follows
\bea
\mu \frac{d g}{d \mu} &=& \frac{4-D}{2} g \CR
\mu \frac{d z_\ord{\thalf} }{d \mu} &=& (D-8) z_\ord{\thalf} + \beta_\ord{\thalf} g^{2 (n_\ord{\thalf} - 1)} \CR
 \mu \frac{d z^\prime_\ord{\thalf} }{d \mu} &=& (D-8) z^\prime_\ord{\thalf} + \beta^\prime_\ord{\thalf} g^{2 (n_\ord{\thalf} - 1)} \CR
\mu \frac{d z_\ord{\tKo} }{d \mu} &=& (D-10) z_\ord{\tKo} + \beta_\ord{\tKo} g^{2 (n_\ord{\tquarter} - 1)}  + \beta_\ord{\thalf\tKo} z_\ord{\thalf} g^{2 n_\ord{\thalf\tquarter}} +  \beta_\ord{\thalf^\prime \tKo} z^\prime_\ord{\thalf} g^{2 n_\ord{\thalf\tquarter}}\CR
 \mu \frac{d z_\ord{\tquarter} }{d \mu} &=& (D-10) z_\ord{\tquarter} + \beta_\ord{\tquarter} g^{2 (n_\ord{\tquarter} - 1)}  + \beta_\ord{\thalf\tquarter} z_\ord{\thalf} g^{2 n_\ord{\thalf\tquarter}} +  \beta_\ord{\thalf^\prime \tquarter} z^\prime_\ord{\thalf} g^{2 n_\ord{\thalf\tquarter}}
\eea
where the $\beta$ parameters are dimensionless constants that can be computed perturbatively. For instance, if one uses dimensional regularization in the minimal scheme, they occur as the coefficients of  simple poles in $\varepsilon$ of the four point functions. These equations can be easily solved, and give the renormalization scale dependence of the renormalized coupling constants exactly:
\bea
g(\mu) &=& g(1) \,  \mu^{\frac{4-D}{2}} \CR
z_\ord{\thalf}(\mu) &=&  g(\mu)^{2(n_\ord{\thalf}-1)} \Scal{ \bar z_\ord{\thalf} + \beta_\ord{\thalf} \ln \mu } \\*
z^\prime_\ord{\thalf} (\mu)&=&   g(\mu)^{2(n_\ord{\thalf}-1)} \Scal{ \bar z^\prime_\ord{\thalf} + \beta^\prime_\ord{\thalf} \ln \mu } \CR
z_\ord{\tKo} (\mu)&=&   g(\mu)^{2(n_\ord{\tquarter}-1)} \Scal{ \bar z_\ord{\tKo} + \scal{ \beta_\ord{\tKo} + \beta_\ord{\thalf\tKo} \bar z_\ord{\thalf} + \beta_\ord{\thalf^\prime \tKo} \bar z^\prime_\ord{\thalf} } \ln \mu  + \sfrac{1}{2} \scal{ \beta_\ord{\thalf\tKo} \beta_\ord{\thalf} + \beta_\ord{\thalf^\prime \tKo} \beta^\prime_\ord{\thalf} } \ln^2 \mu } \CR
z_\ord{\tquarter} (\mu)&=&  g(\mu)^{2(n_\ord{\tquarter}-1)} \Scal{ \bar z_\ord{\tquarter} + \scal{ \beta_\ord{\tquarter} + \beta_\ord{\thalf \tquarter} \bar z_\ord{\thalf} + \beta_\ord{\thalf^\prime \tquarter} \bar z^\prime_\ord{\thalf} } \ln \mu  + \sfrac{1}{2} \scal{ \beta_\ord{\thalf \tquarter} \beta_\ord{\thalf} + \beta_\ord{\thalf^\prime \tquarter} \beta^\prime_\ord{\thalf} } \ln^2 \mu }\ . \nonumber
\eea
These solutions exhibit the well known fact that the theory is ill-defined in the ultra-violet limit.

The Callan--Symanzik functional operator acts on any functional $\mathscr{F}$ of the fields as the derivative with respect to the renormalisation scale $\frac{d\, }{d \mu}$, that is
\begin{multline}
{\cal C} \, \mathscr{F} = \mu \frac{\partial \mathscr{F}}{\partial \mu} +  \frac{4-D}{2} g \frac{\partial \mathscr{F}}{\partial g} \\* + \Scal{ (D-8) z_\ord{\thalf} + \beta_\ord{\thalf} g^{2 (n_\ord{\thalf} - 1)}}  \frac{\partial \mathscr{F}}{\partial z_\ord{\thalf} }  + \Scal{  (D-8) z^\prime_\ord{\thalf} + \beta^\prime_\ord{\thalf} g^{2 (n_\ord{\thalf} - 1)}}  \frac{\partial \mathscr{F}}{\partial z^\prime_\ord{\thalf} } \\*+ \Scal{ (D-10) z_\ord{\tKo} + \beta_\ord{\tKo} g^{2 (n_\ord{\tquarter} - 1)}  + \beta_\ord{\thalf\tKo} z_\ord{\thalf} g^{2 n_\ord{\thalf\tquarter}} +  \beta_\ord{\thalf^\prime \tKo} z^\prime_\ord{\thalf} g^{2 n_\ord{\thalf\tquarter}} } \frac{\partial \mathscr{F}}{\partial z_\ord{\tKo} } \\*+ \Scal{  (D-10) z_\ord{\tquarter} + \beta_\ord{\tquarter} g^{2 (n_\ord{\tquarter} - 1)}  + \beta_\ord{\thalf\tquarter} z_\ord{\thalf} g^{2 n_\ord{\thalf\tquarter}} +  \beta_\ord{\thalf^\prime \tquarter} z^\prime_\ord{\thalf} g^{2 n_\ord{\thalf\tquarter}} }  \frac{\partial \mathscr{F}}{\partial z_\ord{\tquarter} } + \cdots  \label{CS}
\end{multline}
where the $\cdots$ stand for terms involving partial derivatives with respect to higher-order coupling constants or for field and gauge parameter anomalous dimensions that can be written as BRST-exact terms and which will be disregarded. The independence of the generating functional of one-particle-irreducible  graphs $\Gamma$ in the renormalization scale is equivalent to the Callan-Symanzik equation
\be {\cal C} \, \Gamma = 0\ee
We will also consider insertions of composite operators in $\Gamma$. Supersymmetry and BRST invariance imply that the insertion of the classical action $S$ involves only supersymmetric functionals as BRST non-trivial counterterms. Power counting then determines the following action of the Callan--Symanzik functional operator upon the insertion of the classical action into $\Gamma$
\begin{multline} {\cal C}\bigl[ S \cdot \Gamma\bigr] = - \bigl[ S \cdot \Gamma\bigr]  + \gamma_\ord{\thalf} g^{2 n_\ord{\thalf}} \bigl[ I_{\stfrac{1}{2}} \cdot \Gamma\bigr] +  \gamma^\prime_\ord{\thalf} g^{2 n_\ord{\thalf}} \bigl[ I^\prime_{\stfrac{1}{2}} \cdot \Gamma\bigr] + \gamma_\ord{\tKo} g^{2 n_\ord{\tquarter}} \bigl[ I_\Ko \cdot \Gamma\bigr] \\*+ \gamma_\ord{\tquarter} g^{2 n_\ord{\tquarter}} \bigl[ I_{\stfrac{1}{4}} \cdot \Gamma\bigr] + \scal{ \gamma_\ord{{\rm c} \thalf \tKo} z_\ord{\thalf} + \gamma_\ord{{\rm c} \thalf^\prime \tKo} z^\prime_\ord{\thalf}  }  g^{2(n_\ord{\thalf\tquarter} +1)} \bigl[ I_\Ko \cdot \Gamma\bigr]  \\*+ \scal{\gamma_\ord{{\rm c} \thalf\tquarter} z_\ord{\thalf} + \gamma_\ord{{\rm c} \thalf^\prime \tquarter} z^\prime_\ord{\thalf}  }   g^{2(n_\ord{\thalf\tquarter} +1)} \bigl[ I_{\stfrac{1}{4}} \cdot \Gamma\bigr] +\cdots
\end{multline}
where the $\cdots$ stand for higher order insertions and BRST-exact insertions. The anomalous dimension constants $\gamma$ have been defined to be dimensionless. We consider also the insertions of the invariant counterterms, which satisfy
\bea
{\cal C}\bigl[ I_{\stfrac{1}{2}} \cdot \Gamma\bigr] &=& (8-D) \bigl[ I_{\stfrac{1}{2}} \cdot \Gamma\bigr]  + \gamma_\ord{\thalf\tKo} g^{2 n_\ord{\thalf\tquarter}} \bigl[ I_\Ko \cdot \Gamma\bigr] + \gamma_\ord{\thalf \tquarter} g^{2 n_\ord{\thalf\tquarter}} \bigl[ I_{\stfrac{1}{4}} \cdot \Gamma\bigr]+\cdots \CR
{\cal C}\bigl[ I^\prime_{\stfrac{1}{2}} \cdot \Gamma\bigr] &=& (8-D) \bigl[ I^\prime_{\stfrac{1}{2}} \cdot \Gamma\bigr]  + \gamma_\ord{\thalf^\prime \tKo} g^{2 n_\ord{\thalf\tquarter}} \bigl[ I_\Ko \cdot \Gamma\bigr] + \gamma_\ord{\thalf^\prime \tquarter} g^{2 n_\ord{\thalf\tquarter}} \bigl[ I_{\stfrac{1}{4}} \cdot \Gamma\bigr]+\cdots \CR
{\cal C}\bigl[ I_\Ko \cdot \Gamma\bigr] &=& (10-D) \bigl[ I_\Ko \cdot \Gamma\bigr] + \cdots \CR
{\cal C}\bigl[ I_{\stfrac{1}{4}} \cdot \Gamma\bigr] &=& (10-D) \bigl[ I_{\stfrac{1}{4}} \cdot \Gamma\bigr] + \cdots
\eea
We are also going to need to understand the quantum version of the obvious classical equation
\be \frac{ \partial \Sigma }{\partial g }  = - \frac{2}{g^3} S  + \cdots \ee
where $\cdots$ stands for terms involving the sources. The quantum action principle implies that the partial derivative of the 1PI generating functional $\Gamma$ with respect to any coupling constant is given by the insertion of a local functional into $\Gamma$. The supersymmetry Ward identities and power counting then imply
\begin{multline}
 \frac{ \partial \Gamma }{\partial g } = - \frac{2}{g^3} \bigl[S\cdot \Gamma] + a_\ord{\thalf}  g^{2 n_\ord{\thalf}-3} \bigl[ I_{\stfrac{1}{2}} \cdot \Gamma\bigr] +  a^\prime_\ord{\thalf} g^{2 n_\ord{\thalf}-3} \bigl[ I^\prime_{\stfrac{1}{2}} \cdot \Gamma\bigr] + a_\ord{\tKo} g^{2 n_\ord{\tquarter}-3} \bigl[ I_\Ko \cdot \Gamma\bigr] \\*+ a_\ord{\tquarter} g^{2 n_\ord{\tquarter}-3} \bigl[ I_{\stfrac{1}{4}} \cdot \Gamma\bigr] + ( a_\ord{{\rm c} \thalf \tKo} z_\ord{\thalf} + a_\ord{{\rm c} \thalf^\prime \tKo} z^\prime_\ord{\thalf}  )   g^{2 n_\ord{\thalf\tquarter} -1} \bigl[ I_\Ko \cdot \Gamma\bigr]  \\*+ ( a_\ord{{\rm c} \thalf \tquarter} z_\ord{\thalf} + a_\ord{{\rm c} \thalf^\prime \tquarter} z^\prime_\ord{\thalf}  )   g^{2 n_\ord{\thalf\tquarter} -1} \bigl[ I_{\stfrac{1}{4}} \cdot \Gamma\bigr] +\cdots
\end{multline}
In the same way one has
\bea
 \frac{ \partial \Gamma }{\partial z_\ord{\thalf} } &=& \bigl[ I_{\stfrac{1}{2}} \cdot \Gamma\bigr]  + b_\ord{\thalf\tKo} g^{2 n_\ord{\thalf\tquarter}} \bigl[ I_\Ko \cdot \Gamma\bigr] + b_\ord{\thalf\tquarter} g^{2 n_\ord{\thalf\tquarter}} \bigl[ I_{\stfrac{1}{4}} \cdot \Gamma\bigr]+\cdots \CR
 \frac{ \partial \Gamma}{\partial z^\prime_\ord{\thalf} } &=&  \bigl[ I^\prime_{\stfrac{1}{2}} \cdot \Gamma\bigr]  + b_\ord{\thalf^\prime \tKo} g^{2 n_\ord{\thalf\tquarter}} \bigl[ I_\Ko \cdot \Gamma\bigr] + b_\ord{\thalf^\prime \tquarter} g^{2 n_\ord{\thalf\tquarter}} \bigl[ I_{\stfrac{1}{4}} \cdot \Gamma\bigr]+\cdots \CR
 \frac{ \partial \Gamma }{\partial z_\ord{\tKo} }  &=& \bigl[ I_\Ko \cdot \Gamma\bigr] + \cdots \CR
 \frac{ \partial  \Gamma}{\partial z_\ord{\tquarter} } &=& \bigl[ I_{\stfrac{1}{4}} \cdot \Gamma\bigr] + \cdots
\eea
Now that we have defined our data, we want to relate the $\beta$ functions to the corresponding anomalous dimensions. The reason, which will become clear shortly, is that it will be easier to determine algebraically the properties of the anomalous dimensions than the properties of the $\beta$ functions. The relation is obtained by acting on the 1PI generating functional $\Gamma$ with the functional operators defined by the commutators of the Callan--Symanzik operator with the derivatives with respect to the various coupling constants. We obtain from the definition (\ref{CS}) of the Callan--Symanzik operator that
\begin{multline}
\Bigl[ \, \frac{\partial\, }{\partial g} \, , \, {\cal C} \, \Bigr] = \frac{4-D}{2}   \frac{\partial\, }{\partial g} + 2 (n_\ord{\thalf} - 1) \beta_\ord{\thalf} g^{2 n_\ord{\thalf} - 3} \frac{\partial\, }{\partial z_\ord{\thalf}} + 2 (n_\ord{\thalf} - 1) \beta^\prime_\ord{\thalf} g^{2 n_\ord{\thalf} - 3} \frac{\partial\, }{\partial z^\prime_\ord{\thalf}} \\*
+ \Scal{ 2 (n_\ord{\tquarter}-1) \beta_\ord{\tKo} g^{2 n_\ord{\thalf} - 3} + 2 n_\ord{\thalf\tquarter} \beta_\ord{\thalf\tKo} z_\ord{\thalf} g^{2n_\ord{\thalf\tquarter} - 1} + 2 n_\ord{\thalf\tquarter} \beta_\ord{\thalf^\prime \tKo} z^\prime_\ord{\thalf} g^{2n_\ord{\thalf\tquarter} - 1} }  \frac{\partial\, }{\partial z_\ord{\tKo}} \\*
+ \Scal{ 2 (n_\ord{\tquarter}-1) \beta_\ord{\tquarter} g^{2 n_\ord{\thalf} - 3} + 2 n_\ord{\thalf\tquarter} \beta_\ord{\thalf\tquarter} z_\ord{\thalf} g^{2n_\ord{\thalf\tquarter} - 1} + 2 n_\ord{\thalf\tquarter} \beta_\ord{\thalf^\prime \tquarter} z^\prime_\ord{\thalf} g^{2n_\ord{\thalf\tquarter} - 1} }  \frac{\partial\, }{\partial z_\ord{\tquarter}} + \cdots
\end{multline}
as well as
\bea
\Bigl[ \, \frac{\partial\, }{\partial z_\ord{\thalf} } \, , \, {\cal C} \, \Bigr] &=& (D-8)  \frac{\partial\, }{\partial z_\ord{\thalf}} + \beta_\ord{\thalf\tKo} g^{2 n_\ord{\thalf\tquarter}} \frac{\partial\, }{\partial z_\ord{\tKo}} + \beta_\ord{\thalf\tquarter} g^{2n_\ord{\thalf\tquarter}} \frac{\partial\, }{\partial z_\ord{\tquarter}} + \cdots \CR
\Bigl[ \, \frac{\partial\, }{\partial z^\prime_\ord{\thalf} } \, , \, {\cal C} \, \Bigr] &=& (D-8)  \frac{\partial\, }{\partial z^\prime_\ord{\thalf}} + \beta_\ord{\thalf^\prime \tKo} g^{2 n_\ord{\thalf\tquarter}} \frac{\partial\, }{\partial z_\ord{\tKo}} + \beta_\ord{\thalf^\prime \tquarter} g^{2n_\ord{\thalf\tquarter}} \frac{\partial\, }{\partial z_\ord{\tquarter}} + \cdots
\eea
It follows trivially from the Callan--Symanzik equation that
\be \biggl( \Bigl[ \, \frac{\partial\, }{\partial g} \, , \, {\cal C} \, \Bigr]  + {\cal C} \, \frac{\partial\, }{\partial g}  \biggr) \Gamma = 0 \label{CommuC}\ee
We assume that each independent invariant counterterm (\ie not BRST equivalent and thus not equal modulo the equations of motion) defines an independent operator at the quantum level, in such way that the corresponding insertions into the 1PI generating functional are linearly independent. The expansion of the equation (\ref{CommuC}) gives the following equations
\begin{gather}
(n_\ord{\thalf} - 1) \beta_\ord{\thalf} = \gamma_\ord{\thalf} \hspace{10mm} (n_\ord{\thalf} - 1) \beta^\prime_\ord{\thalf} = \gamma^\prime_\ord{\thalf} \CR
(n_\ord{\tquarter} - 1) \beta_\ord{\tKo}  + (n_\ord{\thalf} - 1) b_\ord{\thalf\tKo} \beta_\ord{\thalf} +(n_\ord{\thalf} - 1) b_\ord{\thalf^\prime \tKo} \beta^\prime_\ord{\thalf}   =  \gamma_\ord{\tKo} - \frac{a_\ord{\thalf}}{2} \gamma_\ord{\thalf\tKo} - \frac{a^\prime_\ord{\thalf}}{2} \gamma_\ord{\thalf^\prime \tKo} \CR
(n_\ord{\tquarter} - 1) \beta_\ord{\tquarter}  + (n_\ord{\thalf} - 1) b_\ord{\thalf\tquarter} \beta_\ord{\thalf} +(n_\ord{\thalf} - 1) b_\ord{\thalf^\prime \tquarter} \beta^\prime_\ord{\thalf}   =  \gamma_\ord{\tquarter} - \frac{a_\ord{\thalf}}{2} \gamma_\ord{\thalf\tquarter} - \frac{a^\prime_\ord{\thalf}}{2} \gamma_\ord{\thalf^\prime \tquarter} \CR
n_\ord{\thalf\tquarter} \beta_\ord{\thalf\tKo} = \gamma_\ord{{\rm c} \thalf \tKo} \hspace{10mm}  n_\ord{\thalf\tquarter} \beta_\ord{\thalf^\prime \tKo} = \gamma_\ord{{\rm c} \thalf^\prime \tKo} \CR
n_\ord{\thalf\tquarter} \beta_\ord{\thalf\tquarter} = \gamma_\ord{{\rm c} \thalf\tquarter} \hspace{10mm}  n_\ord{\thalf\tquarter} \beta_\ord{\thalf^\prime \tquarter} = \gamma_\ord{{\rm c} \thalf^\prime \tquarter} \label{CallanRelations}
\end{gather}
and the expansions of the equations
\be \biggl( \Bigl[ \, \frac{\partial\, }{\partial z_\ord{\thalf} } \, , \, {\cal C} \, \Bigr]  + {\cal C} \, \frac{\partial\, }{\partial z_\ord{\thalf} }  \biggr) \Gamma = 0   \hspace{10mm}  \biggl( \Bigl[ \, \frac{\partial\, }{\partial z^\prime_\ord{\thalf} } \, , \, {\cal C} \, \Bigr]  + {\cal C} \, \frac{\partial\, }{\partial z^\prime_\ord{\thalf} }  \biggr) \Gamma = 0  \ee
give that
\be\begin{split}
\beta_\ord{\thalf\tKo} &= - \gamma_\ord{\thalf\tKo} \\*
\beta_\ord{\thalf\tquarter} &= - \gamma_\ord{\thalf\tquarter}
\end{split}\hspace{10mm}\begin{split}
\beta_\ord{\thalf^\prime \tKo} &= - \gamma_\ord{\thalf^\prime \tKo} \\*
\beta_\ord{\thalf^\prime \tquarter} &= - \gamma_\ord{\thalf^\prime \tquarter}
\end{split}\ee
The main result is that one can study the $\beta$ functions for a potential counterterm via the corresponding anomalous dimensions $\gamma$ for mixing of the counterterm operator with the classical Lagrangian operator. Thus, the cancellation of the anomalous dimensions corresponding to the mixing of the classical Lagrangian with insertions of $I_{\stfrac{1}{2}}$ and $I^\prime_{\stfrac{1}{2}}$ implies cancellation of the associated $\beta$ functions in dimensions five and six. Similarly, in seven dimensions the cancellation of the anomalous dimensions corresponding to the mixing of the classical Lagrangian with the insertions $I_\Ko$ and $I_{\stfrac{1}{4}}$ {\em would} lead to the conclusion that the associated $\beta$ function vanishes. However it is known that logarithmic divergences do occur in this case and we will see that in this case one fails to prove that the corresponding anomalous dimensions vanish. In eight and ten dimensions, the first potential logarithmic divergences occur at one loop, and the algebraic renormalisation method does not give any information in these one-loop cases. Indeed, maximally supersymmetric Yang--Mills theory diverges at one loop in eight and ten dimensions.

\subsection{Invariant counterterms and the descent equations}

We define a commuting supersymmetry parameter $\epsilon$. On gauge invariant functions of the fields, the differential $\epsilonb Q$ is nilpotent modulo a derivative term and the equations of motion
\be \scal{ \epsilonb Q }^2 =  - i (\epsilonb \gamma^\mu \epsilon ) \partial_\mu \ee
Thanks to the introduction of the shadow fields as well as sources coupled to the supersymmetry and BRST transformations of the fields, this differential can be promoted to a linearised Slavnov--Taylor operator $\Qsla_{|\Sigma}$ which verifies this nilpotency exactly on any field \cite{shadow}. This is still true if one considers the all-order classical action $\Sigma$ with the all-order supersymmetry transformations.

At the tree level we have in $D\le 8$ dimensions that
\bea
\Qsla_{|\Sigma}\,  A_\mu  \hspace{-2mm} &=& i \scal{ \epsilonb \gamma_\mu \lambda}  + D_\mu c \hspace{20mm}\Qsla_{|\Sigma} \ \phi^I = - \scal{\epsilonb \tau^I \lambda }\  [ \, \phi^I ]\CR
\Qsla_{|\Sigma} \ \lambda &=& \bigl[\saa F + i \tau_I  \saaa D \phi^I - \tau_{IJ} [ \phi^I , \phi^J ]\bigr] \epsilon - [ c , \lambda]  +  g^2 M(\epsilon) \Scal{\lambda^\q - [\lambda^\qs , \Omega ] } \CR
\Qsla_{|\Sigma}\  c &=& (\epsilonb \tau_I \epsilon ) \phi^I - i ( \epsilonb \gamma^\mu \epsilon ) A_\mu - c^2  \label{algebraTree}
\eea
where the $\gamma^\mu$ and the $\tau^I$ are the gamma matrices associated to $Spin(1,D-1)$ and $Spin(10-D)$ respectively. $c$ is the so-called shadow field, which is an anticommuting scalar field in the adjoint representation of the gauge group. The source dependent term in the right hand side of the transformation of the fermion field $\lambda$ is defined in such a way that $\Qsla_{|\Sigma}$ defines a functional representation of the supersymmetry algebra (\ie without involving the equations of motion) and
\be M(\epsilon) \equiv \epsilon \epsilonb - \frac{1}{2} ( \epsilonb \gamma^\mu \epsilon) \gamma_\mu - \frac{1}{2} ( \epsilonb \tau_I \epsilon ) \tau^I \ee
The theory is invariant with respect to the $U(1)$ symmetry associated to shadow number, where $\Qsla_{|\Sigma}$, $\epsilon$ and $c$ have shadow number one, and the fields $A_\mu,\, \lambda$ and $\phi^I$ have shadow number zero.

It is a remarkable fact that this algebra can be derived from the Baulieu--Singer like extended curvature definition
\be
( d + \Qsla_{|\Sigma} + i_{i(\epsilonb \gamma \epsilon)} ) \scal{A + c } + \scal{A + c }^2  = F  + i \scal{\epsilonb \gamma_1 \lambda } +  ( \epsilonb \tau_I \epsilon ) \phi^I  \label{Curvature}
 \ee
and its Bianchi identity, implied by the nilpotency of the extended differential $( d + \Qsla_{|\Sigma} + i_{i(\epsilonb \gamma \epsilon)} )$, by decomposing it with respect to form degree \cite{shadow}. We use the convention that $\gamma_p$ is the $p$-form obtained as the antisymmetrised product of $p$ gamma matrices with a factor of $\scal{ \frac{1}{p!} }^2$.

An allowed counterterm is invariant under the action of the linearised Slavnov--Taylor operator $\Qsla_{|\Sigma}$. It is therefore the integral of a $D$-form $\cL_D$ which is invariant under the action of $\Qsla_{|\Sigma}$ modulo a total derivative.
\be \Qsla_{|\Sigma} \cL_D + d \cL_{D-1} = 0\ . \ee
Applying $\Qsla_{|\Sigma}$ to this equation, we obtain
\be  {\Qsla_{|\Sigma} }^2 \cL_D +  \Qsla_{|\Sigma} d \cL_{D-1} = - d \scal{ i_{i(\epsilonb \gamma \epsilon)} \cL_D +  \Qsla_{|\Sigma}  \cL_{D-1} } = 0\ .  \label{first} \ee
The algebraic Poincar\'e lemma implies that the de Rham cohomology restricted to the considered complex is given by the wedge product of constant forms constructed from the constant parameter $\epsilon$ and invariant polynomials of the Yang--Mills curvature. We are interested in the non-trivial cocycles of the de Rham cohomology of extended form degree  (\ie form degree plus shadow number) $D$ and $D+1$. The only non-trivial elements of extended form degree $D$ of the de Rham cohomology within maximally supersymmetric Yang--Mills theory, are the $D$-forms invariant polynomials of the Yang--Mills curvature (\ie $\trace F^{\scriptscriptstyle \frac{D}{2}}$, $\trace F^{{\scriptscriptstyle \stfrac{D}{2}}-2}\,  \trace F^2$ for even $D$ and nothing otherwise). However there is always a non-trivial cocycle of extended form degree $D+1$ and shadow number $2$ in dimensions $D>4$. Indeed in ten dimensions, one has the three $9$-forms of shadow number $2$
\be ( \epsilonb \gamma_5 \epsilon)_{\, \wedge} \trace   F_{\, \wedge} F \hspace{10mm} ( \epsilonb \gamma_1 \epsilon)_{\, \wedge}\trace   F_{\, \wedge} F_{\, \wedge} F_{\, \wedge} F  \hspace{10mm} ( \epsilonb \gamma_1 \epsilon)_{\, \wedge}\trace   F_{\, \wedge} F_{\, \wedge}  \trace F_{\, \wedge} F  \label{DeRham} \ee
The two last do not dimensionally reduce to non-trivial cocycles in $D\le 8$ dimensions but the first gives rise to a  non trivial cocycle in any dimension $D> 4$.

Except for this special example, we thus obtain from (\ref{first}) that there exists a gauge invariant $\cL_{D-2}$ such that
\be  i_{i(\epsilonb \gamma \epsilon)} \cL_D +  \Qsla_{|\Sigma}  \cL_{D-1}  + d \cL_{D-2} = 0\ . \ee
The same operation permits one to define forms of all degrees until $\cL_p = 0$ or until one reaches $\cL_0$, with the subtlety that they can be of a Chern--Simons-like structure if $\Qsla_{|\Sigma}  \cL_{D-2}$ generates a non-trivial de Rham cohomology class of the form (\ref{DeRham}). We define an extended form $\tilde \cL$ of form-degree plus shadow number $D$ as the formal sum of all the forms $\cL_D + \cL_{D-1} + \cL_{D-2} + \cdots$; this extended form then defines a cocycle of the extended nilpotent differential $d + \Qsla_{|\Sigma}+ i_{i(\epsilonb \gamma \epsilon)} $,
\be (d + \Qsla_{|\Sigma}+ i_{i(\epsilonb \gamma \epsilon)} ) \tilde \cL =0 \ee
We consider the cohomology of this extended differential in the complex of extended forms built from functions of the fields defining non-trivial elements in the cohomology of the BRST differential modulo the extended differential itself. This cohomology is isomorphic to the cohomology of the extended differential $d + (\epsilonb Q)  + i_{i(\epsilonb \gamma \epsilon)}$ in the complex of gauge-invariant extended form functions of the fields modulo the equations of motion \cite{Henneaux}, enlarged by the Chern--Simons-like cochain that can occur owing to the non-triviality of the cocycle (\ref{DeRham}). The latter can be derived from the extended curvature definition (\ref{Curvature}) and the usual Chern--Simons formula\footnote{Note that the shadow field $c$ plays the role of the gauge potential in the Grassmann odd direction in the ``ectoplasm'' formalism.}
\begin{multline}
( d + \Qsla_{|\Sigma} + i_{i(\epsilonb \gamma \epsilon)} )  \trace   \biggl( \scal{A + c } \Scal{ F  + i \scal{\epsilonb \gamma_1 \lambda } +  ( \epsilonb \tau_I \epsilon ) \phi^I } - \frac{1}{3} \scal{A + c }^3 \biggr) \\*
= \trace   \Scal{ F  + i \scal{\epsilonb \gamma_1 \lambda } +  ( \epsilonb \tau_I \epsilon ) \phi^I }^2
\label{ChernSimons} \end{multline}

Since the action of $ \Qsla_{|\Sigma}$ on a functional of the physical fields is linear in the supersymmetry parameter, each local functional of the fields left invariant by supersymmetry defines a non-trivial cocycle.

Let us consider a cocycle with a vanishing form of maximal form-degree $D$; the non-zero form of highest form-degree $p < D$ then verifies $d \cL_p= 0$. The algebraic Poincar\'e lemma then implies that $\cL_p$ can be written as $d \Psi_{p-1}$. The next equation then gives
 \be  \Qsla_{|\Sigma}  d \Psi_{p-1} + d \cL_{p-1} = d \scal{\cL_{p-1} -  \Qsla_{|\Sigma} \Psi_{p-1} } = 0 \ee
in such way that there exists $\Psi_{p-2}$ such that $\cL_{p-1} = \Qsla_{|\Sigma} \Psi_{p-1} + d \Psi_{p-2}$. One obtains by iteration a complete extended form $\tilde \Psi  = \Psi_{p-1} + \Psi_{p-2} + \cdots$ that defines an antecedent of the cocycle $\tilde \cL$
\be \tilde \cL =  (d + \Qsla_{|\Sigma}+ i_{i(\epsilonb \gamma \epsilon)} ) \tilde \Psi\ . \ee

We thus conclude that the non-trivial cocycles are in one-to-one correspondence with the supersymmetric counterterms.

We now wish to consider the inclusion of densities corresponding to the possible invariant counterterms together with corresponding sources in the action, in order to study their insertions into the 1PI generating functional. Before doing so, it is important first to state the property of the Chern--Simons like cochains to be renormalised only by gauge invariant cochains which are functions of the physical fields. This can be understood intuitively from the extension of the corresponding result for the ordinary Chern--Simons function \cite{KellyChernSimons} to superspace. In the algebraic method, this can be proven in the Landau gauge by making use of additional so-called ghost Ward identities \cite{GuillaumeN4}. The ghost Ward identities imply that the operators that renormalise the defining operators of the Chern--Simons-like cochain (\ref{ChernSimons}) cannot depend on the shadow field $c$. Then the Ward identities imply that these operators must in fact be gauge invariant functions of the physical fields.

We now define an infinite basis of non-trivial solutions $\tilde \cL^\ord{a}$ associated to each invariant counterterm $I_a$. The action of the Callan--Symanzik operator on an insertion of the cocycle $\tilde \cL^\ord{a}$ is determined by the supersymmetry Ward identity to be of the form
\be {\cal C} \, \bigl[ \tilde \cL^\ord{a} \cdot \Gamma] =  \sum_b \gamma_\gra{a}{b}    \bigl[ \tilde \cL^\ord{b} \cdot \Gamma]  + \bigl[(d+  \Qsla_{|\Sigma}+ i_{i(\epsilonb \gamma \epsilon)} ) \tilde \psi^\ord{a} \cdot \Gamma \bigr]+ \Slav_{|\Gamma} \bigl[ \Omega^\ord{a} \cdot \Gamma \bigr]  \label{CocycleRenormalisation} \ee
in such a way that the renormalisation of each component of the cocycle is related to the renormalisations of all the other components of the cocycle.

One must, however, take care of the fact that the transformations of the sources under the action of the linearised Slavnov--Taylor operator may be modified by quantum corrections. These modifications occur in perturbation theory as anomalies in the supersymmetry Ward identity which are linear in the sources.
\be \Qsla (\Gamma[u]) = g^{2 (n_{\mathcal{A}}-1)} \int \tilde u_{\, \wedge } \tilde \Delta + \mathcal{O}(g^{2 n_{\mathcal{A}}} ) \ee
The consistency conditions then imply that $\tilde \Delta$ defines a cocycle of form-degree plus shadow number equal to $(D+1)$,
\be (d + \Qsla_{|\Sigma}+ i_{i(\epsilonb \gamma \epsilon)} )  \tilde \Delta = 0 \ .\ee
If $\tilde \Delta$ turns out to be trivial, then the anomaly can be absorbed into an appropriate non-invariant counterterm. In the same way that non trivial cocycles of form-degree plus shadow number $D$ are in one-to-one correspondence with the invariant counterterms, the non-trivial cocycles of form-degree plus shadow number $(D+1)$ are in one-to-one correspondence with the non-trivial anomalies of supersymmetry, -- these would be local functionals linear in $\epsilon$ that are invariant modulo the equations of motion under the action of $(\epsilonb Q)$. We have assumed from the start that such anomalies do not exist.

The meaning of this result is that, although the realisation of the supersymmetry algebra on a supermultiplet of composite operators is generally modified perturbatively by quantum corrections, the realisation of the supersymmetry algebra on the part of a multiplet associated to a particular supersymmetry invariant defined by a given cocycle is exact at the tree level. This property allows one to relate the various anomalous dimensions of the composite operators despite the fact that the supersymmetry realisation is non-linear.

\subsection{Allowed counterterms}

We will first describe the cocycle associated to the classical action $S$, which turns out to involve the Chern--Simons-like cochain (\ref{ChernSimons}) in $D>4$ dimensions. Indeed, the $(D-1)$-form of the cocycle involves then an element of the form
\be \cL^\ord{\rm c}_{D-1} = \frac{3i}{2 } \star \trace   \scal{\epsilonb  \gamma_{\mu\nu\sigma} \lambda}^{\mu\nu} dx^\sigma + \cdots \ee
in such a way that
\be   i_{i(\epsilonb \gamma \epsilon)} \cL_D^\ord{\rm c} +  \Qsla_{|\Sigma}  \cL_{D-1}^\ord{\rm c} =  \frac{i}{2}  (\epsilonb \gamma_{D+1} \gamma_{D-5}  \epsilon) \, \trace   F_{\, \wedge} F + d \, ( \cdots)  \ee
where $\gamma_{D+1}$ is the identity for odd $D$ or the product of all gamma's for even $D$. $\gamma_{D-5}$ is the $(D-5)$-form built from the antisymmetric product of $D-5$ gamma matrices. This term gives rise to a cochain of the form  (\ref{ChernSimons}) in dimension $4<n \le 8$. Moreover, by power counting the last $(D-5)$-form associated to (\ref{ChernSimons}) defines the last form of the cocycle $\tilde \cL^\ord{\rm c}$. So as a result
\be \cL^\ord{\rm c}_{D-5} = \frac{i}{2}  (\epsilonb \gamma_{D+1} \gamma_{D-5}  \epsilon) \,  \trace   \Scal{ (\epsilonb \tau_I \epsilon ) \, c \, \phi^I - \frac{1}{3} c^3 } \ee
The Ward identities allow the operator  $\trace   \Scal{ (\epsilonb \tau_I \epsilon ) \, c \, \phi^I - \frac{1}{3} c^3 }$ to be renormalised by any scalar gauge-invariant operator cubic in the spinor parameter that has the right power counting. Its renormalisation is in fact directly related to the renormalisation of the first component of the supercurrent (\ref{3.33}), \ie the $\ft12$ BPS operator \cite{GuillaumeN4}
\be \trace   \Scal{\phi^I \phi^J  - \frac{1}{n}  \delta^{IJ} \phi_K \phi^K } \ee
which always appears in the component $ \cL^\ord{\rm c}_{D-4}$.

It turns out that in order to be eligible to renormalise the cocycle associated to the classical action, a cocycle must admit a cohomologically equivalent representative whose lowest form degree is at least $D-5$, and such that the corresponding form (if non-zero) includes an $(\epsilonb \gamma_{D+1} \gamma_{D-5}  \epsilon)$ factor needed to renormalise $\cL^\ord{\rm c}_{D-5}$.

Let us now consider the cocycle associated to the $\ft12$ BPS invariants $I_{\stfrac{1}{2}}$. It is much easier to consider the eight-dimensional case, although the result will be valid in any dimension. In eight dimensions the R-symmetry group reduces to an axial $U(1)$ and one considers chiral and antichiral components of $Q$ written $Q_+$ and $Q_-$. Then $I_{\stfrac{1}{2}}$ reduces to
\be  I_{\stfrac{1}{2}} = \int d^D x \Scal{ Q_+^8 \trace   \bar \Phi^4 +  Q_-^8 \trace   \Phi^4 } \ee
where
\be Q_+^8 \equiv \frac{1}{8!} \varepsilon_{{\alpha} {\beta} { \gamma} { \delta} { \eta} {\chi} { \xi} {\zeta}} Q^\alpha Q^{{\beta}} Q^{{\gamma}} Q^{{\delta}} Q^{{\eta}} Q^{{\chi}} Q^{{\xi}} Q^{{\zeta}} \hspace{10mm} Q_-^8 \equiv \frac{1}{8!} \varepsilon_{\dot{\alpha} \dot{\beta} \dot{ \gamma} \dot{ \delta} \dot{ \eta} \dot{\chi} \dot{ \xi} \dot{\zeta}} Q^{\dot{\alpha}} Q^{\dot{\beta}} Q^{\dot{\gamma}} Q^{\dot{\delta}} Q^{\dot{\eta}} Q^{\dot{\chi}} Q^{\dot{\xi}} Q^{\dot{\zeta}} \ee

Let us first consider the right-hand-side of the action
\be (\epsilonb Q) \,  Q_-^8 \trace   \Phi^4 =  (\epsilonb_+ Q_+) \,  Q_-^8 \trace    \Phi^4 = - \partial_\mu  [ i \epsilonb_+ \gamma^\mu ]^{\dot{\alpha}} \frac{1}{7!} \varepsilon_{\dot{\alpha} \dot{\beta} \dot{ \gamma} \dot{ \delta} \dot{ \eta} \dot{\chi} \dot{ \xi} \dot{\zeta}} Q^{\dot{\beta}} Q^{\dot{\gamma}} Q^{\dot{\delta}} Q^{\dot{\eta}} Q^{\dot{\chi}} Q^{\dot{\xi}} Q^{\dot{\zeta}} \   \trace    \Phi^4\ . \ee
Applying then $(\epsilonb Q) $ to the next component, one has
\begin{multline}  (\epsilonb Q) [ i \epsilonb_+ \gamma^\mu ]^{\dot{\alpha}} \frac{1}{7!} \varepsilon_{\dot{\alpha} \dot{\beta} \dot{ \gamma} \dot{ \delta} \dot{ \eta} \dot{\chi} \dot{ \xi} \dot{\zeta}} Q^{\dot{\beta}} Q^{\dot{\gamma}} Q^{\dot{\delta}} Q^{\dot{\eta}} Q^{\dot{\chi}} Q^{\dot{\xi}} Q^{\dot{\zeta}} \   \trace    \Phi^4 \\ = i ( \epsilonb \gamma^\mu \epsilon ) Q_-^8 \trace   \Phi^4 - \partial_\nu   [ i \epsilonb_+ \gamma^\mu ]^{\dot{\alpha}}   [ i \epsilonb_+ \gamma^\nu ]^{\dot{\beta}} \frac{1}{6!} \varepsilon_{\dot{\alpha} \dot{\beta} \dot{ \gamma} \dot{ \delta} \dot{ \eta} \dot{\chi} \dot{ \xi} \dot{\zeta}}  Q^{\dot{\gamma}} Q^{\dot{\delta}} Q^{\dot{\eta}} Q^{\dot{\chi}} Q^{\dot{\xi}} Q^{\dot{\zeta}} \   \trace    \Phi^4 \ .\end{multline}
Iteratively, one thus deduces the complete cocycle,
\be \star^* [ Q_- + 2  i \gamma_\mu dx^\mu  \epsilon_+  ]^8 \, \trace   \Phi^4 \label{cocycle8} \ ,\ee
where $\star^*$ is defined to act as $(-1)^p$ times the Hodge star operator $\star$ on a $p$-form. This is equivalent to saying that $[ Q_- + 2 i \gamma_\mu dx^\mu  \epsilon_+  ]^8 \, \trace   \Phi^4$ is invariant under the Hodge dual extended differential $d^\star + ( \epsilonb Q) + i ( \epsilonb \gamma_\mu \epsilon ) dx^\mu $, where $d^\star$ is defined in such a way that
\be \{ d , d^\star \} = \partial_\mu \partial^\mu\ . \ee
To prove that (\ref{cocycle8}) is actually a cocycle, we decompose the extended differential into
\be d^\star + ( \epsilonb Q) + i ( \epsilonb \gamma_\mu \epsilon ) dx^\mu = d^\star + (\epsilonb_+ Q_+) + \scal{ \epsilonb_- [ Q_- + 2  i \gamma_\mu dx^\mu  \epsilon_+ ] } \ ;\ee
then the right-hand-side trivially cancels $[ Q_- + 2 i \gamma_\mu dx^\mu  \epsilon_+  ]^8 \, \trace   \Phi^4$ because there are only eight independent anticommuting operators $Q_- + 2  i \gamma_\mu dx^\mu  \epsilon_+$. Because $\trace   \Phi^4$ is a scalar chiral operator, one has
\be \scal{ d^\star + (\epsilonb_+ Q_+) } \trace   \Phi^4 = 0 \ee
and moreover
\be \bigl\{ d^\star + (\epsilonb_+ Q_+)  \, , \, Q_- + 2  i \gamma_\mu dx^\mu  \epsilon_+ \bigr\} = \bigl\{  d^\star \, , \,  2  i \gamma_\mu dx^\mu  \epsilon_+ \bigr\} + \bigl\{  (\epsilonb_+ Q_+)  \, , \, Q_- \bigr\} = 0 \ee
as a consequence of the supersymmetry algebra.

The last $0$-form of the cocycle (\ref{cocycle8}) can't be cancelled by the addition of a cohomologically trivial cocycle to (\ref{cocycle8}), because it is a $0$-form and $\trace \Phi^2$ is a chiral primary operator (\ie  $\trace \Phi^2$ does not appear in the $Q_-$ variation of any other gauge-invariant operator). 

The cocycle (\ref{cocycle8}) is complex, and the cocycle associated to the $I_{\stfrac{1}{2}}$ invariant is its real part, whereas its imaginary part corresponds to the fourth Chern character. The arguments go exactly the same way for the double trace invariant $I^\prime_{\stfrac{1}{2}}$.

For the non-renormalisation theorems, we are interested in the $D=5$ and $D=6$ cocycles. As a matter of fact, the properties of the cocycles associated to $\ft12$ BPS operators that we have just exhibited in eight dimensions are valid in any dimension.

The last form associated to the classical action in six dimensions is a $1$-form, whereas the last forms associated to the invariants $I_{\stfrac{1}{2}}$ and $ I_{\stfrac{1}{2}}^\prime$ are $0$-forms. The irreducible component of the $0$-form associated to $I_{\stfrac{1}{2}}$ depends on the descendants of the corresponding $\ft12$ BPS operator
\bea Q \, \ft14 \trace   \Scal{\phi^{(I} \phi^{J} \phi^{K} \phi^{L)}}  &=& - \epsilonb \tau^{(I} \, \traceS  \Scal{\lambda \phi^J \phi^K \phi^{L)} }_{{\bf20} \oplus{\bf 20}} \\*
Q \, \traceS \Scal{\lambda \phi^I \phi^J \phi^{K} }_{{\bf20} \oplus{\bf 20}}   &=& i \gamma^\mu \tau_d \epsilon \, \partial_\mu \ft14 \trace   \Scal{\phi^{(I}  \phi^{J} \phi^K \phi^{L)} } + \gamma^{\mu\nu} \epsilon \, \cO_{\mu\nu}^{IJK} +  \gamma^\mu \tau^{L} \epsilon  \, \cO_{\mu\, L}^{IJK} - \gamma_7 \epsilon \, \cO_7^{IJK} \nn
\eea
where the subscript ${{\bf20} \oplus{\bf 20}} $ denotes a restriction to the ${ \Yboxdim8pt  {\yng(3)}} \otimes { \Yboxdim8pt {\young(\bulletup \bulletup \bulletup \bulletup)}}
\oplus { \Yboxdim8pt  {\yng(4)}} \otimes  { \Yboxdim8pt {\young(\bulletup \bulletup \bulletup)}} $ representation of $Sp(1)\times Sp(1)$, the subscript $\mathfrak{S}$ stands for the symmetrised trace, and
\bea
{\cal O}^{IJK}_{\mu\nu} &\equiv& \traceS \Scal{ F_{\mu\nu} \phi^{(I} \phi^J \phi^{K)} + \sfrac{3}{4} \overline{\lambda} \gamma_{\mu\nu} \tau^{(I} \lambda \, \phi^J \phi^{K)} } \CR
{\cal O}_{\mu\, L}^{IJK} &\equiv& \traceS\Scal{  \scal{ i \phi_L D_\mu \phi^{(I} - i \phi^{(I} D_\mu \phi_L + \sfrac{3}{4} \overline{\lambda} \gamma_\mu {\tau_L}^{(I} \lambda} \phi^J \phi^{K)} }\CR
{\cal O}_7^{IJK} &\equiv& \traceS \Scal{ \scal{ \sfrac{1}{3} {\varepsilon^{(I}}_{LMN} \phi^L  \phi^M  \phi^N  + \frac{1}{8} \overline{\lambda} \gamma_7 \tau^{(I} \lambda } \phi^J \phi^{K)}} \ .
\eea
The $0$-form of the cocycle can therefore be written as
\begin{multline} \cL^\ord{\thalf}_0 = \mathpzc{a}\,  ( \epsilonb \gamma_7 \gamma^{\mu\nu} \tau_I \epsilon) ( \epsilonb \tau_J \epsilon) ( \epsilonb \tau_K \epsilon) {\cal O}^{IJK}_{\mu\nu}  \\*+ \mathpzc{b}\,  ( \epsilonb \gamma^{\mu\nu\sigma} {\tau_{I}}^L \epsilon) ( \epsilonb \gamma_7 \gamma_{\nu\sigma} \tau_J \epsilon) ( \epsilon \tau_K \epsilon) \cO^{IJK}_{\mu\, L} + \mathpzc{c}\,  (\epsilonb \tau_I \epsilon)  (\epsilonb \tau_J \epsilon)  (\epsilonb \tau_K \epsilon)  \cO^{IJK}_7 \end{multline}
for some coefficients $\mathpzc{a},\, \mathpzc{b}$ and $\mathpzc{c}$. It follows from $\tau^{(I} \tau^{J)}= 0$ that such a form can neither be written as a contraction with $i(\epsilonb \gamma^\mu \epsilon)$ nor as the $Q$ variation of a form built from the operator $ \traceS \scal{\lambda \phi^I \phi^J \phi^{K} }_{{\bf20} \oplus{\bf 20}} $. This can be understood directly from the irreducible representations of $Sp(1)\times Sp(1)$. The $0$-form
$\cL^\ord{\thalf}_0$  is built from operators in the $( { \Yboxdim8pt  {\yng(3)}} \otimes { \Yboxdim8pt {\young(\bulletup \bulletup \bulletup)}})_+$ representation\footnote{The subscript $+$ means that it is the self-dual complex representation which corresponds to a real representation of $SO(10-D)$.} and the $({ \Yboxdim8pt  {\yng(2)}} \otimes { \Yboxdim8pt {\young(\bulletup \bulletup \bulletup \bulletup)}})_+  \oplus ({ \Yboxdim8pt  {\yng(4)}} \otimes { \Yboxdim8pt {\young(\bulletup \bulletup)}})_+$ representation, from which it follows that $\cL^\ord{\thalf}_0$ is not built from a contraction with $i(\epsilonb \gamma^\mu \epsilon)$. In order for the $0$-form possibly  to be $Q$-exact, these operators would have to appear in the supersymmetry variation  of an operator in the ${ \Yboxdim8pt  {\yng(2)}} \otimes { \Yboxdim8pt {\young(\bulletup \bulletup \bulletup)}}  \oplus { \Yboxdim8pt  {\yng(3)}} \otimes { \Yboxdim8pt {\young(\bulletup \bulletup)}}$, whereas they appear only in the supersymmetry variation of an operator in the ${ \Yboxdim8pt  {\yng(3)}} \otimes { \Yboxdim8pt {\young(\bulletup \bulletup \bulletup \bulletup)}}
\oplus { \Yboxdim8pt  {\yng(4)}} \otimes  { \Yboxdim8pt {\young(\bulletup \bulletup \bulletup)}} $.

In a similar way, the $0$-form associated with $I_\stfrac{1}{2}$ in five dimensions is built from operators in the $ { \Yboxdim6pt  {\yng(3,2)}}_+$ and the ${ \Yboxdim6pt  {\yng(4,1)}}_+$ of $Sp(2)$. Once again this implies that $I_\stfrac{1}{2}$ cannot be written as a contraction with $i(\epsilonb \gamma^\mu \epsilon)$. In order for the $0$-form to be $Q$-exact, these operators would have to appear in the supersymmetry variation of an operator in the ${ \Yboxdim6pt  {\yng(3,1)}}_+$, whereas they only appear in the supersymmetry variation of operators in the ${ \Yboxdim6pt  {\yng(3,3)}}_+$ and the ${ \Yboxdim6pt  {\yng(4,2)}}_+$. On the other hand,  the last form associated to the classical action is given by the $0$-form
\be \cL^\ord{\rm c}_0 = \frac{1}{2} (\epsilonb \epsilon) \trace   \Scal{ (\epsilonb \tau_I \epsilon ) \, c \, \phi^I - \frac{1}{3}c^3 } \ee
and it is a property of the representations $ { \Yboxdim6pt  {\yng(3,2)}}_+$ and ${ \Yboxdim6pt  {\yng(4,1)}}_+$ that one can not extract a scalar $(\epsilonb  \epsilon)$ from $ \cL^\ord{\thalf}_0$.

There is therefore no way for the cocycles $\tilde \cL^\ord{\thalf}$ or $\tilde \cL^{\ord{\thalf}\prime}$ to contribute to the renormalisation of $\tilde \cL^\ord{\rm c}$ and we conclude that the logarithmic divergences associated to the $\ft 12$ BPS invariants are forbidden by the supersymmetry Ward identities in dimensions five and six.

Note that although this discussion of the $\ft12$ BPS invariants is rather general, it does not extend to invariants associated to lesser BPS operators such as the $\ft14$ BPS invariant $I_{\stfrac{1}{4}}$. One can formally understand why this is so from the representations associated to the invariants. One may use Young tableaux for the representations of the internal symmetry group in which the operators defining a given form $\cL_p$ of a cocycle $\tilde \cL$ lie, where each box corresponds to the fundamental representation carried by the spinor parameter \cite{superactions}. It then follows that one box is either removed or added at each step of the descent equations (\ie by getting $\cL_{p\pm1}$ from $\cL_{p}$). It turns out that the representations in which the $\ft12$ BPS operators associated to $I_\stfrac{1}{2}$ and $I_\stfrac{1}{2}^\prime$ lie correspond to Young Tableaux with too many boxes to be related to the Chern--Simons operators associated to the classical action. However, one does not find such obstruction for the lesser BPS invariants. The idea is that there is a subset of the operators defining the $\ft14$ BPS multiplet which transform into each other in the same way as the operators defining $\tilde \cL^\ord{\rm c}$ do, which implies in turn that the supersymmetry Ward identities do not protect $\tilde \cL^\ord{\rm c}$ from being renormalised by $\cL^\ord{\tquarter}_0$.

In five dimensions, the Chern--Simons operator defining the $0$-form $\cL^\ord{\rm c}_0$ is in an unspecified representation of rank three of $Sp(2)$, \ie ${ \Yboxdim8pt  {\yng(1)}} \otimes { \Yboxdim8pt  {\yng(1)}}  \otimes { \Yboxdim8pt  {\yng(1)}}$, whereas the $\ft12$ BPS operator from which $ \cL^\ord{\thalf}_0$ is derived can only lead to a representation of rank five. However, the $\ft14$ BPS operator lies in the representation ${ \Yboxdim5pt  {\yng(4,3,1)}}_+$, which can lead in principle to a $0$-form $\cL^\ord{\tquarter}_0$ which is in the fundamental representation ${ \Yboxdim8pt  {\yng(1)}} $.

Similarly in six dimensions, the Chern--Simons operator defining the $1$-form $\cL^\ord{\rm c}_1$ is also in an unspecified representation of rank three of $Sp(1)\times Sp(1)$, whereas the $\ft12$ BPS operator from which $ \cL^\ord{\thalf}_1$ is derived can only lead to a representation of rank five. The $\ft14$ BPS operators lie in the representations ${ \Yboxdim8pt  {\yng(4)}}_+$ and $ { \Yboxdim8pt {\young(\bulletup \bulletup \bulletup \bulletup)}}_+$ which can in principle lead to  $1$-forms $\cL^\ord{\tquarter}_1$ which are in the fundamental representations ${ \Yboxdim8pt  {\yng(1)}}$ and $ { \Yboxdim8pt {\young(\bulletup)}} $, respectively.

The seven-dimensional case is a bit trickier to discuss because there is no available tool such as the Young tableaux to discuss tensor products of the spinor representation of $Spin(6,1)$, and the representations of $Sp(1)$ do not permit one to characterise the cocycle $\tilde \cL^\ord{\tquarter}$. Indeed, the latter cocycle is no longer related to a Lorentz scalar $\ft14$ BPS operator in dimensions $7$ and $8$, but rather to $\ft38$ BPS primary operators in the $2$-form representation of the Lorentz group. For simplicity, we will discuss these operators in eight dimensions, in which case they are defined as follows
\be  \mathcal{O}^\stfrac{3}{8}_{\mu\nu} \equiv  \trace \overline{\lambda}_- \gamma_{\mu\nu} \lambda_-  \, \trace \Phi^2 \hspace{10mm}  \overline{\mathcal{O}}^\stfrac{3}{8}_{\mu\nu} \equiv  \trace \overline{\lambda}_+ \gamma_{\mu\nu} \lambda_+  \, \trace {\bar \Phi}^2 \ee
One checks that $\overline{Q}_+ \gamma^{\mu\nu} Q_+  \mathcal{O}^\stfrac{3}{8}_{\mu\nu}  $ is a chiral operator
\be Q_-  \   \overline{Q}_+ \gamma^{\mu\nu} Q_+  \mathcal{O}^\stfrac{3}{8}_{\mu\nu} = 0 \ee
and so one can get a representative of the corresponding cocycle just as in the case of the $\ft12$ BPS operator (\ref{cocycle8}), \ie
\be \tilde \cL^\ord{\tquarter}  =  \star^* [ Q_- + 2  i \gamma_\mu dx^\mu  \epsilon_+  ]^8 \, \trace \overline{Q}_+ \gamma^{\mu\nu} Q_+  \mathcal{O}^\stfrac{3}{8}_{\mu\nu}   +  \star^* [ Q_+ + 2  i \gamma_\mu dx^\mu  \epsilon_-  ]^8 \, \trace \overline{Q}_- \gamma^{\mu\nu} Q_-  \overline{\mathcal{O}}^\stfrac{3}{8}_{\mu\nu}  \ee
The invariant $I_0$ is related to similar $\ft38$ BPS (non-primary) operators, \ie
\be  \mathcal{O}^{\stfrac{3}{8}\prime}_{\mu\nu} \equiv  \trace \overline{\lambda}_- \gamma_{\mu\nu} [ \Phi , [ \Phi, \lambda_- ] ] = [{Q_+}]^6_{\mu\nu}\,  \trace \Phi \bar \Phi
\ee
and its complex conjugate. Note that these manifestly vanish in the abelian case. The cocycle $\tilde \cL^\ord{\tKo}$ can thus be written as well as
\be \tilde \cL^\ord{\tKo}  =  \star^* [ Q_- + 2  i \gamma_\mu dx^\mu  \epsilon_+  ]^8 \, \trace \overline{Q}_+ \gamma^{\mu\nu} Q_+  \mathcal{O}^{\stfrac{3}{8}\prime}_{\mu\nu}   +  \star^* [ Q_+ + 2  i \gamma_\mu dx^\mu  \epsilon_-  ]^8 \, \trace \overline{Q}_- \gamma^{\mu\nu} Q_-  \overline{\mathcal{O}}^{\stfrac{3}{8}\prime}_{\mu\nu}  \ee
One may then conclude too naively that such invariants are protected, but these representatives of the cocycles $\cL^\ord{\tquarter}$ and $\cL^\ord{\tKo}$ are in fact cohomologically equivalent to much shorter cocycles.

In order to exhibit this fact, let us first introduce Young tableaux for the tensor products of the Weyl spinor representation of $Pin(7,1)$. We want to keep the $\mathds{Z}_2$ chiral symmetry because it plays a rather important role in this case. The centre of $Spin(7,1)$ is $\mathds{Z}_2 \times \mathds{Z}_2$, and one can obtain two different real forms of $SO(8,\mathds{C})$ by taking the quotient with respect to one $\mathds{Z}_2$ or to the other, namely $SO(7,1)$ and $SO^*(8)_\ord{-14}$. $SO^*(8)_\ord{-14}$ is the real form of $SO(8,\mathds{C})$ defined in such a way that its maximal compact subgroup is $Spin(7)\subset SO(8)\subset SO(8,\mathds{C})$. By triality, a Weyl spinor of $Spin(7,1)$ is a complex vector of $SO^*(8)_\ord{-14}$. The Young tableaux of $O^*(8)_\ord{-14} \cong Pin(7,1)/ \mathds{Z}_2 $ are pretty much the same as the ones of $O(8,\mathds{C})$, the only difference being that there is a complex self-duality condition on the ${ \Yboxdim6pt  {\yng(1,1)}}$ that involves the $Spin(7)$-invariant octonionic $4$-form of $O^*(8)_\ord{-14}$.

The Chern--Simons-like operator defining  the $3$-form $\cL^\ord{\rm c}_3$ is in an unspecified  representation of rank three of $O^*(8)_\ord{-14}$, whereas the $\ft12$ BPS chiral operators are in the ${ \Yboxdim3pt  {\yng(1,1,1,1,1,1,1,1)}}$ representation, and can only lead to an operator in the ${ \Yboxdim3pt  {\yng(1,1,1,1,1)}}$ representation for the $3$-form  $ \cL^\ord{\thalf}_3$. On the other hand, the $\ft38$ operators  $\mathcal{O}^\stfrac{3}{8}_{\mu\nu}$ and $\mathcal{O}^{\stfrac{3}{8}\prime}_{\mu\nu}$ are in the ${ \Yboxdim3pt  {\yng(1,1,1,1,1,1,1)}}$ representation, and can lead in principle to $3$-forms $\cL^\ord{\tquarter}_3$ and $\cL^\ord{\tKo}_3$ respectively, which corresponding operators are in the fundamental ${ \Yboxdim8pt  {\yng(1)}}$.

The conclusion of this section is that, independently of the dimension $5\le D \le 8$, the supersymmetry Ward identities imply that the cocycle $\tilde \cL^\ord{\rm c}$ can not be renormalised in perturbation theory by the cocycles $\tilde \cL^\ord{\thalf}$ and $\tilde \cL^{\ord{\thalf}\prime}$, but that they do not prevent the cocycle $\tilde \cL^\ord{\rm c}$ to be renormalised by the cocycle $\tilde \cL^\ord{\tquarter}$ and $\tilde \cL^\ord{\tKo}$. It then follows from (\ref{CallanRelations}) and (\ref{CocycleRenormalisation}) that the supersymmetry Ward identities rule out the potential $4$-loop (respectively $2$-loop) logarithmic divergences in five (respectively six) dimensions, but they are not in contradiction with any other logarithmic divergences allowed by power counting and with the existence of a corresponding supersymmetry invariant (\eg $I_\stfrac{1}{4}$ or $I_\Ko$).

\subsection{Maximal supergravity}

Most of this discussion can be generalised to maximal supergravity, although the method has to be improved since supergravity fields do not simply lie in representations of the superPoincar\'e algebra. Maximal supergravity admits as a gauge symmetry the $\cN=8$ superalgebra, which has a bosonic subalgebra containing the direct sum of infinitesimal four-dimensional diffeomorphisms together with $\mathfrak{sl}(2,\mathds{C}) \oplus \mathfrak{su}(8)\oplus \bigoplus_{n=1}^{28} \mathfrak{u}(1)$. The fields of the theory are given by the vierbeins $e^a$, the gravitino fields  $\psi^i_\alpha$ which are $1$-form valued in the $SL(2,\mathds{C}) \times SU(8)$ fundamental representation, $28$ abelian gauge fields $A^{ij}$, the dilatino fields $\chi^{ijk}$ in the product of the rank three antisymmetric $SU(8)$ tensor representation and the fundamental representation of $SL(2,\mathds{C})$, as well as $70$ scalar fields $\mathcal{V}$ lying in the coset space $E_{7(7)} / ( SU(8) / \mathds{Z}_2)$. All the gauge invariances of supergravity, including local supersymmetry, can be represented by a single BRST operator $\BRST$. Any invariant local functional  $S^\ord{n}$ which does not involve a Chern--Simons-like term with respect to the internal gauge symmetry (\ie $SL(2,\mathds{C}) \times SU(8) \times U(1)^{28}$), leads to an extended cocycle satisfying \cite{mieg,marc}
\be ( d + \BRST- \cL_\xi - i_{i(\epsilonb \gamma \epsilon) } ) \tilde \cL = 0 \ee
where $\xi^\mu$ is the anticommuting ghost associated to the diffeomorphisms and $\epsilon^i$ is the commuting ghost associated to local supersymmetry. Their BRST variation are given by
\be \BRST\xi^\mu = \xi^\nu \partial_\nu \xi^\mu  - 2 i  e_a^\mu \scal{ \epsilonb_i  \gamma^a \epsilon^i } \hspace{10mm} \BRST\epsilon^i = \cL_\xi \epsilon^i - \frac{1}{2} \saa \Omega \epsilon^i - {C^i}_j \epsilon^j + i_{i (\epsilonb  \gamma \epsilon) } \psi^i  + ( \epsilonb_j \epsilon_k ) \chi^{ijk} \ee
where $\Omega_{ab}$ and $ {C^i}_j$ are the Faddeev--Popov ghosts associated to Lorentz  and $SU(8)$ invariances respectively. If one inserts the composite operators appearing in a given cocycle $\tilde \cL$ by the introduction of sources $\tilde u$ as in (\ref{SourceCoupling}), one obtains that the sources must also transform as a cocycle with respect to the Slavnov--Taylor BRST operator
\be ( d + \Slav_{|\Sigma}  - \cL_\xi - i_{i(\epsilonb \gamma \epsilon) } ) \tilde u = 0 \ee
The main difference between Yang--Mills theory and supergravity is that the transformations of the sources depend on the quantum fields in supergravity. The proof of the absence of deformations of the Ward identity at the quantum level is then more tricky, but we will nonetheless assume this to be the case. Since this is related to the renormalisation of the ghosts, it is probably only true in particular linear gauges subject to additional ghost Ward identities. Assuming the absence of such deformations, it follows that the composite operators that define a cocycle $\tilde \cL$ only mix under renormalisation with composite operators defining themselves cocycles of the same shape (\ie which have same number of components and the same tensor structure of the last component), and with composite operators that define trivial cocycles. Since trivial cocycles give no contribution to the associated invariant counterterms, the invariant counterterms that are consistent with the supersymmetry Ward identity define cocycles which are of the same shape as the one obtained from the classical action.

At tree level, the operator $\Slav_{|\Sigma}  - \cL_\xi$ acts on composite operators invariant with respect to the internal gauge symmetry as a supersymmetry transformation with parameter given by the supersymmetry ghost $\epsilon^i$, and it acts on the latter as
\be ( \Slav_{|\Sigma} - \cL_\xi )  \epsilon^i =  i_{i (\epsilonb  \gamma \epsilon) } \psi^i + ( \epsilonb_j \epsilon_k ) \chi^{ijk}+ \cdots \ee
up to internal gauge transformations.

The gravity coupling constant $\kappa$ is of dimension $-1$, with the consequence that the theory is strictly non-renormalisable. We consider that all fields have been rescaled by a factor of $\kappa$, in such a way that all bosonic fields are of canonical dimension zero and all fermionic fields are of canonical dimension $\ft12$. Then all the non-linear supersymmetry transformations do not depend upon $\kappa$, and the coupling constant only appears as an overall factor in front of the classical action as
\be \Sigma = \frac{1}{\kappa^2} S + z_\ord{\thalf} \kappa^4 I_{\stfrac{1}{2}} + \cdots \ee
where the dots stand for the source terms and possible higher order terms required to renormalise the theory. $S$ is the classical action, and $I_{\stfrac{1}{2}}$ is the invariant of canonical dimension eight that corresponds by power counting to the $3$-loop logarithmic divergences. At the linearised level, the $\cN=8$ superfield is a scalar superfields in the $ { \Yboxdim4pt  {\yng(1,1,1,1)}}_+$ representation of $SU(8)$
\be W^{ijkl} = \frac{1}{24} \varepsilon^{ijklmnop} W_{mnop} \ .\ee

The algebraic method is a bit more subtle when dealing with gravity theories. The main difference is that, although the Lagrange density itself is gauge invariant in Yang--Mills theory, the Lagrange density is only gauge invariant up to an exact derivative in a theory invariant under diffeomorphisms. The same is true for local supersymmetry. It follows that the densities $\cL^\ord{n}$ that we shall consider as insertions are not strictly speaking BRST invariant, but only satisfy
\be \Slav_{|\Sigma} \, \cL^\ord{n} = d  \Omega^\ord{n} \ .\ee
In fact, the classical action $S$ evaluated on field configurations satisfying the equations of motion is known to vanish in pure supergravity theory. It follows that, if the corresponding $4$-form density were BRST invariant, it would be BRST-exact and could not be renormalised by non-trivial composite operators. However, the Lagrangian density is not in fact BRST invariant, so this unduely strong conclusion does not obtain.

We now explain why the cocycle associated to the classical action is short. Power counting shows that the $0$-form component of the cocycle $\tilde \cL^\ord{\rm c}$ associated to the classical action is of canonical dimension zero. There is no non-trivial cohomology class of the algebraic de Rham complex of form degree $5$ and therefore, $\cL^\ord{\rm c}_0$ must be invariant with respect with the internal gauge symmetry.  It follows that it can only depend on the scalar fields $\mathcal{V}$ and the supersymmetry ghosts $\epsilon$. Because of the descent equations
\be (   \Slav_{|\Sigma} - \cL_\xi) \, \cL^\ord{\rm c}_0 = - i_{i(\epsilonb \gamma \epsilon)} \cL^\ord{\rm c}_1 \label{LastForm} \ee
the 0-form component must be the contraction of a Lorentz invariant quartic in the spinor parameters $\epsilon$ together with a $\ft12$ BPS primary operator function of the scalar fields. In order for this to be non-trivial, it cannot possibly be written as a contraction with $(\epsilonb_i \gamma^a \epsilon^i)$. The only independent combinations are then
\be ( \epsilonb_i \epsilon_j) ( \epsilonb_k \epsilon_l ) \hspace{10mm} ( \epsilonb_{\{i} \gamma^{ab} \epsilon_j ) ( \epsilonb_k \gamma_{ab} \epsilon_{l\}} ) \hspace{10mm} ( \epsilonb_{[i} \gamma^a \epsilon^{[k} ) ( \epsilonb_{j]} \gamma_a \epsilon^{k]} ) \ee
The last term would be in contradiction with the parity of the Lagrange density, unless one were to include the determinant of the vierbein. However inspection of the supersymmetry transformations shows that such a composite operator cannot satisfy (\ref{LastForm}).  If one considers composite operators defined by the contraction of one of the two remaining combinations of supersymmetry ghosts together with a function of the scalar fields in the corresponding representation of $SU(8)$ (\ie ${ \Yboxdim6pt  {\yng(2,2)}}$ or ${ \Yboxdim6pt  {\yng(4)}}$), inspection of equation (\ref{LastForm}) shows that the corresponding operators should at least satisfy
\be (   \Slav_{|\Sigma} - \cL_\xi) \, { \cal X}^{ij,kl} = \epsilonb_\alpha^{[i} \Psi^{ j], kl\ \alpha} + \epsilonb^{\dot{\alpha}}_m \Psi^{[ijm], kl}_{\dot{\alpha}} \hspace{10mm}
(   \Slav_{|\Sigma} - \cL_\xi) \, { \cal X}^{ijkl} = \epsilonb_\alpha^{(i} \Psi^{ jkl) \ \alpha}  \label{ConBPS}
\ee
that is, schematically,
\be Q \  { \Yboxdim8pt  {\yng(2,2)}} =  { \Yboxdim8pt  {\yng(2,1)}} \oplus { \Yboxdim8pt  {\yng(2,2,1)}} \hspace{15mm}  Q \  { \Yboxdim8pt  {\yng(4)}} =  { \Yboxdim8pt  {\yng(3)}} \label{ConBPS2} \ee
The first constraint corresponds to the constraint for the ultra-short $\cN=4$ supercurrent and there is no such primary operator in $\cN=8$. Neither is there a chiral operator satisfying the second constraint. We conclude that there is no solution of (\ref{LastForm}) with the appropriate canonical dimension and the  last form of the cocycle associated to the classical Lagrangian is at least of form degree one.

The $I_{\stfrac{1}{2}}$ invariant is the integral of the $4$-form density $\cL^\ord{\thalf}$, which can be written at the linearised level as an integral of the $\ft12$ BPS operator $  { \Yboxdim4pt  {\yng(4,4,4,4)}}_+$ over the corresponding half superspace \cite{superactions},
\be \cL^\ord{\thalf}_{\rm Linear} = \star [ D^{16} ]_{{\Yboxdim2pt  {\yng(4,4,4,4)}}_+}  \, \scal{ W^4  }_{{\Yboxdim2pt  {\yng(4,4,4,4)}}_+} \ .\ee
It follows that the $0$-form of the cocycle associated to $I_{\stfrac{1}{2}}$ is given by operators in the ${ \Yboxdim6pt  {\yng(2,2)}}$ and the  ${ \Yboxdim6pt  {\yng(4)}}$ of $SU(8)$ which descend from the $\ft12$ BPS operator $  { \Yboxdim4pt  {\yng(4,4,4,4)}}_+$. In order for such a term to be $(\Slav_{|\Sigma} - \cL_\xi )$-exact, the corresponding operators would have to appear in the supersymmetry variations of operators in the ${ \Yboxdim6pt  {\yng(2,1)}}$ and the  ${ \Yboxdim6pt  {\yng(3)}}$\,, whereas they only appear in the supersymmetry variations of operators in the  ${ \Yboxdim6pt  {\yng(2,2,1)}}$, the  ${ \Yboxdim6pt  {\yng(3,2)}}$ and the  ${ \Yboxdim6pt  {\yng(4,1)}}$ of $SU(8)$ by properties of the $\ft12$ BPS supermultiplet. The cocycle associated to the $I_{\stfrac{1}{2}}$ invariant thus has non-trivial components of all form degrees, and it is not cohomologically equivalent to any shorter representative.

We thus conclude that the insertion of the classical action cannot be renormalised by the quartic invariant $I_{\stfrac{1}{2}}$, implying via the Callan--Symanzik equation that there is no three-loop logarithmic divergence associated with this invariant.

The arguments of the $\cN=8$ supergravity proof are very close to the ones for maximally supersymmetric Yang--Mills theory, so it is likely that the $I_{\stfrac{1}{2}}$ invariant is the only one to be eliminated as a counterterm candidate by the full supersymmetry Ward identities.

%%%%%%%%%%%%%%%%%%%%%%%%%%%%%%%%%%%%%%%%%%%%%%%%%%%%%%%%%%%%%%%%%%%

%%%%%%%%%%%%%%%%%%%%%%%%%%%%%%%%%%%%%%%%%%%%%%%%%%%%%%%%%%%%%%%%%%%

\section{Conclusions}

We have seen that quite different approaches to analysing the ultraviolet divergence structure of quantised maximal super Yang--Mills and maximal supergravity theories lead to similar expectations for the first loop order at which ultraviolet divergences may occur. Both approaches lead to the expectation that counterterm candidates corresponding to $\ft12$ BPS operator integrands are ruled out by the supersymmetric structure, but no further: $\ft14$ or lesser BPS operators, or non BPS operators (requiring full superspace integrals), have no apparent claim to being ruled out. This can be seen from various traditional ``off-shell'' approaches involving superspace Feynman rules, or equally well from the algebraic ``on-shell'' formalism which employs an outgrowth of the Batalin-Vilkovisky formalism to deal with Slavnov-Taylor identities for the full nonlinear supersymmetry of the theory. Although quite different in detail, both the off-shell and on-shell approaches agree completely with the known patterns of super Yang--Mills and supergravity ultraviolet divergences \cite{Bern:2007hh,Bern:1998ug,Bern:2008pv}.

In this article, we have reviewed the constraints on counterterms that arise in the various types of off-shell superspace formalisms. It is expected that any formalism that employs more than half of the full on-shell supersymmetry should be enough to rule out the $\ft12$ BPS counterterms. The minimal such formalisms have a ``half supersymmetry plus one'' structure with finite numbers of component fields:  9 supercharges for the maximal $\cN=4$ SYM case and 17 supercharges for the maximal $\cN=8$ supergravity theory. There exist also formalisms with yet more off-shell linearly realised supersymmetries, \eg the 12 supercharge $\leftrightarrow$ $\cN=3$ harmonic superspace formalism for maximal super Yang--Mills \cite{Galperin:1985uw,Delduc:1988cp}. Although full construction of the Feynman rules for maximal super Yang--Mills and maximal supergravity theories in a number of such off-shell formalisms remains to be completed, straightforward counting of the fermionic dimension structure leads to the expectation that these formalisms should be sufficient to rule out the $\ft12$ BPS counterterms. Moreover, the constraints on allowed divergences always appear to be the same: protection for $\ft12$ BPS operators, but no further. In the algebraic renormalisation approach, based on Slavnov-Taylor type identities for the full nonlinear extended supersymmetry plus gauge invariances of the maximal theories, one comes to a exactly the same conclusion. Thus, barring yet more ``miracles'' (which one may define technically as cancellations not explained by non-renormalisation theorems), the expectation must remain that counterterm candidates with less than $\ft12$ BPS structure or non-BPS candidates will correspond to actual divergences in the theories. This expectation is fully borne out by explicit computation in the case of maximal super Yang--Mills theory.

This leaves the subject with an apparent paradox. There are indications 
from string theory \cite{Berkovits:2006vc} which would, if applicable in 
the supergravity limit \cite{Green:2006yu}, suggest that $\cN=8$ 
supergravity might diverge first at nine loops. (This was foreshadowed 
by earlier string non-renormalisation theorems which made use of duality 
symmetries to restrict the loop order at which certain $R^4$ terms can 
appear \cite{Green:1997tv,Berkovits:1997pj,Green:1998by}.) One way of 
obtaining the nine-loop bound directly in field theory would be to 
suppose that counterterms have to be full
superspace integrals, \ie non-BPS, of gauge-invariant integrands, and 
that they have to respect gauge-invariance in the highest-possible 
dimension \cite{Stelle:1985nu}, \ie in $D=11$, although it is difficult to 
see how such a conclusion could be justified using known
field-theoretic methods. The stronger suggestion \cite{Green:2006gt}, that
supergravity might be finite to all orders, could also be argued if the 
divergence structure of maximal Yang-Mills and supergravity theories were
the same in all dimensions, a possibility that would be consistent with 
currently known explicit calculations (but so are the divergence expectations shown in Table \ref{tab1}).
Similar suggestions have also  been obtained from the {\em finite} structure of one-loop amplitudes in 
maximal supergravity \cite{Bern:2007hh}. How to square this with the 
rather unanimous results of the various off-shell and on-shell 
non-renormalisation theorems? One possibility might be a special status for 
diagrams with four external legs, which is the context of all the recent explicit 
UV divergence calculations. Could there be stronger results for this 
class of diagram? A possibly related observation is that BPS 
counterterms all start out at lowest order with four external fields 
\cite{Drummond:2003ex}. But this would still leave an infinite class of 
non-BPS counterterms to contend with, requiring from the Feynman diagram 
point of view an infinite set of miracles if full UV finiteness of $D=4$ 
maximal supergravity were to be achieved.

\subsection*{Acknowledgements}
We would like to thank Nathan Berkovits, Zvi Bern, Lance Dixon, Michael Green and Renata Kallosh  for discussions. The research of P.S.H.\ was supported in part by the EU under contract MRTN-2004-512194 and by STFC under rolling grant PP/C5071745/1. The research of K.S.S.\ was supported in part by the EU under contract MRTN-CT-2004-005104, by the STFC under rolling grant PP/D0744X/1 and by the Alexander von Humboldt Foundation through the award of a Research Prize. G.B. would like to thank Imperial College London and K.S.S. would like to thank the Albert Einstein Institute and CERN for hospitality during the course of the work.

%%%%%%%%%%%%%%%%%%%%%%%%%%%%%%%%%%%%%%%%%%%%%%%%%%%%%%%%%%%%%%%%%%%%

\end{document}